%% file: axnabla.tex
\documentclass{LMCS} 

\def\doi{8 (3:02) 2012}
\lmcsheading%
{\doi}
{1--76}
{}
{}
{Jan.~21, 2011}
{Jul.~31, 2012}
{}

\input{preamble}
\usepackage{enumerate}

\begin{document}
\title[Completeness for the coalgebraic cover modality]{Completeness for the coalgebraic cover modality}

\author[C.~Kupke]{Clemens Kupke\rsuper a}
\address{{\lsuper a}Department of Computer Science, University of Oxford, Parks Road, Oxford, United Kingdom.}
\email{ckupke@cs.ox.ac.uk}
\thanks{{\lsuper a}Supported by grants EP/F031173/1 and EP/H051511/1 from the UK EPSRC}

 \author[A.~Kurz]{Alexander Kurz\rsuper b}
 \address{{\lsuper b}Department of Computer Science, University of Leicester,
   United Kingdom.}
 \email{kurz@mcs.le.ac.uk}
 \thanks{{\lsuper b}Partially supported by grant EP/C014014/1 from EPSRC}
\author[Y.~Venema]{Yde Venema\rsuper c}
\address{{\lsuper c}Institute for Logic, Language and Computation, 
   Universiteit van Amsterdam, Science Park 904, 1098 XH Amsterdam, The Netherlands.}
\email{y.venema@uva.nl}
\thanks{{\lsuper c}The research of this author has been made possible by  VICI grant 639.073.501 of
   the Netherlands Organization for Scientific Research (NWO)}
   


\begin{abstract}
We study the finitary version of the coalgebraic logic introduced by
L.~Moss.  The syntax of this logic, which is introduced uniformly
with respect to a coalgebraic type functor, required to preserve
weak pullbacks, extends that of classical
propositional logic with a so-called coalgebraic cover modality
depending on the type functor.  Its semantics is defined in terms
of a categorically defined relation lifting operation.

As the main contributions of our paper we introduce a derivation
system, and prove that it provides a sound and complete
axiomatization for the collection of coalgebraically valid
inequalities.  Our soundness and completeness proof is algebraic,
and we employ Pattinson's stratification method, showing that our
derivation system can be stratified in countably many layers,
corresponding to the modal depth of the formulas involved.

In the proof of our main result we identify some new concepts and
obtain some auxiliary results of independent interest.  We survey
properties of the notion of relation lifting, induced by an
arbitrary but fixed set functor.  We introduce a category
of Boolean algebra presentations, and establish an adjunction
between it and the category  of Boolean algebras.

Given the fact that our derivation system  involves only
formulas of depth one, it can be encoded as a endo-functor on
Boolean algebras. We show that this functor is finitary and preserves
embeddings, and we prove that the Lindenbaum-Tarski algebra of our
logic can be identified with the initial algebra for this functor.

\end{abstract}

\maketitle

\EnableBpAbbreviations

\newpage
\newcommand{\invoeg}[1]{\input{#1}}

\input{sec-introduction}

\input{sec-preliminaries}
\input{sec-relationlifting}
\input{sec-boolean}
\input{sec-nabla}
\input{sec-derivation}
\input{sec-onestep}
\input{sec-completeness}
\input{sec-conclusion}
\bibliographystyle{plain}
\bibliography{nabla}
\input{sec-appendix}

\end{document}

%% file: preamble.tex
\usepackage{amssymb,latexsym,amsmath,bbm}
\usepackage{graphicx}
\usepackage[all]{xypic}
\usepackage{bussproofs}
\usepackage{hyperref}
\usepackage{url}

\makeatletter
\DeclareRobustCommand*\cal{\@fontswitch\relax\mathcal}
\makeatother

\newenvironment{tbs}{%
   \begin{itemize}\tt }{\end{itemize}}
\newcommand{\btbs}{\begin{tbs}}                                                                      
\newcommand{\etbs}{\end{tbs}}

\newcommand{\hide}[1]{}

\def\mvsquareforeod{\hbox{$\lhd$}}
\def\mveod{\ifmmode\mvsquareforeod\else{\unskip\nobreak\hfil
\penalty50\hskip1em\null\nobreak\hfil\mvsquareforeod
\parfillskip=0pt\finalhyphendemerits=0\endgraf}\fi}

\newenvironment{definition}{\begin{defi}\rm}{\hfill$\mveod$\end{defi}}

\newenvironment{convention}{\begin{conv}\rm}{\end{conv}}

\def\mvsquareforqed{\hbox{\textsc{qed}}}
\let\mvsquareforqed=\qed
\def\mvqed{\ifmmode\mvsquareforqed\else{\unskip\nobreak\hfil
\penalty50\hskip1em\null\nobreak\hfil\mvsquareforqed
\parfillskip=0pt\finalhyphendemerits=0\endgraf}\fi}

\newenvironment{proofof}[1]{\begin{trivlist}\item[\hskip\labelsep{\bf
Proof~of~{#1}.\ }]}{\hspace*{\fill} \mvqed\end{trivlist}}

\newtheorem{claim2}{Claim}

\newcommand{\emdef}[1]{\emph{#1}}

\newcommand{\class}[1]{\mathsf{#1}}
\newcommand{\Set}{\mathsf{Set}}
\newcommand{\BA}{\mathsf{BA}}
\newcommand{\Boole}{\mathsf{Boole}}
\newcommand{\Boolenb}{\Boole_{\nb}}
\newcommand{\Prs}{\mathsf{Pres}}
\newcommand{\Rel}{\mathsf{Rel}}
\newcommand{\Coalg}{\mathsf{Coalg}}
\newcommand{\Alg}{\mathsf{Alg}}
\newcommand{\Kl}{\mathsf{Kl}}

\newcommand{\acal}{\class{A}}
\newcommand{\clC}{\class{C}}

\newcommand{\funP}{P}             
   \newcommand{\Pow}{\funP}       
\newcommand{\Pom}{P_{\om}}        
\newcommand{\funQ}{\breve{P}}     
\newcommand{\funaP}{\mathbb{P}}   
\newcommand{\PowMo}{\mathcal{P}}  
\newcommand{\funaQ}{\breve{\mathbb{P}}} 
\newcommand{\Id}{\mathit{Id}}     
\newcommand{\Bag}{\mathit{B}}
\newcommand{\Btree}{\mathit{B}_{C}}

\newcommand{\T}{T}           
\newcommand{\Tom}{T_{\om}}   
\newcommand{\Tomnb}{T_{\om}^{\nb}} 
\newcommand{\Base}{\mathit{Base}}

\newcommand{\funU}{U}             
\newcommand{\funaF}{\mathbb{F}}   
\newcommand{\Tba}{\mathcal{L}_{0}}   
\newcommand{\funB}{B}             
\newcommand{\funC}{C}             
\newcommand{\unitBC}{\eta}
\newcommand{\counitBC}{\epsilon}

\newcommand{\Mossfun}{A_{M}}
\newcommand{\mng}{\mathit{mng}}

\newcommand{\funpM}{M}
\newcommand{\funaM}{\mathbb{M}}
\newcommand{\funV}{V}

\newcommand{\rl}[1]{\overline{#1}}   
\newcommand{\ti}[1]{\widetilde{#1}}  

\newcommand{\Tb}{\rl{\T}}
\newcommand{\Tomb}{\rl{\T_\om}}
\newcommand{\Pb}{\rl{\funP}}

\newcommand{\id}{\mathit{id}}

\newcommand{\ntrto}{\mathrel{\dot{\rightarrow}}}

\newcommand{\pb}{\mathit{pb}}
\newcommand{\prs}[2]{\langle #1 ; #2 \rangle}
\newcommand{\Tnb}{\mathcal{L}_{1}}   
\newcommand{\Graph}{\mathit{Gr}}

\newcommand{\struc}[1]{\langle #1 \rangle}

\newcommand{\bbA}{\mathbb{A}}
\newcommand{\bbB}{\mathbb{B}}

\newcommand{\bbD}{\mathbb{D}}

\newcommand{\bbF}{\mathbb{F}}

\newcommand{\bbM}{\mathbb{M}}
\newcommand{\bbN}{\mathbb{N}}

\newcommand{\bbX}{\mathbb{X}}
\newcommand{\bbZ}{\mathbb{Z}}

\newcommand{\bbtwo}{\mathbbm{2}}

\newcommand{\At}{\mathit{At}}
\newcommand{\Minit}{\mathcal{M}}

\renewcommand{\phi}{\varphi} 
\newcommand{\bw}{\bigwedge}
\newcommand{\bv}{\bigvee}
\newcommand{\bcsmall}{\mbox{$\bigcup$}}
\newcommand{\bwsmall}{\mbox{$\bigwedge$}}
\newcommand{\bvsmall}{\mbox{$\bigvee$}}
\newcommand{\bwlarge}{\displaystyle \bigwedge}
\newcommand{\bvlarge}{\displaystyle \bigvee}
\newcommand{\dia}{\Diamond}

\newcommand{\nb}{\nabla}
\newcommand{\hs}{\heartsuit}

\newcommand{\is}{\approx}
\newcommand{\isleq}{\preccurlyeq}
\newcommand{\sqleq}{\sqsubseteq}
\newcommand{\sqgeq}{\sqsupseteq}
\newcommand{\Lmoss}{\mathcal{L}}

\newcommand{\Sfor}{\mathit{Sfor}}

\newcommand{\Prop}{\mathsf{Prop}}

\newcommand{\isbnf}{\mathrel{\;::=\;}}
\newcommand{\divbnf}{\mathrel{\mid}}
\newcommand{\isdef}{\mathrel{:=}}

\newcommand{\forces}{\Vdash}

\newcommand{\finsem}{s}

\newcommand{\nax}{\mathbf{M}}
\newcommand{\vdnax}{\vdash_{\nax}}
\newcommand{\quot}[1]{\rho_{#1}}
\newcommand{\D}{\mathcal{D}}  

\newcommand{\SRD}{\mathit{SRD}}
\newcommand{\nbsem}{\lambda\!^{\T}}
\newcommand{\semone}[1]{[\![#1]\!]_{1}}
\newcommand{\eqsem}{\equiv_{sem}}
\newcommand{\eqax}{\equiv_{ax}}

\newcommand{\nada}{\varnothing}

\newcommand{\sse}{\subseteq}

\newcommand{\size}[1]{|#1|}
\newcommand{\coloneqq}{\mathrel{:=}}

\newcommand{\Dom}{\mathsf{dom}}
\newcommand{\Ran}{\mathsf{rng}}
\newcommand{\converse}[1]{#1\breve{\hspace*{1mm}}} 
\newcommand{\cv}[1]{\converse{#1}}   
\newcommand{\corel}{\mathbin{;}}       
\newcommand{\cof}{\mathbin{\circ}}   

\newcommand{\rel}[1]{\mathrel{#1}}
\newcommand{\Tin}{\rel{\Tb{\in}}}
\newcommand{\Pin}{\rel{\Pb{\in}}}

\newcommand{\rst}[1]{\!\upharpoonright_{#1}\,}

\newcommand{\yiff}{\mbox{ iff }}

\newcommand{\al}{\alpha}
\newcommand{\be}{\beta}
\newcommand{\ga}{\gamma}
\newcommand{\de}{\delta}
\newcommand{\la}{\lambda}
\newcommand{\si}{\sigma}
\newcommand{\om}{\omega}

\newcommand{\pGR}{\prs{G}{R}}
\newcommand{\pGRp}{\prs{G'}{R'}}
\newcommand{\wh}[1]{\widehat{#1}}

\newcommand{\semzero}[1]{[\![#1]\!]_{0}}
\usepackage{color,graphics,xcolor}
  \usepackage{ulem}

\definecolor{darkgreen}{rgb}{0,0.5,0}
\definecolor{darkblue}{rgb}{0,0,0.8}
\definecolor{darkred}{rgb}{0.9,0,0}
\newcommand{\akk}[1]{{\color{black}#1}}


%% file: sec-introduction.tex
\section{Introduction}
\label{s:introduction}

\newcommand{\Lang}{\mathcal{L}}

Coalgebra, introduced to computer science by Aczel in the late
1980s~\cite{acze:nonw88,acze:fina89}, is rapidly gaining ground as a general 
mathematical framework for many kinds of state-based evolving systems.
Examples of coalgebras include data streams, (infinite) labelled trees, Kripke
structures, finite automata, (probabilistic/weighted) transition systems, 
neighborhood models, and many other familiar structures.
As emphasized by Rutten~\cite{rutt:univ00}, who developed, in analogy with
Universal Algebra, the theory of Universal Coalgebra as a general theory of 
such transition systems, the coalgebraic viewpoint combines wide applicability
with mathematical simplicity.
In particular, one of the main advantages of the coalgebraic approach is 
that a substantial part of the theory of systems can be developed
\emph{uniformly} in a functor $\T$ which represents the \emph{type} of the 
coalgebras we are dealing with.
Here we restrict attention to \emph{systems}, where $\T$ is an endofunctor 
on the category $\Set$ of sets with functions, so that a $\T$-coalgebra is a
pair of the form
\[
\bbX = \struc{X,\xi: X \to \T X}
\]
with the set $X$ being the carrier or state space of the coalgebra, and the 
map $\xi$ its unfolding or transition map.
Many important notions, properties, and results of systems can be explained 
just in terms of properties of their type functors.
As a key example, any set functor $\T$ canonically induces a notion of
observational or \emph{behavioural equivalence} between $\T$-coalgebras;
this notion generalizes the natural notions of bisimilarity that were
independently developed for each specific type of system.

In order to describe and reason about the kind of behaviour modelled by 
coalgebras, there is a clear need for the design of coalgebraic specification 
languages and derivation systems, respectively.
The resulting research programme of \emph{Coalgebraic Logic} naturally
supplements that of Coalgebra by searching for logical formalisms that, next 
to meeting the usual desiderata such as striking a good balance between
expressive power and computational feasibility, can be defined and studied
uniformly in the functor $\T$.
Given the fact that Kripke models and frames are prime examples of coalgebras,
it should come as no surprise that in search for suitable coalgebraic logics,
researchers looked for inspiration to \emph{modal logic}~\cite{blac:moda01}.

This research direction was inititiated by Moss~\cite{moss:coal99}; roughly
speaking, his idea was to take the functor $\T$ \emph{itself} as supplying a 
modality $\nb_{\T}$, in the sense that for every element $\al \in \T\Lang$
(where $\Lang$ is the collection of formulas), the object $\nb_{\T}\al$ is a
formula in $\Lang$.
While Moss' work was recognized to be of seminal conceptual importance in
advocating modal logic as a specification language for coalgebra, his particular 
formalism did not find much acclaim, for at least two reasons.
First of all, the semantics of his modality is defined in terms of relation 
lifting, and for this to work smoothly, Moss needed to impose a restriction on
the functor (the coalgebra type functor $\T$ is required to preserve weak
pullbacks).
Thus the scope of his work excluded some interesting and important coalgebras
such as neighborhood models and frames.
And second, for practical purposes, the syntax of Moss' language was considered
to be rather unwieldy, with the nonstandard operator $\nb_{\T}$ looking 
strikingly different from the usual $\Box$ and $\dia$ modalities.

Following on from Moss' work, attention turned to the question how
to obtain modal languages for $\T$-coalgebras which use more standard modalities
\cite{kurz:cmcs98-j,roes:coal00,jaco:many01}, and how to find
derivation systems for these formalisms.
This approach is now usually described in terms of predicate
liftings~\cite{patt:coal03,schr:expr05} or, equivalently,
Stone duality~\cite{bons:dual05,kurz:coal06}.
Other approaches towards coalgebraic logic, such as the one using 
co-equations~\cite{adam:logi05} until now have received somewhat less attention.
For a while, this development directed interest away from Moss' logic, and
the relationship between various approaches towards coalgebraic logic was not 
completely clear.

In the mean time, however, it had become obvious that even in standard modal
logic, a nabla-based approach has some advantages.
In this setting the coalgebra type $\T$ is instantiated by the power set
functor $\funP$, so that (the finitary version of) the nabla operator 
$\nb_{\!\funP}$, takes a (finite) \emph{set} $\al$ of formulas and returns a 
single formula $\nb_{\!\funP}\al$.
The semantics of this so-called \emph{cover modality} can be
explicitly formulated as follows, for an arbitrary Kripke structure
$\bbX$ with accessibility relation $R$:
\begin{equation}
\label{eq:1a}
\begin{array}{lll}
\bbX,x \Vdash \nb_{\!\funP}\al & \mbox{ if } & \mbox{ for all $a \in
\al$ there is a
  $t \in R[x]$ with $\bbX,t \Vdash a$, and}
\\&& \mbox{ for all $t \in R[x]$ there is an $a \in \al$ with
  $\bbX,t \Vdash a$}.
\end{array}
\end{equation}
In short: $\nb_{\!\funP}\al$ holds at a state $x$ iff the formulas in
$\al$ and the set $R[x]$ of successors of $x$ `cover' one
another.
Readers familiar with classical first-order logic will recognize the
quantification pattern underlying (\ref{eq:1a}) from the theory of 
Ehrenfeucht-Fra\"{\i}ss\'e games, Scott sentences, and the like, 
see for instance~\cite{hodg:mode93}.
In
modal logic, related ideas made an early appearance in Fine's work
on normal forms~\cite{fine:norm75}. 

Using the standard modal language, $\nb_{\!\funP}$ can be seen as a defined
operator:
\begin{equation}
\label{eq:1} \nb_{\!\funP}\al = \Box\bvsmall\al \land \bwsmall
\dia \al,
\end{equation}
where $\dia\al$ denotes the set $\{ \dia a \mid a \in
\al \}$. But is in fact an easy exercise to prove that with
$\nb_{\!\funP}$ defined by (\ref{eq:1a}), we have the following semantic
equivalences:
\begin{equation}
\label{eq:2}
\begin{array}{lll}
   \dia\al &\equiv& \nb_{\!\funP} \{ \al, \top \}
\\ \Box\al &\equiv& \nb_{\!\funP}\varnothing \lor \nb_{\!\funP} \{ \al \}
\end{array}
\end{equation}
In other words, the standard modalities $\Box$ and $\dia$ can be defined in
terms of the nabla operator (together with $\lor$ and $\top$).
When combined, (\ref{eq:1}) and (\ref{eq:2}) show that the language based on
the nabla operator offers an alternative formulation of standard modal logic.

In fact, independently of Moss' work, Janin \& Walukiewicz~\cite{jani:auto95}
had already made the much stronger observation that the set of connectives
$\{ \Box,\dia,\land,\lor \}$ may in some sense be replaced by the connectives
$\nb_{\!\funP}$ and $\lor$, that is, without the conjunction operation.
This fact, which is closely linked to fundamental automata-theoretic 
constructions, lies at the heart of the theory of the modal $\mu$-calculus,
and has many applications, see for instance~\cite{dago:logi00,sant:comp10}.
These observations naturally led Venema~\cite{vene:auto06} to introduce, 
parametric in the coalgebraic type functor $\T$, a finitary version of Moss' 
logic, extended with fixpoint operators, and to generalize the link between 
fixpoint logics and automata theory to the coalgebraic level of generality.
Subsequently, Kupke \& Venema~\cite{kuve08:coal} showed that many fundamental
results in automata theory and fixpoint logics are really theorems of universal
coalgebra.
The key role of the nabla modality in these results revived interest in 
Moss' logic.

Our paper addresses the main problem left open in the literature on $\nb$-based
coalgebraic logic, namely that of providing a \emph{sound and complete 
derivation system} for the logic.
Moss' approach is entirely semantic, and does not provide any kind of syntactic
calculus.
As a first result in the direction of a derivation system for nabla modalities,
Palmigiano \& Venema~\cite{palm:nabl07} gave a complete axiomatization for the
cover modality $\nb_{\!\funP}$.
This calculus was streamlined into a formulation that admits a straightforward
generalization to an arbitrary set functor $\T$, by B{\'\i}lkov\'a, Palmigiano
\& Venema~\cite{bilk:proo08}, who also provided suitable Gentzen systems for 
the logic based on $\nb_{\!\funP}$.
In this paper we will prove the soundness and completeness of this
axiomatization in the general case.
\medskip

In the remaining part of the introduction we briefly survey the paper, its
main contributions, and its proof method.
Throughout the paper we let $\T$ denote the coalgebraic type functor; usually
we make the proviso that $\T$ preserves weak pullbacks and inclusions (all of
this will be discussed further on in detail).
Our key instrument in making Moss' language more standard is to base its
syntax on the \emph{finitary} version $\Tom$ of the functor $\T$ which is
defined on objects as follows: for a set $X$, $\Tom X \isdef \bigcup\{ \T Y \mid 
Y \sse_{\om} X \}$.
As we will discuss in detail, for each object $\al \in \Tom X$ there is a
\emph{minimal} finite set $\Base_{X}(\al) \sse_{\om} X$ such that $\al \in \T 
\Base(\al)$, and the maps $\Base_{X}$ provide a natural transformation
\[
\Base: \Tom \ntrto \Pom.
\]
The formulas of our coalgebraic language $\Lmoss$ can now be defined by the
following grammar:
\[
a \isbnf \neg a \divbnf \bwsmall\phi \divbnf \bvsmall\phi
  \divbnf \nb_{\T} \alpha.
\]
where $\phi \in \Pom \Lmoss$ and $\alpha \in \Tom\Lmoss$.
That is, the propositional basis of our coalgebraic language $\Lmoss$ takes the
finitary conjunction ($\bw$) and disjunction ($\bv$) connectives as
primitives, and
to this we add the coalgebraic modality $\nb_{\T}$, which returns a formula
$\nb_{\T}\al$ for every object $\al \in \Tom\Lmoss$.
The point of restricting Moss' modality to the set $\Tom\Lmoss$ is that the 
formula $\nb_{\T}\al$ has a finite, clearly defined set of immediate
subformulas, namely the set $\Base(\al)$; thus every formula has a finite
set of subformulas.

The key observation of 
Moss~\cite{moss:coal99} was that the \emph{semantics}
\eqref{eq:1a} of $\nb$ can be expressed in terms of the so-called Egli-Milner
\emph{lifting} of the satisfaction relation ${\forces} \sse X \times \Lang$.
Generalizing this observation from the Kripke functor $\funP$ to the arbitrary 
type $\T$, he uniformly defined the semantics of $\nb_{\T}$ in a $\T$-coalgebra
$\bbX = \struc{X,\xi}$ as follows:
\[
\bbX,x \forces \nb_{\T}\al \mbox{ iff } \xi(x) \rel{\Tb{\forces}}\al.
\]
Here $\Tb{\forces}$ denotes a categorically defined \emph{lifting} of the 
satisfaction relation ${\forces} \sse X \times \Lang$ between states and 
formulas to a relation $\Tb{\forces} \sse \T X \times \T\Lang$.
Given the importance of the \emph{relation lifting} operation $\Tb$ in Moss'
logic, we include in this paper a fairly detailed survey of its properties 
and related concepts.

The coalgebraic \emph{validities}, that is, the formulas that are true at
every state of every $\T$-coalgebra thus constitute a semantically defined
coalgebraic \emph{logic}, and it is this logic that we will axiomatize in 
this paper.
Our approach will be \emph{algebraic} in nature, and so it will be convenient
to work with equations, or rather, inequalities (expressions of the form
$a \isleq b$, where $a$ an $b$ are terms/formulas of the language).

We obtain our derivation system for Moss' logic by extending a sound and 
complete derivation system for propositional logic with three rules for the 
$\nb$-operator.
The first rule, denoted by $(\nb 1)$, can be seen as a combined montonicity 
and congruence rule.
Rule $(\nb 2)$ is a distributive law  that expresses that any conjunction of
$\nb$-formulas is equivalent to a (possibly infinite) disjunction of 
$\nb$-formulas built from conjunctions.
Finally, rule $(\nb 3)$ expresses that $\nb$ distributes over disjunctions. 
In the case that the functor $\T$ under consideration
maps finite sets to finite sets, the rules $(\nb 2)$ and $(\nb 3)$ take the
form of axioms. 

The proof of our soundness and completeness theorem is based on the
\emph{stratification method} of Pattinson~\cite{patt:coal03}.
We will show that not only the \emph{language} of our system, but also
its \emph{semantics} and our \emph{derivation system} can be stratified in 
$\om$ many layers corresponding to the modal depth of the formulas involved.
(This means for instance that if two formulas of depth $n$ are provably 
equivalent, this can be demonstrated by a derivation involving only formulas
of depth at most $n$.)
What glues these layers nicely together can be formulated in terms of 
properties of a one-step version of the derivation system $\nax$.

In our algebraic approach, this one-step version of $\nax$ is incarnated as
a \emph{functor} on the category of Boolean algebras:
\[
\funaM: \BA \to \BA.
\]
To mention a few interesting properties of this functor, of which the 
definition is uniformly parametrized by the functor $\T$: $\funaM$ is finitary,
and preserves atomicity of Boolean algebras, and injectivity of homomorphisms.
We will be interested in algebras for the functor $\funaM$, and in particular,
we will see that the \emph{initial} $\funaM$-algebra can be seen as the 
Lindenbaum-Tarski algebra of our derivation system $\nax$.

For the definition of $\funaM$, we need to go into quite a bit of detail
concerning the theory of \emph{presentations} of (Boolean) algebras.
In particular, we define a \emph{category} $\Prs$ of presentations by 
introducing a suitable notion of presentation morphism, and establish an
adjunction between the categories $\Prs$ and $\BA$:
\begin{equation}
\xymatrix{
\BA \ar@/_/[r]_{C} \ar@{}[r] |{\bot}
  & \Prs \ar@/_/[l]_{B} 
}
\end{equation}
This adjunction (which is almost an equivalence) is the instrument that
allows us to turn the modal rule and axioms of $\nax$ into the functor 
$\funaM$; the key property that makes
this work is that all modal rules and axioms of $\nax$ are formulated in 
terms of \emph{depth-one} formulas.

What is left to do, in order to prove the soundness and completeness of our
logic, is connect the algebra functor $\funaM: \BA \to \BA$ (that is, the
`logic') to the coalgebra functor $\T: \Set \to \Set$ (the `semantics').
Here we will apply a well-known method in coalgebraic 
logic~\cite{bons:dual05,kurz:coal06}
which is often described in terms of \emph{Stone duality} because its aim is
to link functors on two different base categories that are connected themselves 
by a Stone-type duality or adjunction.

%
In our case, to make the connection between $\funaM$ and $\T$ we invoke the
already existing link on the level of the base logic, provided by the
(contravariant) power set functor $\funaQ$ from $\Set$ to $\BA$
(we do not need its adjoint functor sending a Boolean algebra to its
set of ultrafilters):
\begin{equation}
\label{diag:duality}
\xymatrix{
\BA \ar@(dl,ul)[]^{\funaM} 
  & \Set \ar@/_/[l]_{\funaQ} \ar@(dr,ur)[]_{T}
}
\end{equation}
The key remaining step in the completeness proof involves the definition of 
a natural transformation
\begin{equation*}
\label{eq:delta1}
\de: \funaM\funaQ \ntrto \funaQ\T.
\end{equation*}
As usual in the Stone duality approach towards coalgebraic logic, the
\emph{existence} of $\de$ corresponds to the \emph{soundness} of the logic.
To get an idea of why this is the case, observe that the existence of $\de$
enables us to see a $\T$-coalgebra $\bbX = \struc{X,\xi}$ as an
$\funaM$-algebra, namely its \emph{complex algebra} $\bbX^{*} \isdef
\struc{\funaQ X, \funaQ\xi\cof\de_{X}}$.
Finally, as we will see in the final part of our stratification-based proof,
the \emph{completeness} of $\nax$ is based on the observation that 
\begin{equation}
\label{eq:delta2}
\de \text{ is injective},
\end{equation}
that is, for each set $X$, the $\BA$-homomorphism $\de_{X}: \funaM\funaQ X \to
\funaQ\T X$ is an \emph{embedding}.
The proof of \eqref{eq:delta2}, which technically forms the heart of our 
proof, is based on the fact that the nabla-axioms allow us to write 
depth-one formulas into a certain normal form, and on the earlier mentioned
properties of the functor $\funaM$.
\medskip

This paper replaces, extends and partly corrects (c.q.\ clarifies, see 
Remark~\ref{r:aiml}) an earlier version~\cite{kupk:comp08}.
The main differences with respect to~\cite{kupk:comp08} are the following.
First of all, we provide a detailed, self-contained overview of the notion
of relation lifting and its properties (which was only covered as Fact~3
in the mentioned paper).
Second, our categorical treatment of presentations and the algebras they
present (which is novel to the best of our knowledge) clarifies and 
substantially extends the treatment in~\cite{kupk:comp08}.
Third, our axiomatization simplifies the earlier one; in particular, we show
here in detail that we do not need axioms or rules specifically dealing with
negation (more specifically, we prove that an earlier rule ($\nb4$) is
derivable in the system here.
Fourth, we provide a more precise definition and a more detailed discussion 
of the functor $\funaM$; for instance, the result that $\funaM$ preserves
atomicity is new.
Fifth and final, we show here in much more detail and precision how the 
soundness and completeness of our axiomatization follows from the one-step
soundness and completeness.

\paragraph{Overview}

In the next section we fix our notation, introduce the necessary basic
(co-)alge\-braic terminology and discuss properties of functors on the category
of sets that will play an important role in our paper.
After that, in Section~\ref{s:relationlifting}, we recall the notion of a 
relation lifting $\Tb$ induced by a set functor $\T$ and give an overview of
its properties. 
Section~\ref{s:boolean} and Section~\ref{s:moss} introduce the terminology 
that we need concerning Boolean algebras and their presentations, and
concerning Moss' coalgebraic logic, respectively.

After that we move to the main results of our paper.
First, in Section~\ref{s:derivation} we introduce the derivation system for
Moss' coalgebraic logic and we define the algebra functor $\funaM:\BA \to \BA$. 
In Section~\ref{s:onestep} we prove that our derivation system is one-step 
sound and complete. 
Within the above described categorical framework this is equivalent to 
establishing the existence of a natural transformation $\delta: \funaM \funaQ
\ntrto \funaQ \T$ (one-step soundness) and proving that this transformation
$\delta$ is injective (one-step completeness).
Finally, in Section~\ref{s:completeness} 
we prove our main result, namely soundness and completeness
of our derivation system with respect to the coalgebraic semantics. 
We conclude with an overview of related work and open questions.

Finally, since this paper features a multitude of categories, functors and 
natural transformations, for the reader's convenience we list these in an
appendix.

\paragraph{\bf Acknowledgement}
We thank the anonymous referee for many useful comments.

%% file: sec-preliminaries.tex
\section{Preliminaries}
\label{s:preliminaries}

The purpose of this section is to fix our notation and terminology, and to 
introduce some concepts that underlie our work in all other parts of the
paper.

\subsection{Basic mathematics and category theory}
\label{ss:basics1}

First we fix some basic mathematical issues.
Given a set $X$, we let $\funP X$ and $\Pom X$ denote the power set and
the finite power set of $X$, respectively.
We write $Y \sse_{\om} X$ to indicate that $Y$ is a finite subset of $X$.

Given a relation $R \sse X \times X'$, we denote the \emph{domain} and 
\emph{range} of $R$ by $\Dom(R)$ and $\Ran(R)$, respectively, and we denote
by $\pi^R_1:R \to X$ its first projection and by $\pi^R_2:R \to X'$ its
second projection map.
Given subsets $Y \sse X$, $Y' \sse X'$, the \emph{restriction} of $R$ to  
$Y$ and $Y'$ is given as
\[
R\rst{Y \times Y'} \isdef R \cap (Y \times Y').
\]
The \emph{converse} of a relation $R \sse X \times X'$ is denoted as 
$\converse{R} \sse X' \times X$.

The \emph{composition} of two relations $R \sse X \times X'$ and $R' \sse X'
\times X''$ is denoted by $R\corel R'$, while the composition of two functions
$f: X \to X'$ and $f': X' \to X''$ is denoted by $f'\cof f$. 
That is, we denote function composition by $\cof$ and write it from right 
to left and we denote relation composition of relations by $\corel$ and write
it from left to right.

It is often convenient to identify a function $f:X \to X'$ with its
\emph{graph}, that is, the relation $\Graph(f)=\{(x,f(x))\mid x\in X\}
\subseteq X\times X'$.
For example given a relation $R \subseteq X \times X'$  and a function
$f: X' \to X''$ we write $R\corel f$ to denote the composition of relations 
$R\corel\Graph(f)$.

We will assume familiarity with basic notions from category theory, including those
of categories, functors, natural transformations, (co-)monads and
(co-)limits; see for instance~\cite{macl:cate98}.
We denote by $\Set$ the category of sets and functions, and by $\Rel$ the 
category of sets and binary relations.
$\BA$ is the category with Boolean algebras as objects and homomorphisms
as arrows.

Endofunctors on $\Set$ will simply be called \emph{set functors}.
We denote by $\Pow $ the \emph{power set functor} which maps a set $X$ to
its power set $\Pow X$ and a function $f:X\to X'$ to its \emph{direct image}
$\Pow f: \Pow X \to \Pow X'$, given by $\Pow(X) \ni Y \mapsto \{f(y)\mid 
y\in Y\}$.
Similarly, $\Pom X$ denotes the finite power set functor.
$\funP$ is in fact (part of) a monad $(\funP,\mu,\eta)$, with  $\eta_{X}: 
X \to \funP(X)$ denoting the singleton map $\eta_{X}: x \mapsto \{ x\}$, and
$\mu_{X}: \funP\funP X \to \funP X$ denoting union, $\mu_{X}(\mathcal{A}) 
\isdef \bigcup \mathcal{A}$.
The contravariant power set functor will be denoted as $\funQ$; this
functor maps a set $X$ to its power set $\funQ X = \funP X$, and a function 
$f: X \to X'$ to its \emph{inverse image} $\funQ f: \funQ X' \to \funQ X$
given by $\funQ X' \ni Y' \mapsto \{ x \in X \mid fx \in Y' \}$.

\subsection{(Co-)algebras}
\label{ss:coalgebras}

We provide some details concerning the notions of an algebra and a
coalgebra for a functor.
We start with coalgebras since these provide the semantic structures of the
logics considered in this paper.

\begin{definition}
Given a functor $\T$ on a category $\clC$, a $\T$-coalgebra $(X,\xi)$
is an arrow $\xi:X\to \T X$ in $\clC$; a $T$-coalgebra morphism
$f:(X,\xi)\to(X',\xi')$ is an arrow $f:X\to X'$ such that $\T f\cof\xi
=\xi'\cof f$, in a diagram:
\begin{equation*}
\xymatrix{
    X  \ar[d]_{\xi}   \ar[r]^{f} 
  & X'     \ar[d]^{\xi'} 
\\
   \T X \ar[r]^{\T f} 
& \T X'
}
\end{equation*}
The functor $\T$ is called the \emph{type} of the coalgebra $(X,\xi)$, 
The category of $\T$-coalgebras is denoted by $\Coalg(\T)$ and we denote 
coalgebras by capital letters $\mathbb{X},\mathbb{Y},\dots$ in blackboard bold. 

In the case of a set coalgebra (that is, a coalgebra for a set functor),
elements of the (carrier of the) coalgebra will be called \emph{states}
of the coalgebra, and a \emph{pointed coalgebra} is a pair consisting
$(\bbX,x)$ consisting of a coalgebra $\bbX=(X,\xi)$ and a state $x$ of $\bbX$.
\end{definition}

Here are some simple, standard examples of coalgebras for set functors.

\begin{exa}
\label{ex:1}\hfill
\begin{enumerate}[(1)]
\item
We let $\Id$ denote the \emph{identity} functor on $\Set$.
Given a set $C$, we let $C$ itself also denote the \emph{constant} functor,
mapping every set $X$ to $C$, and every function $f$ to the identity map
$\id_{C}$ on $C$.
Coalgebras for this functor are called \emph{$C$-colorings}; in case $C$
is of the form $\Pow(\Prop)$ for some set $\Prop$ of proposition letters, 
we may think of a coloring $\xi: X \to C$ as a \emph{$\Prop$-valuation}
(in the sense that $\xi$ says of every proposition letter $p$ and every
state $x$ whether $p$ is true of $x$ or not).
\item
A \emph{Kripke frame} $\struc{S,R}$ can be represented as a coalgebra
$\struc{S,\si_{R}}$ for the power set functor $\funP$, with $\si_{R}: 
S \to \funP S$ mapping a point $s$ to its collection of successors.
It is left as an exercise for the reader to verify that the coalgebra 
morphisms for this functor precisely coincide with the \emph{bounded 
morphisms} of modal logic.
\item
Coalgebras for the functor $\funQ\cof\funQ$ (that is, the contravariant 
power set functor composed with itself) can be identified with the
\emph{neighborhood frames} known from the theory of modal logic as 
structures that generalize Kripke frames.
As a special case of this, but also generalizing Kripke frames, the
\emph{monotone neighborhood functor} $N$ maps a set $X$ to the collection 
$N(X) \isdef \{ \al \in \funQ\funQ X \mid \al \text{ is upward closed 
}\}$, and a function $f$ to the map $\funQ\funQ f$.
\item
For a slightly more involved example, consider the finitary \emph{multiset}
or \emph{bag} functor $\Bag_{\om}$.
This functor takes a set $X$ to the collection $B_{\om}X$ of maps $\mu: X
\to \bbN$ of finite support (that is, for which the set $\mathit{Supp}(\mu)
:= \{ x \in X \mid \mu(x) > 0 \}$ is finite), while its action on arrows is 
defined as follows.
Given an arrow $f: X \to X'$ and a map $\mu \in \Bag_{\om}X$, we define
$(\Bag_{\om}f)(\mu): X' \to \bbN$ by putting
\[
(\Bag_{\om}f)(\mu)(x') := \sum \{ \mu(x) \mid f(x) = x' \}.
\]
\item
As a variant of $\Bag_{\om}$, consider the finitary probability functor
$D_{\om}$, where $D_{\om} X = \{ \de: X \to [0,1] \mid
\mathit{Supp}(\de) \text{ is finite and } \sum_{x\in X}\de(x) = 1 \}$,
while the action of $D_{\om}$ on arrows is just like that of $\Bag_{\om}$.
\end{enumerate}
\end{exa}

\begin{exa}
\label{ex:2}
Many examples of coalgebraically interesting set functors are obtained by
\emph{composition} of simpler functors.
Inductively define the following class $\mathit{EKPF}$ of \emph{extended
Kripke polynomial functors}:
\[
\T \isdef \Id  \mid C \mid \funP \mid \Bag_{\om} \mid D_{\om} \mid 
   \T_{0} \cof \T_{1} \mid \T_{0} + \T_{1} \mid \T_{0} \times \T_{1} \mid
   \T^{D},
\]
where $\cof$, $+$ and $\times$ denote functor composition, coproduct (or 
disjoint union) and product, respectively, and $(-)^{D}$ denotes 
exponentiation with respect to some set $D$.
Examples of such functors include:
\begin{enumerate}[(1)]
\item
Given an alphabet-color set $C$, the \emph{$C$-streams} are simple
specimens of coalgebras for the functor $C \times \Id$; similarly,
$C$-labelled binary trees are coalgebras for the functor $\Btree =
C \times \Id \times \Id$.
\item
\emph{Labelled transition systems} over a set $A$ of atomic actions can 
be seen as coalgebras for the functor $\funP(-)^{A}$.
\item
\emph{Deterministic automata} are coalgebras for the functor
$(-)^{\Sigma} \times 2$ where $\Sigma$ is the finite alphabet.
\item
\emph{Kripke models} over a set $\Prop$ of proposition letters can be
identified with coalgebras for the functor $\funP(\Prop) \times \funP(-)
= \funP \cof C_{\Prop} \times \funP\cof\Id$.
\item
Generalizing the previous example, viewing $\T$-coalgebra as \emph{frames},
we can define \emph{$\T$-models} over a set $\Prop$ of proposition letters 
as coalgebras for the functor $\T_{\Prop} = \funP(\Prop) \times \T(-)$.
\end{enumerate}
\end{exa}

\noindent As running examples through this paper we will often take
the binary tree functor over a set $C$ of colors, and the power set
functor.

The key notion of equivalence in coalgebra is of two states in two coalgebras
being \emph{behaviorally equivalent}.
In case the functor $\T$ admits a final coalgebra $\bbZ = \struc{Z,\zeta}$
the elements of $Z$ often provide an
intuitive encoding of the notion of behaviour, and the unique coalgebra
homomorphism $!_{\bbX}$ can be seen as a map that assigns to a state $x$ in
$\bbX$ its \emph{behaviour}.
In this case we call two states, $x$ in $\bbX$ and $x'$ in $\bbX'$,
\emph{behaviorally equivalent} if $!_{\bbX}(x) = !_{\bbX'}(x')$.
In the general case, when we may not assume the existence of a final
coalgebra, we define the notion as follows.

\begin{definition}
\label{d:beheq}
Two elements (often called states) $x,x'$ in two coalgebras $\bbX$ and
$\bbX'$, respectively, are \emph{behaviorally equivalent} iff there
are coalgebra morphisms $f,f'$ with a common codomain such that
$f(x)=f'(x')$.
\end{definition}

\medskip Turning to the dual notion of algebra, we shall use algebras
mainly to describe logics for coalgebras, and the notion of an algebra
`for a functor' will provide us with an elegant way to exploit the
duality with coalgebras.

\begin{definition}
  Given a functor $L$ on a category $\acal$, an $L$-algebra
  $(A,\alpha)$ is an arrow $\alpha:LA\to A$ in $\acal$ and an
  $L$-algebra morphism $f:(A,\alpha)\to(A',\alpha')$ is an arrow
  $f:A\to A'$ such that $f\cof\alpha=\alpha'\cof Lf$. The category
  of $L$-algebras is denoted by $\Alg(L)$.
\end{definition}

\begin{exa}\hfill
\begin{enumerate}[(1)]
\item If $\acal=\Set$, then every signature (or similarity type)
  induces a functor $LX=\coprod_{n<\omega} \mathit{Op_n}\times X^n$
  where $\mathit{Op_n}$ is the set of operation symbols of arity
  $n$. Then $\Alg(L)$ is (isomorphic to) the category of algebras for
  the signature.

\item If $\acal=\BA$, then we can define a functor $L:\BA\to\BA$ to
  map an algebra $A$ to the algebra $LA$ generated by $\Box a$, $a \in
  A$, and quotiented by the relation stipulating that $\Box$ preserves
  finite meets. Then $\Alg(L)$ is isomorphic to the category of modal
  algebras \cite{kupk:ston04}.
\end{enumerate}
\end{exa}

\noindent As the second example above shows, functors on $\BA$ give
rise to modal logics extending Boolean algebras with operators.

\subsection{Properties of set functors}
\label{ss:setfunctors}

As mentioned in the introduction, in this paper we will restrict our
attention to set functors satisfying certain properties.
The first one of these is crucial.

\paragraph*{Weak pullback preservation}
Recall that a set $P$ together with functions $p_1:P \to X_1$ and $p_2: 
P \to X_2$ is a \emph{pullback} of two functions $f_1:X_1 \to X$
    and $f_2:X_2 \to X$ if $f_1 \cof p_1= f_2 \cof p_2$ and for
    all sets $P'$ and all functions $p_1':P' \to X_1$, $p_2': P' \to X_2$
    such that $f_1 \cof p_1' = f_2 \cof p_2'$ there exists a {\em unique} function
    $e:P' \to P$ such that $p_i \cof e = p_i'$ for  $i=1,2$. 
\\
\centerline{
\xymatrix{%
    P' \ar@/_/[ddr]_{p'_1}
      \ar@/^/[drr]^{p'_2} \ar@{-->}[dr]^e & & 
\\  & P \ar[r]^{p_2} \ar[d]_{p_1} 
    & X_2 \ar[d]^{f_2}
\\  & X_1 \ar[r]_{f_1} & X 
}}
\\      
If the function $e$ is not necessarily unique we call $(P,p_1,p_2)$ a 
\emph{weak pullback}.
Furthermore we call a relation $R \subseteq X_1 \times X_2$ a (weak) pullback 
of $f_1$ and $f_2$ if $R$ together with the projection maps $\pi_1^R$ and
$\pi_2^R$ is a (weak) pullback of $f_1$ and $f_2$.

In the category of sets, (weak) pullbacks have a straightforward 
characterization 
\begin{fact}\label{f:wp}%
\cite{gumm01:func}.
Given two functions $f_1:X_1 \to X_3$ and $f_2:X_2 \to X_3$, let 
\[
\pb(f_1,f_2) \coloneqq \{ (x_1,x_2) \mid f_1(x_1) = f_2(x_2) \}.
\]
Furthermore, given a set $P$ with functions $p_1:P \to X_1$ and $p_2:P 
\to X_2$, let 
\[
e: y  \mapsto (p_1(y),p_2(y)).
\]
define a function $e: P \to \pb(f_1,f_2)$.
Then 
\begin{enumerate}[(1)]
\item $(P,p_1,p_2)$ is a pullback of $f_1$ and $f_2$ iff 
   $f_1 \cof p_1=  f_2 \cof p_2$ and $e$ is an isomorphism.

\item $(P,p_1,p_2)$ is a weak pullback of $f_1$ and $f_2$
    iff $f_1 \cof p_1=  f_2 \cof p_2$ and $e$ is surjective.
\end{enumerate}
\end{fact}

A functor $\T$ \emph{preserves weak pullbacks} if it transforms every
weak pullback $(P,p_{1},p_{2})$ for $f_{1}$ and $f_{2}$ into a weak
pullback $(\T P,\T p_{1}, \T p_{2})$ for $\T f_{1}$ and $\T f_{2}$.
An equivalent characterization is to require $\T$ to \emph{weakly
  preserve pullbacks}, that is, to turn pullbacks into weak pullbacks.
Further on in Corollary~\ref{cor:extT}, we will see yet another, and
probably more motivating, characterization of this property.

\begin{exa}
All the functors of Example~\ref{ex:1} preserve weak pullbacks, except for
the neighborhood functor and its monotone variant.
It can be shown that the property of preserving weak pullbacks is preserved
under the operations $\cof,+,\times$ and $(-)^{D}$, so that all extended 
polynomial Kripke functors (Example~\ref{ex:2}) preserve weak pullbacks.
\end{exa}

\paragraph{Standard functors}
The second property that we will impose on our set functors is that of
standardness. 
Given two sets $X$ and $X'$ such that $X \sse X'$, let $\iota_{X,X'}$ denote
the inclusion map from $X$ into $X'$.
A weak pullback-preserving set functor $\T$ is \emph{standard} if it 
\emph{preserves inclusions}, that is, if $\T\iota_{X,X'} = \iota_{\T X,\T X'}$ 
for every inclusion map $\iota_{X,X'}$.

\begin{rem}
Unfortunately the definition of standardness is not uniform throughout the
literature.
Our definition of standardness is taken from Moss~\cite{moss:coal99},
while for instance Ad\'{a}mek \& Trnkov\'{a}~\cite{adam:auto90} have an 
additional condition involving so-called distinguished points.
Fortunately, the two definitions are equivalent in case the functor preserves 
weak pullbacks, see Kupke~\cite[Lemma A.2.12]{kupk:fini06}.
Since we almost exclusively consider standard functors that also preserve
weak pullbacks, we have opted for the simpler definition. 

For readers who are interested in some more details, fix sets 0,1 and 2 of 
of the corresponding sizes (0,1 and 2), respectively, and let $e,o$ denote
the two maps $e,o: 1\to 2$.
Then the second condition of standardness in the sense of~\cite{adam:auto90}
can be phrased as the requirement that $\T 0 = \{x\in \T 1 \mid \T i(x)=
\T o(x)\}$, in words: all distinguished points are standard.
\end{rem}

In any case the restriction to standard functors is for convenience only, 
since every set functor is `almost 
standard'~\cite[Theorem~III.4.5]{adam:auto90}.
That is, given an arbitrary set functor $\T$, we may find a standard set
functor $\T'$ such that the restriction of $\T$ and $\T'$ to all non-empty 
sets and non-empty functions are naturally isomorphic.
The important observation about $T'$ is that $\Alg(T)\cong\Alg(T')$ and
$\Coalg(T)\cong\Coalg(T')$.
Consequently, in our work we can assume without loss of generality that our 
functors are standard and we will do so whenever convenient.

\begin{exa}
The finitary bag functor $\Bag_{\om}$ of Example~\ref{ex:1} is
not standard, but we may `standardize' it by representing any map $\mu: X
\to \bbN$ of finite support by its  `positive graph' $\{ (x,\mu x) \mid
\mu x > 0 \}$. Similarly, the finite distribution functor $D_\om$ 
can be standardized by identifying
a probability distribution $\mu: X \to [0,1] \in D_\om X$ with the (finite) set $\{(x,\mu x) \mid \mu x > 0 \}$.
\end{exa}

\paragraph{Finitary functors}
Let $\T$ be a set functor that preserves inclusions.
Then $\T$ is \emph{finitary} or \emph{$\om$-accessible} if, for all sets
$X$,
\[
TX = \bigcup \{TY \mid
Y\subseteq X, \textrm{finite}\}.
\] 
Generalizing the construction of $\Pom$ from $\funP$, we can define, for
any set functor $\T$ that preserves inclusions, its \emph{finitary version}
\label{page:Tom}
$\Tom: \Set \to \Set$ by putting
\begin{eqnarray*}
\Tom(X) &\isdef&  \bigcup \{\T Y \mid Y\sse_{\om} X \},
\\
\Tom(f) &\isdef& \T f.
\end{eqnarray*}
It is easy to verify that $\Tom$ preserves inclusions, is finitary and a
subfunctor of $\T$ as we have a natural transformation $\tau_{X}: \Tom X
\hookrightarrow \T X$.
Given the definition of the action of $\Tom$ on arrows, we shall often write
$\T f$ instead of $\Tom f$.

In order to avoid confusion, we already mention the following fact, but we 
postpone its proof until subsection~\ref{ss:standard}.

\begin{prop}
\label{p:Tomwp}
Let $\T$ be a standard set functor that preserves weak pullbacks.
Then $\Tom$ is also a standard functor that preserves weak pullbacks.
\end{prop}

The reason that we are interested in finitary functors is that we want our
language to be \emph{finitary}, in the sense that a formula has only finitely
many subformulas.
The key property of finitary functors that will make this possible, is that
every $\al \in \T X$ is supported by a finite subset of $X$, and in fact,
there will always be a \emph{minimal} such set.

\begin{definition}
\label{d:base}
Given a finitary functor $\T$ and an element $\al \in \T X$, we define
\[
\Base^{\T}_{X}(\al) \isdef \bigcap \{ Y \sse_{\om} X \mid \al \in \T Y
\}.
\]
\end{definition}

We write $\Base^{\T}$ rather than $\Base^{\Tom}$, and in fact omit the
superscript whenever possible.

\begin{exa}
  The following examples are easy to check: $\Base^{\Id}_{X}: X \to
  \Pom X$ is the singleton map, $\Base^{\funP}_{X}: \Pom X \to \Pom X$
  is the identity map on $\Pom X$, $\Base^{\Btree}_{X}: C \times X
  \times X \to \Pom X$ maps the triple $(c,x_{1},x_{2})$ to the set
  $\{ x_{1}, x_{2} \}$, and $\Base^{D_{\om}}$ maps a finitary
  distribution to its support.
\end{exa}

\begin{prop}
\label{fact:basenatural} 
Let $\T: \Set \to \Set$ be a standard functor that preserves weak
pullbacks.
\begin{enumerate}[\em(1)]
\item For any $\al \in \Tom X$, $\Base^{\T}_{X}(\al)$ is the smallest
set $Y$ such that $\al \in \T Y$.

\item $\Base^{\T}$ provides a natural transformation $\Base: \Tom \to
\Pom$.
\end{enumerate}
\end{prop}

\begin{proof}
Part~(1) is proved in~\cite{vene:auto06}.

For the second part, consider a map $f:X\to X'$.
We have to show $\Pom f\cof \Base_X=\Base_{X'}\cof T_\omega f$.
Fix $\alpha\in T_\omega X$ and write $B=\Base_X(\alpha)$ and $B'=
\Base_{X'}(T_\omega f(\alpha))$.
We need to prove $B'=f[B]$. 
  
For the inclusion ``$\subseteq$'', from 
$$\xymatrix{
    T_\omega B \ar@{^{(}->}[d]\ar[r] & T_\omega(f[B]) \ar@{^{(}->}[d] \\
    T_\omega X \ar[r]^{T_\omega f} & T_\omega X'}
$$
we see that $f[B]$ supports $T_\omega f(\alpha)$ and, as $B'$ is the
smallest such, $B'\subseteq f[B]$ follows. 

For the opposite inclusion ``$\supseteq$'', since $T_\omega$ preserves weak
pullbacks, the dotted arrow in
$$\xymatrix{
1 \ar@/_/[ddr]_{\alpha} \ar@/^/[drr]^{T_\omega f(\alpha)}
   \ar@{.>}[dr]|-{}            \\    
& T_\omega(f^{-1}(B')) \ar@{^{(}->}[d]\ar[r] & T_\omega(B') \ar@{^{(}->}[d] \\
&    T_\omega X \ar[r]^{T_\omega f} & T_\omega X'}
$$
exists and shows that $\alpha\in T_\omega(f^{-1}(B'))$. By minimality
of the base, it follows $B\subseteq f^{-1}(B')$, that is, $B'\supseteq
f[B]$.
\end{proof}

\begin{rem}
   A stronger version of the previous proposition follows from results 
in~\cite{gumm05:from}. Let us briefly sketch 
the details using the terminology of~\cite{gumm05:from}.
First of all note that it is not difficult to see that
all finitary set functors preserve intersections. Therefore \cite[Theorem 7.4] {gumm05:from}
implies that $\Base$ is sub-cartesian (not
necessarily natural) and this implies together with \cite[Theorem 8.1]{gumm05:from} that $\T$ preserves preimages iff $\Base$ is natural.
Any weak pullback preserving functor preserves preimages and thus
this statement implies Proposition~\ref{fact:basenatural}.
\end{rem}

%% file: sec-relationlifting.tex
\section{Relation Lifting} 
\label{s:relationlifting}

Given the key role that the lifting of binary relations plays in the 
semantics of Moss' logic, we need to discuss the notion in some detail.
After giving the formal definition, we mention some of the basic properties
of relation lifting: first the ones that hold for any functor, then the ones
for which we require the functor to preserve weak pullbacks, and finally,
we see important technical properties of relation lifting that rest on the 
fact that the set functor under consideration is standard. 
We discuss the connection of the relation lifting with categorical 
distributive laws: as we will see later on, this connection plays an important
role in the axiomatization of $\nb$. 
Finally we introduce the notion of a slim redistribution, which is needed to
formulate one of our axioms.

\subsection{Basics}
\label{ss:basics2}

First we give the formal definition of relation lifting.

\begin{definition} \label{d:rellift}
Let $\T$ be a set functor.
Given a binary relation $R$ between two sets $X_1$
and $X_2$, we define the relation $\Tb R \subseteq 
\T X_1 \times \T X_2$
as follows:
\[
\Tb R := \{ ((\T\pi^R_{1}) \rho, (\T\pi^R_{2})\rho) \mid
 \rho \in \T R \}.
\]
The relation $\Tb R$ will be called the \emph{$\T$-lifting} of $R$.
\end{definition}

\noindent In other words, we apply the functor $\T$ to the relation $R$, seen as a
\emph{span} $\xymatrix{X_{1}  & R \ar[l]_{\pi_{1}} \ar[r]^{\pi_{2}} & X_{2}}$,
and define $\Tb R$ as the image of $\T R$ under the product map $\langle
\T\pi_{1},\T\pi_{2}\rangle$ obtained from the lifted projection maps
$\T\pi$ and $\T\pi'$.
In a diagram:
\\ \centerline{\xymatrix{%
X_{1}  & R \ar[l]_{\pi_{1}} \ar[r]^{\pi_{2}} & X_{2}
\\
\T X_{1} & \T R \ar[l]_{\T \pi_{1}} \ar@{>>}[d]
\ar@/_5mm/[dd]_{\langle\T\pi_{1},\T\pi_{2}\rangle}
\ar[r]^{\T\pi_{2}} & \T X_{2}
\\
   & \Tb R \ar@{^{(}->}[d]
\\
   & \T X_{1} \times \T X_{2} \ar[ruu] \ar[luu]
}}

Let us first see some concrete examples.

\begin{exa}\label{ex:rellift}
Fix two sets $X$ and $X'$, and a relation $R \sse X \times X'$.
For the identity and constant functors, we find, respectively:
\begin{eqnarray*}
   \rl{\Id}  R &=& R
\\ \rl{C} R &=& \id_{C}.
\end{eqnarray*}
The relation lifting associated with the power set functor $\funP$ can be 
defined concretely as follows:
\[
\Pb R = \{ (A,A') \in \funP X \times \funP X' \mid 
   \forall a \in A\, \exists a' \in A'. aRa'
   \text{ and }
   \forall a' \in A'\, \exists a \in A. aRa'
   \}.
\]
This relation is known under many names, of which we mention that of the 
\emph{Egli-Milner} lifting of $R$.
Relation lifting for the finitary multiset functor is slightly more involved: 
given two maps $\mu\in \Bag_{\om}X,
\mu'\in \Bag_{\om}X'$, we put
\begin{align*}
\mu \rel{\rl{\Bag}_{\om}R} \mu' \text{ iff there is some map } \rho: R\to\bbN 
   \text{ such that }
 & \forall x \in X.\, \textstyle{\sum} \{ \rho(x,x') \mid x' \in X' \} = \mu(x)
\\ \text{and } 
 & \forall x' \in X'.\, \textstyle{\sum} \{ \rho(x,x') \mid x \in X \} = \mu'(x').
\end{align*}
The definition of $\rl{D_{\om}}$ is similar.

Finally, relation lifting interacts well with various operations on 
functors~\cite{herm98:stru}.
In particular, we have
\begin{eqnarray*}
   \rl{T_{0}\cof T_{1}}R   &=& \rl{T_0}(\rl{T_1} R)
\\ \rl{T_{0}+T_{1}}R       &=& \rl{T_0}R \cup \rl{T_1}R
\\ \rl{T_{0}\times T_{1}}R &=& 
      \left\{ \left((\xi_{0},\xi_{1}),(\xi'_{0},\xi'_{1})\right) \mid
      (\xi_{i},\xi'_{i}) \in \rl{T_i}R, \text{ for } i \in \{0,1\} 
      \right\}
\\ \rl{T^{D}}R &=& \{ (\phi,\phi') \mid (\phi(d),\phi'(d)) \in \rl{T}R
      \text{ for all } d \in D \}.
\end{eqnarray*}
From this one may easily calculate the relation lifting of all extended 
Kripke polynomial functors of Example~\ref{ex:2}.
\end{exa}

\begin{rem}
\label{r:rl-wd}
Strictly speaking, when defining the $\T$-lifting of a relation $R \sse X_{1}
\times X_{2}$, we should explicitly mention the type of $R$, that is, the 
pair of sets $X_{1}$ and $X_{2}$.

To see this, let $X_{1},X_{2},Y_{1}$ and $Y_{2}$ be sets such that $Y_{i} 
\sse X_{i}$, for $i \in \{ 1,2 \}$.
Now any relation $R \sse Y_{1} \times Y_{2}$ can also be seen as a relation
between $X_{1}$ and $X_{2}$.
But in general we do not have $\T Y_{i} \sse \T X_{i}$, and so the relation
$\Tb R \sse Y_{1}\times  Y_{2}$ is not necessarily a relation between 
$X_{1}$ and $X_{2}$.
It is easy to see that if $\T$ preserves inclusions, then this problem
evaporates. 
Since we will assume $\T$ to be standard almost throughout the paper, we
ignore this subtlety for the time being.
Readers who are worried about this may add the condition that $\T$ preserves
inclusions throughout the subsections~\ref{ss:basics2} and~\ref{ss:wpp}.
\end{rem}

\begin{rem}
\label{r:bisi}
Relation lifting can be used to define the notion of a \emph{bisimulation}
between two coalgebras. 
Recall that, given two coalgebras $\bbX_{1} = \struc{X_{1},\xi_{1}}$ and
$\bbX_{2} = \struc{X_{2},\xi_{2}}$, a relation $Z \times X_{1} \times X_{2}$
is a bisimulation if there is a coalgebra map $\zeta: Z \to \T Z$ making the
two projection functions $\pi_{1}: Z \to X_{1}$ and $\pi_{2}: Z \to X_{2}$ 
into coalgebra morphisms.
It can be shown that this is equivalent to requiring that $\xi_{1}(x_{1})
\rel{\Tb Z} \xi(x_{2})$ whenever $x_{1} \rel{Z} x_{2}$. 
\end{rem}

As mentioned, in this section we will discuss some important properties
of relation lifting. 
We start with listing a number of properties that $\T$-lifting
has for {\em any} given set functor $\T$.
The proof of the fact below is elementary.
\begin{fact}
\label{fact:basiclift}
Let $\T$ be an arbitrary set functor. 
Then the relation lifting $\Tb$
\begin{enumerate}[(1)]
\item extends $\T$: $\Tb f = \T f$ for all functions $f:X_{1} \to X_{2}$, 
\item preserves the diagonal: $\Tb \Id_{X} = \Id_{\T X}$ for any set $X$;
\item is monotone: $R \subseteq Q$ implies $\Tb {R} \subseteq \Tb {Q}$ for 
   all relations $R,Q \subseteq X_{1} \times X_{2}$;
\item commutes with taking converse: $\Tb \converse{R}=\converse{(\Tb R)}$ 
  for all relations $R \subseteq X_{1} \times X_{2}$.
\end{enumerate}
\end{fact}

\subsection{Weak pullback preserving functors}
\label{ss:wpp}

Fact~\ref{fact:basiclift} states a number of operations on relations that
interact well with relation lifting.
Conspicuously absent in that list is \emph{relational composition}: observe
that $\Tb$ would be a \emph{functor} on the category $\Rel$ if it would 
satisfy $\Tb(R\corel Q) = \Tb R \corel \Tb Q$.
Here we arrive at the main reason why we are interested in functors that
preserve weak pullbacks: as we will see now, that property is a necessary
and sufficient condition on $\T$ for $\Tb$ to be functorial.

In fact, given the characterisation of (weak) pullbacks in the category $\Set$,
in terms of the relation $\pb$ (see Fact~\ref{f:wp}), it is easy to formulate 
the composition $R\corel Q$ of two relations $R$ and $Q$  as a pullback of the
projection maps $\pi_2^R$ and $\pi_1^Q$. 
Therefore it is not surprising that the question whether the $\T$-lifting
of a relation commutes with the composition of relations is tightly connected 
with the preservation of weak pullbacks by $\T$. 
The following fact was first proved in~\cite{trnk80:gene}.

\begin{fact}\label{fact:char_wpp}
A functor $\T:\Set \to \Set$ weakly preserves pullbacks iff for all relations 
$R \subseteq X_1\times X_2$ and $Q \subseteq X_2 \times X_3$ we have
\begin{equation}
\label{eq:char-wpp}
\Tb (R \corel  Q) = \Tb R \corel  \Tb Q. 
\end{equation}
\end{fact}

\begin{proof}
First, assume that $\T$ preserves weak pullbacks and let $R\subseteq X_1\times
X_2$ and $Q \subseteq X_2 \times X_3$ be two binary relations. 
The pullback of  $\pi_2^R$ and $\pi_1^Q$ is given  by the following set:
\[
\pb \coloneqq 
  \{ \left\langle(x_1,x_2),(x_3,x_4)\right\rangle \mid
   (x_1,x_2) \in R, (x_3,x_4) \in Q \; \mbox{and} \;
   x_2 = x_3\} ,
\]
   and there is a surjective map $e: \pb( \pi_2^R,\pi_1^Q)
   \twoheadrightarrow 
   R \corel  Q$ given by $e(\left\langle(x_1,x_2),(x_3,x_4)
   \right\rangle) = (x_1,x_4)$ with the property that
   \begin{equation}\label{equ:e}
   		\pi_1^{R\corel Q} \cof e = \pi_1^R \cof \pi_1^{\pb} \quad
		\mbox{and} \quad \pi_2^{R\corel Q} \cof e = 
		\pi_2^Q \cof \pi_2^{\pb}.
   \end{equation}

The situation is depicted in Figure~\ref{fig:compo}.
\begin{figure}
\centerline{\xymatrix{
      & & \pb \ar@/_{.7cm}/[ldd]_{\pi_1^\pb} \ar@/^{.7cm}/[rdd]^{\pi_2^\pb}
      \ar@{-->>}[d]^e & & \\
     & &  R\corel Q \ar@/_{1cm}/[lldd]_{\pi_1^{R\corel Q}} \ar@/^{1cm}/[rrdd]^{\pi_2^{R\corel Q}} & & \\
& R \ar[ld]_{\pi^R_1} \ar[rd]^{\pi^R_2} & & Q \ar[ld]_{\pi_1^Q}
\ar[rd]^{\pi_2^Q} & \\
X_1 & & X_2 & & X_3 }}
\caption{Composition of relations \& pullback}
\label{fig:compo}
\end{figure}
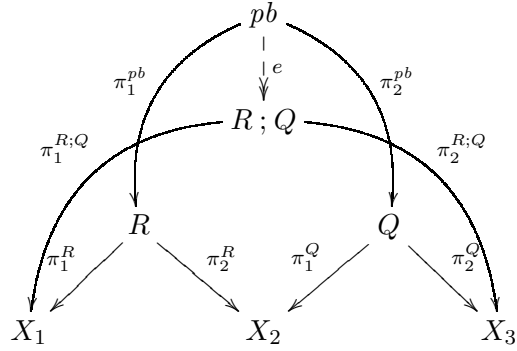

We now prove \eqref{eq:char-wpp}.
For the inclusion ``$\sse$'', let $(x,y) \in \Tb(R\corel Q)$.
By definition there exists some $z \in \T(R\corel Q)$ such that $\T \pi^{R\corel Q}_1(z)
= x$ and $\T \pi^{R\corel Q}_2(z) = y$.
We know that $e$ and thus also $\T e$ is surjective.
Therefore there exists some $z' \in \T(\pb)$ such that $\T e(z')= z$, and
using (\ref{equ:e}) we obtain $\T\pi_1^R(\T \pi_1^\pb(z')) = 
\T\pi_1^{R\corel Q}(\T e(z')) = \T \pi_1^{R\corel Q}(z)= x$ and similarly 
$\T\pi_2^Q(\T \pi_2^\pb(z')) = y$.
   On the other hand, by the definition of $\pb$, we have
   $\T \pi_2^R(\T\pi_1^\pb(z'))=\T \pi_1^Q(\T\pi_1^\pb(z'))=u$.
   This implies that $(x,u) \in \Tb(R)$ and $(u,y) \in \Tb(Q)$
   and we proved $(x,y)\in \Tb(R)\corel \Tb(Q)$ as required.
   
For the converse inclusion suppose that $(x,y)  \in \Tb(R)\corel \Tb(Q)$. 
We want to prove that this implies $(x,y)\in \Tb(R\corel Q)$.
It follows from $(x,y)  \in \Tb(R)\corel \Tb(Q)$ that there is some $u \in
\T X_{2}$ such that $(x,u) \in \Tb(R)$ and $(u,y) \in \Tb(Q)$; 
spelling out the definitions we find a $u_x \in \T R$ and a $u_y \in \T Q$ 
such that $\T \pi_1^R(u_x)=x$, $\T \pi_2^Q(u_y) =y$ and 
$\T \pi_2^R(u_x) =\T\pi_1^Q(u_y) = u$.
By our assumption that $\T$ is weak pullback preserving we have that 
$\T(\pb)$, together with the maps $\T \pi_1^{\pb}$, $\T \pi_2^{\pb}$ is the
weak pullback of $\T \pi^R_2$ and $\T \pi^Q_1$.
Therefore there must be some $z \in \T (\pb)$ such that $\T \pi_1^\pb (z) =
u_x$ and $\T \pi_2^\pb(z) = u_y$.
This implies 
   $$ \T \pi^{R\corel Q}_1 (\T e (z)) = \T \pi_1^R(
   \T \pi_1^\pb (z)) = \T \pi_1^R (u_x) = x$$
   and likewise $\T \pi^{R\corel Q}_2 (\T e(z)) = y$.
By definition this means that $(x,y)\in \Tb(R\corel Q)$ as required.
\medskip

For the converse implication of the statement of the proposition, suppose
that $\T$ does not preserve weak pullbacks and let the following be a 
pullback that is not weakly preserved by $\T$:\\
   \centerline{
   	\xymatrix{ P\ar[d]_{p_1} \ar[r]^{p_2} & X_2 \ar[d]_g\\ 
	         X_1 \ar[r]_f & X_3}
   }
Then it is not difficult to see that the following isomorphic diagram, is
also a pullback diagram that is not weakly preserved by $\T$:\\
   \centerline{
    \xymatrix{ R\ar[d]_{\pi^R_1} \ar[r]^{\pi^R_2} & 
    \converse{\Graph(g)} 
      \ar[d]^{\pi_1^{\converse{g}}}\\ 
	         \Graph(f) \ar[r]_{\pi_2^f} & X_3}}
where $\Graph(f)$ and $\converse{\Graph(g)}$ denote the graph of $f$ and the
converse of the graph of $g$, respectively, and $R \sse \Graph{f} \times 
\converse{\Graph{g}}$ is the pullback of $\pi_2^f$ and $\pi_1^{\converse{g}}$.
We will show the existence of a pair $(x,y) \in 
\Tb f \corel  \Tb \converse{g} \setminus \Tb(f\corel \converse{g})$, which
is a clear counterexample to \eqref{eq:char-wpp}.

As before there is a surjection $e': R \twoheadrightarrow f\corel \converse{g}$
satisfying
\begin{equation}
\label{eq:qq1}
\pi_{1}^{f\corel \converse{g}} \cof e' = \pi_{1}^{f} \cof \pi_{1}^{R}
\text{ and }
\pi_{2}^{f\corel \converse{g}} \cof e' = \pi_{2}^{\converse{g}} \cof \pi_{2}^{R}
\end{equation}
By assumption, $(\T R, \T \pi_{1}^{R}, \pi_{2}^{R})$ is not a weak pullback 
of $\T \pi_2^f$ and $\T \pi_1^{\converse{g}}$.
Hence by Fact~\ref{f:wp}(2), there must be a $z_1 \in \T \Graph(f)$ and a
$z_2 \in \T \converse{\Graph(g)}$ such that $\T \pi_2^f (z_1) = 
\T \pi_1^{\converse{g}}(z_2) = u$, while
\begin{equation}\label{equ:noexist}
\mbox{there is no } z \in \T R \mbox{ such that } \T\pi_1^R(z) = z_1
		\mbox{ and } \T \pi_2^R(z) = z_2 .
\end{equation}
Define $x \coloneqq \T \pi_1^f(z_1)$ and $y\coloneqq\T
\pi_2^{\converse{g}}(z_2)$.
Since $\pi_{2}^{f} = f \cof \pi_{1}^{f}$, we have $\T\pi_{2}^{f} = \T f 
\cof \T\pi_{1}^{f}$, and so we find $u = (\T f) x$; likewise, we obtain
$u = (\T g) y$.
From this it is clear that $(x,y) \in \Tb f\corel \Tb \converse{g}$.
Now suppose for a contradiction that $(x,y) \in \Tb (f\corel \converse{g})$.
By definition this entails the existence of some $z' \in \T(f\corel \converse{g})$
such that $\T \pi_1^{f\corel \converse{g}}(z')=x$ and $\T \pi_2^{f\corel \converse{g}}(z')
=y$.
By surjectivity of $e'$, and hence, of $\T e'$, then there must be some 
$z'' \in \T R$ such that $\T e(z'')=z'$.
Furthermore it follows from \eqref{eq:qq1} that 
$$x = \T \pi_1^{f\corel \converse{g}} (z') 
= \T \pi_1^{f\corel \converse{g}} (\T e'(z'')) 
= \T\pi_1^f( \T\pi_1^R(z''))
$$
and, similarly, $y = \T\pi_2^{\converse{g}}( \T\pi_2^R(z''))$.
Both $\T \pi_1^f$ and $\T \pi_2^{\converse{g}}$ are isomorphisms and thus
we obtain $\T\pi_1^R(z'')=z_1$ and $\T \pi_2^R(z'') = z_2$ - a contradiction
to (\ref{equ:noexist}) above.
\end{proof}

Putting this together with Fact~\ref{fact:basiclift}(2,3) we immediately
obtain the following.

\begin{cor}\label{cor:extT}
Let $\T$ be a set functor and let $\Tb$ be the operation that maps a set $X$
to $\Tb X \coloneqq \T X$ and a relation $R$ to the $\T$-lifting $\Tb R$ of
$R$.
Then the following are equivalent:
\begin{enumerate}[\em(1)]
\item
$\T$ preserves weak pullbacks;
\item
$\Tb$ is a functor on the category $\Rel$ of sets and relations;
\item $\Tb$ is a \emph{relator}, that is, a monotone functor on the
  category $\Rel$.
\end{enumerate}
\end{cor}

Closely related to this is an important consequence of the functor preserving
weak pullbacks, namely that the notions of bisimilarity and behavioral
equivalence coincide.

\begin{rem}
In \cite{rutt:univ00} it is proved that if $\T$ preserves weak pullbacks
then for any pair of coalgebras $\bbX = \struc{X,\xi}$ and $\bbX' =
\struc{X',\xi'}$, two states $x$ and $x'$ are behaviorally equivalent iff
there is a bisimulation (see Remark~\ref{r:bisi}) linking $x$ to $x'$.
\end{rem}

\subsection{Standard functors}
\label{ss:standard}

As mentioned earlier on we will almost exclusively work with $\Set$-functors 
that are standard.
In Remark~\ref{r:rl-wd} we saw that this will ensure that the definition of
the lifting of a relation $R$ is independent of the type of $R$.
Now we will see some further nice consequences of standardness for the
notions of relation lifting.

To start with, in case $\T$ is standard, $\Tb$ commutes with the domain and 
range of a function; and if $\T$ preserves weak pullbacks in addition, then
$\Tb$ also commutes with restrictions.

\begin{prop}
\label{p:st-rl}
Let $\T$ be a standard set functor. Then
\begin{enumerate}[\em(1)]
\item $\Tb$ commutes with taking domains: $\Dom(\Tb R) = \T(\Dom R)$
  for all relations $R \sse X_{1} \times X_{2}$. 
\item $\Tb$ commutes with taking range: $\Ran(\Tb R) = \T(\Ran R)$
  for all relations $R \sse X_{1} \times X_{2}$. 
\item If $\T$ preserves weak pullbacks, then $\Tb$ commutes with taking
  restrictions:
  \[
  \Tb (R\rst{Y_{1}\times Y_{2}}) = (\Tb R)\rst{\T Y_{1} \times \T Y_{2}}
  \]
  for all sets $X_{1},X_{2},Y_{1}$ and $Y_{2}$, with $Y_{1}\sse X_{1}$ and
  $Y_{2}\sse Y_{1}$, and for all relations $R \sse X_{1} \times X_{2}$.
\end{enumerate}
\end{prop}

\begin{proof}
For part~1, we first consider the inclusion $\Dom(\Tb R) = \T(\Dom R)$.
Let $R \subseteq X_1 \times X_2$ be a relation and take an element
$\al \in \Dom(\Tb R)$.
Then $(\alpha,\beta) \in \Tb R$, for some $\be\in \T X_{2}$.
We denote by $\iota: \Dom(R) \to X_1$ the inclusion of $\Dom(R)$ into $X_1$
and by $\pi_1': R \to \Dom(R)$ the restriction of the projection map $\pi_1:
R \to X_1$; then we have $\pi_1 = \iota \cof \pi_1'$.
By definition of $\Tb$ there exists some $\rho \in T R$ such that $\T \pi_1
(\rho) = \alpha$ and hence $\T \iota (\T \pi_1' (\rho))=\alpha$.
As $\T$ is standard this shows that $\alpha = \T \pi_1'(\rho) \in \T \Dom(R)$ 
as required.

For the opposite inclusion, 
let $f: \Dom(R) \to \Ran(R)$ be any map such that $f \sse R$; then it follows
that $\T f \sse \Tb R$.
In other words, for all $\al \in \T(\Dom R)$ we have $\al \rel{\Tb R} \T f(\al)$.
From this it is immediate that $\T(\Dom R) \sse \Dom(\Tb R)$.

The proof of part~2 is completely analogous. 
For part~3, we refer to~\cite[Prop.~6.4]{kuve08:coal}.  
\end{proof}

\noindent
Proposition~\ref{p:st-rl} is particularly useful for linking the relation
lifting of $\T$ to that of its finitary version $\Tom$.

\begin{prop}
\label{lem:tomb}
\label{p:tomb}
Let $\T$ be a standard and weak pullback preserving set functor, let $\Tom$
be its finitary version and let $R \subseteq X_1 \times X_2$ be a relation. 
Then
\[
\Tomb R = \Tb R  \cap (\Tom X_{1} \times \Tom X_{2}).
\]
\end{prop}

\begin{proof}
Let $R \subseteq X_1 \times X_2$ be a relation and take a pair
$(\alpha,\beta) \in \Tom X_1 \times \Tom X_2$. 
By definition of $\Tom$ there must be finite sets $X_1' \subseteq_\omega X_1$
and $X_2 '  \subseteq_\omega X_2$ such that $\alpha \in \Tom X_1' = \T X_1'$ 
and $\beta \in \Tom X_2' = \T X_2'$.

In order to prove the inclusion $\supseteq$, assume that $(\alpha,\beta)
\in \Tb R$.
By Proposition~\ref{p:st-rl} we have
\begin{equation}\label{equ:resricttofiniterel}
    (\alpha,\beta) \in \Tb R \quad \mbox{iff} \quad 
    (\alpha,\beta) \in \Tb (R\rst{X_1' \times X_2'}) 
\end{equation}
and because $\Tomb(R\rst{X_1' \times X_2'}) \subseteq \Tomb(R)$ the inclusion
holds if we can prove that
    $(\alpha,\beta) \in \Tomb R'$ with $R' \coloneqq R\rst{X_1' \times X_2'}$.
    The following diagram commutes: \\
    \centerline{\xymatrix{
    \Tom X_1' \ar@{=}[d] & \Tom R' \ar@{=}[d] \ar[l] \ar[r] & \Tom X_2' \ar@{=}[d]\\
    \T X_1' & \ar[l] \T R' \ar[r] & \T X_2' 
    }}
Therefore we have that $(\alpha,\beta) \in \Tb R'$ iff $(\alpha,\beta) \in 
\Tomb R'$. By (\ref{equ:resricttofiniterel}) we have $(\alpha,\beta) \in \Tb
R'$ and hence $(\alpha,\beta) \in \Tomb R'$ as required.
The proof of the opposite inclusion is similar.
\end{proof}

On the basis of Proposition~\ref{p:tomb} we will often be sloppy and write
$(\al,\be) \in \Tb R$ instead of $(\al,\be) \in \Tomb R$,
for elements $\alpha \in \Tom X_1$ and $\beta \in \Tom X_2$. 
More importantly, Proposition~\ref{p:tomb} allow us to prove our earlier 
claim, that $\Tom$ inherits the properties of standardness and weak pullback
preservation from $\T$.

\begin{proofof}{Proposition~\ref{p:Tomwp}}
     Let $\T$ be a standard, weak pullback preserving set functor.
     In order to see that $\Tom$ is standard consider two sets $X,X'$
     with $X'\subseteq X$ and let $\iota:X' \to X$ be the inclusion of $X'$
     into $X$. By the definition of $\Tom$ for every set $X$ we have that
     $\Tom X$ is a subset of $\T X$ and that the inclusion 
     $\tau_X: \Tom X \to \T X$ is natural. 
It follows by naturality that $\Tom\iota$ is also an inclusion:
$$\xymatrix{
   T_\omega X' \ar[d]_{\Tom \iota} \ar@{^{(}->}[r]^{\tau_{X'}} 
   & T X' \ar@{^{(}->}^{\T\iota}[d] 
\\
   T_\omega X  \ar@{^{(}->}[r]_{\tau_{X}} 
   & T_\omega X}
$$
More precisely, for all $\alpha \in \Tom X$ we have
\begin{eqnarray*}
\Tom \iota (\alpha) 
  & = & 
  \tau_X (\Tom \iota (\alpha)) 
  \stackrel{\mbox{\tiny (nat. of $\tau$)}}{=}  
  \T \iota (\tau_X'(\alpha)) = \T \iota (\alpha) 
		\stackrel{\mbox{\tiny{$\T$ standard}}}{=}  
\alpha
\end{eqnarray*}
which demonstrates that $\Tom \iota$ is the inclusion map from $\Tom X'$
into $\Tom X$, and shows that $\Tom$ is standard indeed.
     
     We now prove that $\Tom$ preserves weak pullbacks.
     By  Fact~\ref{fact:char_wpp} it suffices to prove
     that for arbitrary relations $R \subseteq X_1 \times X_2$ and
     $Q \subseteq X_2 \times X_3$ we have
     $\Tomb(R\corel Q) = \Tomb(R) \corel  \Tomb(Q)$. In order to see this
     we use Proposition~\ref{lem:tomb}. We have
\begin{eqnarray*}
(\alpha,\beta) \in \Tomb(R\corel Q) 
   & \mbox{iff} 
   & (\alpha,\beta) \in \Tb(R\corel Q)\rst{\Tom X_1 \times \Tom X_3} 
\\ & \mbox{iff} 
   &  (\alpha,\beta) \in \Tb(R\corel Q)\rst{\T X_1' \times \T X_3'} 
   \mbox{ for some } X_1' \subseteq_\omega X_1, X_3' \subseteq_\omega X_3 
\\ & \mbox{iff} 
   & (\alpha,\beta) \in \Tb((R\corel Q)\rst{X_1' \times X_3'}) 
   \mbox{ for some } X_1' \subseteq_\omega X_1, X_3' \subseteq_\omega X_3 
\\ & \mbox{iff} 
& (\alpha,\beta) \in \Tb(R\rst{X_1' \times X_2'}\corel Q\rst{X_2' \times X_3'})
\\ & & 
   \mbox{for some} \; X_1' \subseteq_\omega X_1,
       X_2' \subseteq_\omega X_2,
       X_3' \subseteq_\omega X_3 
\\ & \mbox{iff} 
   & (\alpha,\beta) \in \Tb(R\rst{X_1' \times X_2'})\corel \Tb(Q\rst{X_2' \times X_3'})
\\ & & \; \mbox{for some} \; X_1' \subseteq_\omega X_1,
        X_2' \subseteq_\omega X_2, 
	X_3' \subseteq_\omega X_3 
\\ & \mbox{iff} & (\alpha,\beta) \in \Tomb(R)\corel \Tomb(Q)
  \rlap{\hbox to 197 pt{\hfill\qEd}}
\end{eqnarray*}\let\mvsquareforqed=\relax
\end{proofof}\let\mvsquareforqed=\qed

\noindent Finally, we finish this subsection with noting that relation
lifting interacts well with the natural transformation $\Base: \Tom
\to \Pom$.

\begin{prop}
\label{prop:base-hom}
Let $\T$ be a standard functor that preserves weak pullbacks.
Given a relation $R \sse X_{1} \times X_{2}$ and elements $\al_{i} \in
\T X_{i}$, $i \in \{ 1,2 \}$, it follows from $\al_{1} \mathrel{\Tb R}
\al_{2}$ that $\Base(\al_{1}) \mathrel{\Pb R} \Base(\al_{2})$.
In particular, we have that $\Base(\al_{1}) \subseteq \Dom(R)$ and
$\Base(\al_{2}) \subseteq \Ran(R)$.
\end{prop}

\begin{proof}
Let $\pi^{R}_{i}$ be the projection of $R$ to $X_{i}$, then it follows
from $\al_{1} \mathrel{\Tb R} \al_{2}$ that $\al_{i} = \T\pi^{R}_{i}(\rho)$
for some $\rho\in \T R$.
But then by naturality of $\Base$ we find that $\Base(\al_{i}) =
\Base(\T\pi^{R}_{i}(\rho)) = (\funP \pi^{R}_{i})(\Base(\rho))$, and 
so $\Base(\rho) \in \funP R$ is a witness to the fact that
$\Base(\al_{1}) \mathrel{\Pb R} \Base(\al_{2})$.
\end{proof}

\subsection{Relation Lifting \& distributive laws}
\label{ss:distributive-laws}

A relation that plays an important role in our paper is the $\T$-lifting of 
the membership relation $\in$.
If needed, we will denote the element relation, restricted to a given set
$X$, as the relation ${\in_{X}} \sse X \times \funP X$.

\begin{definition}\label{def:elementlift}
Given a standard functor $\T$ that preserves weak pullbacks, we define, for
every set $X$, a function $\nbsem_X:\T \Pow X \to \Pow \T X$ by putting
\[
\nbsem_X(\Phi) \coloneqq \{ \alpha \in \T X \mid \alpha 
\mathrel{\Tb {\in_{X}}}
\Phi \}.
\]
Elements of $\nbsem_X(\Phi)$ will be referred to as \emph{lifted members}
of $\Phi$.
The family $\nbsem=\{\nbsem_X\}_{X \in \Set}$ will be called the {\em 
$\T$-transformation}.
\end{definition} 

Properties of $\Tb$ are intimately related to those of $\nbsem$.
In order to express the connection, we need to introduce the concept of a 
distributive law.

\begin{definition}\label{d:distlaw}
  Let $\T$ be a covariant set functor.  A \emph{distributive law} of
  $\T$ over a (co- or contravariant) set functor $M$ is a natural
  transformation $\theta: \T M \to M \T$; that is, the following
  diagram commutes, for every map $f: X \to Y$:
\[
\xymatrix{
    \T M X \ar[d]_{\T M f} \ar[r]^{\theta_X} 
  & M \T X   \ar[d]^{M \T f} 
\\  \T M Y \ar[r]^{\theta_Y} 
  & M \T Y }
\]
(Clearly, in case $M$ is a contravariant functor the downward arrows have to be
reversed.)
For $\theta$ to be \emph{distributive law} of $\T$ over a set monad $(M,\eta,
\mu)$, we require in addition that $\theta$ is compatible with the monad 
structure, in the sense that the following diagrams commute, for every set
$X$:
\begin{equation}
\label{eq:dl-diag}
\xymatrix{
\T X \ar[rd]_{\eta_{\T X}}
	       \ar[r]^{\T \eta_X} &
	       \T  M X
	       \ar[d]^{\theta_X} \\
	       &  M \T X }
\hspace*{20mm}
\xymatrix{\T  M  M X \ar[d]^{\T \mu_X}
	      	\ar[r]^{\theta_{ M X}} &   M \T  M X
		    \ar[r]^{ M \theta_X}&
	         M  M \T X \ar[d]^{\mu_X} \\
	        \T  M X \ar[rr]_{\theta_X}& &  M \T X
}
\end{equation}
\end{definition}

If the functor $\T$ preserves weak pullbacks, the $\T$-transformation
$\nbsem$ provides a distributive laws of $\T$ over the power set monad
$\PowMo = (\Pow,\{ \cdot\},\bigcup)$.
A detailed proof of this fact can be found in~\cite[Sec.~4]{bart04:trac}.	

\begin{fact}\label{fact:distriblaw}
	If $\T$ preserves weak pullbacks,
	$\nbsem=\{\nbsem_X\}_{X \in \Set}$ 
	is a distributive law of $\T$ over the
	power set monad $\PowMo$.
\end{fact}

What it means, set-theoretically, for $\nbsem$ to be a distributive law
of $\T$ over $\PowMo$ is the following.  The fact that $\nbsem$ is a
natural transformation from $\T\funP$ to $\funP\T$ is another way of
saying that for every map $f: X \to Y$, and every object $\Phi \in
\T\funP X$, we obtain the lifted members of $\T\funP\Phi$ by applying
the operation $\T f$ to the lifted members of $\Phi$.  The diagram on
the left of \eqref{eq:dl-diag}, relating the singleton map $\eta_{X}:
X \to \funP X$ to the $\T$-transformation, states that an object $\al
\in \T X$ is always the \emph{unique} lifted member of the lifted set
$\T\eta_{X}(\al)$.  To understand the diagram on the right, recall
that the multiplication $\mu$ of $\PowMo$ is the union map
$\bigcup_{X}: \funP\funP X \to \funP X$.  Applying the functor to this
we obtain a map $\T\bigcup_{X}: \T\funP\funP X \to \T\funP X$.
Observe that given an object $\Phi \in \T\funP\funP X$, we may thus
take lifted members of $(\T\bigcup_{X})(\Phi)$; however, we may also
take lifted members of $\Phi$ itself, and since each of these will
belong to the set $\T\funP X$, we may repeat the operation of taking
lifted members.  Now the right diagram in \eqref{eq:dl-diag} states
that the lifted members of $(\T\bigcup_{X})(\Phi)$ coincide with the
objects we may obtain as lifted members of lifted members of $\Phi$.

\begin{rem}
  The existence of a distributive law of a set functor $\T$ over the
  power set monad $\PowMo$ corresponds to an extension of the functor
  $\T$ to the Kleisli category $\Kl(\PowMo)$ of $\PowMo$.  Furthermore
  it is easy to see that $\Kl(\PowMo)$ is isomorphic to the category
  $\Rel$ of sets with relations.  Putting these facts together it is
  clear that any distributive law of a set functor $\T$ over $\PowMo$
  corresponds to an extension of $\T$ to a functor on the category
  $\Rel$.  We saw in Corollary~\ref{cor:extT} that the $\T$-lifting of
  a relation can be used to extend $\T$ to a functor $\Tb:\Rel \to
  \Rel$ iff $\T$ preserves weak pullbacks. In this case $\nbsem$ is
  the corresponding distributive law. 
Further remarks and references can be found in Section~\ref{s:rellift:notes}.
%
\end{rem}

Perhaps somewhat surprisingly, the $\T$-transformation can be also seen
as a distributive law over the \emph{contravariant} power set functor.

\begin{prop}
\label{p:nbdlfunQ}
Let $\T:\Set \to \Set$ be a functor that preserves weak pullbacks.
Then $\nbsem$ is a distributive law of $\T$ over the contravariant power
set functor.
\end{prop}

\begin{proof}
Let $f: X \to Y$ be a function. 
We have to show that the following diagram commutes:
\centerline{
\xymatrix{
    \T \funQ Y \ar[r]^{\nbsem_Y} \ar[d]_{\T\funQ f} 
  & \funQ\T Y \ar[d]^{\funQ \T f}
\\  \T \funQ X \ar[r]_{\nbsem_X} 
  & \funQ\T X}}
This can be verified by a straightforward calculation:
\[
\begin{array}{rclcl}
\alpha \in \nbsem_X((\T \funQ f)(\Phi)) 
  & \mbox{iff} & \Phi (\T \funQ f\corel \Tb {\ni_X}) \alpha  
  & \mbox{iff} & \Phi (\Tb(\funQ f\corel {\ni_X})) \alpha 
\\& \mbox{iff} & \Phi (\Tb({\ni_Y}\corel \converse{f})) \alpha 
  & \mbox{iff} & \Phi (\Tb {\ni_Y} \corel \converse{\T  f}) \alpha
\\& \mbox{iff} & \T f( \alpha) \in \lambda_Y(\Phi) 
  & \mbox{iff} & \alpha \in (\funQ  \T f) ( \lambda_Y(\Phi)) 
\end{array}
\]
Here we freely apply properties of relation lifting, and in the third
equivalence we use the easily verified fact that $\funQ f\corel {\ni_X} = 
{\ni_Y}\corel \converse{f}$.
\end{proof}

In our paper both distributive laws play an important role.
The fact that $\nbsem$ is a distributive law over $\funQ$ is essential for
proving that the semantics of Moss' logic is bisimulation invariant, and 
the distributivity of $\T$ over the monad $\PowMo$ is crucial for the 
soundness of our axiomatization.
\medskip

To finish this subsection, we gather some elementary facts on the
$\T$-transformation.

\begin{prop}
\label{p:nbsem}
Let $\T$ be a standard, weak pullback-preserving functor, let $X$ be some
set and let $\Phi \in \Tom\funP X$.
\begin{enumerate}[\em(1)]
\item If $\nada \in \Base(\Phi)$ then $\nbsem(\Phi) = \nada$.
\item \label{item:memberofdistri} 
   If $\Base(\Phi) \sse \{ Y \}$ for some $Y \sse X$, then $\nbsem(\Phi)
   \sse \T Y$.
\item \label{item:singletonredistri}
   If $\Base(\Phi)$ consists of singletons only, then $\size{\nbsem(\Phi)} =1$.
\item If $\T$ maps finite sets to finite sets, then for all $\Phi \in
  \Tom \Pom X$, $|\nbsem(\Phi)| < \omega$.
\item If $\Phi \in \Tom\Pom X$, then $\nbsem(\Phi) \in \funP\Tom X$.
\end{enumerate}
\end{prop}

\begin{proof}
For part~1, assume that $\nada \in \Base(\Phi)$ and assume for contradiction
that $\al$ is a lifted member of $\Phi$.
It follows by Proposition~\ref{prop:base-hom} that $\Base(\al) \Pin 
\Base(\Phi)$.
But from this it would follow, if $\nada \in \Base(\Phi)$, that $\Base(\al)$
contains a member of $\nada$, which is clearly impossible.
Consequently, the set $\nbsem(\Phi)$ must be empty.

In order to prove part~\ref{item:memberofdistri}, assume that $\Phi \in 
\T \{ Y \}$, for some subset $Y$ of $X$, and suppose that $\alpha \Tin \Phi$.
Then by Proposition~\ref{p:st-rl}(3) we have $\alpha \mathrel{\Tb 
{\in_{\rst{X \times \{ Y \}}}}} \Phi$ and so by part~1 of the same 
Proposition we find $\alpha \in \T \Dom (\in_{\rst{X \times \{ Y\}}}) =
   \T Y$.

For part~3, observe that another way of saying that $\Base(\Phi)$ consists
of singletons only, is that $\Phi\in \Tom S_{X}$, where $S_{X} \sse \funP X$
is the collection of singletons from $X$.
Let $\theta_{X}: S_{X} \to X$ be the inverse of $\eta_{X}$, that is,
$\theta_{X}$ is the bijection mapping a singleton $\{ x \}$ to $x$.
Clearly then, the map $\Tom\theta_{X}: \Tom X \to \Tom S_{X}$ is a bijection
as well.
In addition, we have $\cv{\theta_{X}} = {\in_{X}}$, from which 
it follows by elementary properties of relation lifting that 
$\cv{(\T\theta_{X})} = \Tb{\in_{X}}$.
From this it is immediate that if $\Phi \in \Tom S_{X}$, then $(\T\theta_{X})
(\Phi)$ is the unique lifted member of $\Phi$.

Concerning part~4, assume that $\Phi \in \Tom \Pom X$.
Then by definition, $\Phi \in \T \mathcal{Y}$ for some $\mathcal{Y}\sse_{\om}
\Pom X$.
From this it follows that $\mathcal{Y} \sse \funP Y$ for some finite
$Y \sse X$, and this implies that $\Base(\Phi) \sse \funP Y$.
If $\al$ is a lifted member of $\Phi$, then by Proposition~\ref{prop:base-hom}
we obtain $\Base(\al) \Pin \Base(\Phi)$, and so in particular we find 
$\Base(\al) \sse \bigcup \Base(\Phi) \sse Y$.
From this it follows that $\nbsem(\Phi) \sse \T Y$, and so by the assumption
on $\T$, the set $\nbsem(\Phi)$ must be finite.

Finally, we consider part~5.
Take an object $\Phi \in \Tom\Pom X$ and let $\al \in \T X$ be an arbitrary
lifted member of $\Phi$.
Reasoning just as for part~4, we obtain that $\al \in \T Y$ for some finite
$Y \sse X$, and so by definition of $\Tom$ we find that $\al \in \Tom X$.
\end{proof}

\subsection{Slim redistributions}
\label{ss:slim}

The syntax of Moss' logic is built using negations, conjunctions, 
disjunctions and the $\nb$-operator. 
An axiomatisation of the logic  has to specify the interaction of these
operations. 
As we will see, so-called \emph{slim redistributions} are the key to 
understand how conjunction interacts with the $\nb$-operator.

\begin{definition}
\label{d:srd}
Let $\T$ be a 
set functor.
A set $\Phi \in \T \Pow X$ is a {\em redistribution} of a set $A \in\Pow \T X$ 
if $A \subseteq \nbsem_X(\Phi)$, that is, every element of $A$ is a lifted
member of $\Phi$.
In case $A \in \Pom\Tom X$, we call a redistribution $\Phi$ {\em slim} if
$\Phi \in \Tom \Pom (\bigcup_{\alpha \in A} \Base(\alpha))$. 
The set of slim redistributions of $A$ is denoted as $\SRD(A)$. 
\end{definition}

Intuitively, redistributions of $A$ are ways to reorganize the material of 
$A$.
The slimness condition $\Phi \in \Tom\Pom (\bigcup_{\al \in A} \Base(\al))$
should be seen as a minimality requirement, ensuring that $\Phi$ is `built 
from the ingredients of $A$'.

\begin{exa}
\label{ex:srd}
First we consider the binary $C$-labelled tree functor $\Btree$ of 
Example~\ref{ex:2}.
Let $\pi_{C},\pi_{1}$ and $\pi_{2}$ denote the respective projections from
$\Btree X$ to $C$, $X$ and $X$, respectively.
An object $\Phi \in \Btree\funP X$ is of the form $(c,Y,Z)$ with $c\in C$
and $Y,Z \in \funP X$.
Such a $\Phi$ is a redistribution of a set $A = \{ (c_{i},y_{i},z_{i}) 
\mid i \in I \} \sse_{\om} \Btree X$ iff for all $i \in I$ we have
$c_{i} = c$, $y_{i} \in Y$ and $z_{i} \in Z$, and such a redistribution is
slim if in addition,  $Y \cup Z \sse \{ y_{i} \mid i \in I \} \cup \{ z_{i} 
\mid i \in I \}$.
On this basis it is not hard to derive that 
\[
\SRD(A) = \left\{\begin{array}{ll}
      \{(c,\nada,\nada) \mid c \in C \} & \text{if } A = \nada
   \\ \nada & \text{if}\  |\pi_{C}[A]|\ge 2
   \\ \{ (c_{A},S_1,S_2) \mid \pi_j[A] \subseteq S_j \subseteq \pi_1[A] \cup \pi_2[A] \mbox{ for } j=1,2\} & \text{if}\  \pi_{C}[A]=\{c_A\}
\end{array}\right.
\]
\end{exa}

\begin{rem}\label{rem:superslim}
For our purpose it would suffice to consider instead of $\SRD(A)$ a smaller set $\SRD'(A)$ as long as it order-generates $\SRD(A)$ in the sense that for all $\Phi\in\SRD(A)$ there is $\Phi'\in\SRD'(A)$ such that $\Phi' \,\Tb(\subseteq)\, \Phi$. 
Such an $\SRD'(A)$ can replace the $\SRD(A)$ in the rule ($\nb 2$) that will form a crucial part in our derivation system. 
In the example above, $\SRD'(A)$ can be given by simplifying the third clause to 
\[
\begin{array}{ll}
 \{ (c_{A},\pi_1[A],\pi_2[A]) \} & \text{if}\  \pi_{C}[A]=\{c_A\}
\end{array}
\]
We thank Fredrik Dahlqvist for pointing out that this clause does not give $\SRD(A)$.
\end{rem}

\begin{exa}\label{ex:srd2}
In case we are dealing with the power set functor $\funP$, first 
observe that given a set $X$, the relation $\Pb{\in_{X}} \sse \funP X \times
\funP\funP X$ is given by
\[
\al \Pin \Phi \mbox{ iff } \al \sse \bigcup\Phi \text{ and }
\al\cap\be\neq\nada \text{ for all } \be \in \Phi.
\]
%
On the basis of this observation it is easy to check that $\Phi \in \funP X$
is a redistribution of $A \in \funP\funP X$ if $\bigcup A \sse \bigcup\Phi$
and $\al\cap\be\neq\nada$ for all $\al\in A$ and $\be\in \Phi$.
Furthermore, we obtain
\[
\Phi \in \SRD(A) \text{ iff }
\bigcup A = \bigcup\Phi
\text{ and } \al\cap\be\neq\nada \text{ for all $\al\in A$, $\be\in \Phi$}.
\]
Hence, in the case of the power set functor we are dealing with a
\emph{symmetric} relation: $\Phi \in \SRD(A)$ iff $A \in \SRD(\Phi)$.
\end{exa}

The following observation, which is due to M.~B\'ilkov\'a, shows that 
slim redistributions naturally occur in the context of distributive
lattices.

\begin{exa}
\label{ex:bilk}
Let $\bbD$ be a distributive lattice.
The distributive law for $\bbD$ can be formulated as follows. For any set
$A \in \Pom\Pom D$, we have
\begin{equation*}
\label{eq:dl0}
\bw_{\al\in A} \bv \al = \bv_{\ga\in\mathit{CF}(A)} \bw \Ran(\ga),
\end{equation*}
where $\mathit{CF}(A)$ is the set of \emph{choice functions} on $A$, that is, 
$\mathit{CF}(A)$ is the set of maps $\ga: A \to D$ such that $\ga(\al) \in 
\al$, for all $\al \in A$.
Then it is straightforward to verify that the set 
$\{ \Ran(\ga) \mid \ga \in \mathit{CF}(A) \}$
is in fact a slim redistribution of $A$.

In fact, we may prove that 
\begin{equation}
\label{eq:rdl1}
\bw_{\al\in A} \bv \al = \bv_{\Phi\in\SRD(A)}\bv_{\phi\in\Phi} \bw \phi.
\end{equation}
Later on we will see that our axiom governing the interaction of $\nb$ with
conjunctions, generalizes \eqref{eq:rdl1}.
\end{exa}

We finish the section with a proposition for future reference.

\begin{prop}
\label{item:redistriofempty} 
$\SRD(\nada) = \T\{\nada\}$.
\end{prop}

\begin{proof}
If $\Phi$ is a slim redistribution of the empty set, then by definition
$\Phi \in \T\Pom(\nada) = \T \{ \nada \}$.
Conversely, any $\Phi \in \T \{ \nada \}$ satisfies the condition that
$\nada \sse \nbsem(\Phi)$, and so $\Phi \in \SRD(\nada)$.
\end{proof}

\subsection{Notes}\label{s:rellift:notes}

The relation lifting via spans as in Definition~\ref{d:rellift} was
defined by Barr in \cite[Section 2]{barr:rela70}. Without stating it
explicitly, he also proves that the relation lifting $\Tb$ is a
functor on $\Rel$ iff $T$ preserves weak pullbacks;
see also Trnkov\'a~\cite{trnk80:gene} and, for a generalisation beyond
set functors, Carboni, Kelly and Wood~\cite[4.3]{carb:2cat91} and
Hermida~\cite[Theorem~2.3]{hermida:relational-modalities}.
\cite{carb:2cat91} also studies the question which functors
$\Rel\to\Rel$
arise from functors $\Set\to\Set$. 
Closely related notions of relator, also accounting for simulation as opposed
to only bisimulation, are studied by Thijs~\cite{thijs:diss} and in the 
context of coalgebraic logic by
\cite{baltag:cmcs00,cirstea:cmcs04,hugh-jaco:simulations}. 
The connection between coalgebraic logic and relation lifting goes
back to the original paper by Moss~\cite{moss:coal99} which introduced
$\nabla$ and defined its semantics by using relation liftings, albeit
without making this notion explicit. Independently, essentially the
same notion of relation lifting was studied in a fibrational setting
by Hermida and Jacobs~\cite{herm98:stru}. For a comparison of the
notions of bisimulation arising from relation lifting and related
definitions see Staton~\cite{staton:calco09}.

The relation lifting can also be obtained via a distributive law
between a functor and a monad as in Definition~\ref{d:distlaw}, which
is a slight, commonly used variant of the notion of a distributive law
between monads \cite{beck:dist69}. As shown in \cite{beck:dist69},
there is a 1-1 correspondence between distributive laws and liftings
of functors to the category of algebras. Similarly, distributive laws
$\lambda:TM\to MT$ between a functor $T$ and a monad $M$, or monad
op-functors $(T,\lambda):(\Set,M)\to(\Set,M)$ in the terminology of
Street~\cite{Street:Monads}, are in 1-1 correspondence with
liftings $\Tb$ of $T$ to the Kleisli category of $M$.

We thank Dirk Hofmann, Ji{\v r}{\'\i} Velebil and Steve Vickers for
pointing out various references and their significance.

%% file: sec-boolean.tex
\section{Boolean algebras and their presentations}
\label{s:boolean}

\subsection{Boolean-type algebras}

It will be convenient for us to work with a syntax for Boolean logic and 
Boolean algebras, in which the finitary meet and join symbols, $\bw$ and
$\bv$, respectively, are the \emph{primitive} symbols for the conjunction
and disjunction operation, respectively.

\begin{definition}
\label{d:BAsyntax}
Given a set $X$, we let $\Tba(X)$ denote the set of Boolean terms/formulas
over $X$, defined by the following grammar:
\[
a \isbnf x \in X \divbnf \neg a\divbnf \bvsmall\phi \divbnf \bwsmall\phi, 
\]
where $\phi$ is a finite set of Boolean terms.
We abbreviate $\bot \isdef \bv\nada$ and $\top \isdef \bw\nada$, and if no
confusion is likely we will write $\Tba \isdef \Tba(\nada)$.
\end{definition}

Observe that each $\Tba(X)$ is non-empty, always containing the elements
$\top$ and $\bot$.

The above definition can be brought in coherence with the categorical 
perspective of section~\ref{s:preliminaries}, as follows.

\begin{definition}
\label{d:Boole}
We define the category $\Boole$ of Boolean-type algebras as the algebras 
for the functor $\Set\to\Set$, $X\mapsto X + \Pom X + \Pom X$.  
A Boolean-type algebra will usually be introduced as a quadruple $\bbB =
\struc{B,\neg^{\bbB},\bw^{\bbB},\bv^{\bbB}}$, where $B$ is the \emph{carrier}
of the algebra, and $\neg^{\bbB}: B \to B$, and $\bw^{\bbB},\bv^{\bbB}: 
\Pom(B) \to B$ the Boolean \emph{operations}.
\end{definition}

Note that this perspective has built in that both conjunction and disjunction
are commutative, associative and have a neutral element.

We let $\funU: \Boole \to \Set$ 
\label{d:funU}
denote the forgetful functor, and 
$\funaF: \Set \to \Boole$ 
\label{d:funaF}
its left adjoint;
that is, given a set $X$, $\funaF X$ denotes the absolutely free
Boolean-type algebra, or Boolean term algebra, over $X$.  Note that
$\funaF X$ is not a Boolean algebra.  
Given a set $X$, observe that $\funU\funaF(X)$ consists of the set $\Tba(X)$
of all Boolean terms/formulas using the elements of $X$ as variables.
In fact, we may extend $\Tba$ to the set functor $\Tba: \Set \to \Set$ given
by 
\begin{equation}
\label{eq:Tba}
\Tba \coloneqq \funU\funaF.
\end{equation}
In this way we obtain the well-known term monad for the Boolean
signature with the usual unit $\eta :\Id \to \Tba$ (`variables are terms') 
and multiplication $\mu: \Tba \Tba \to \Tba$ (`terms built from terms are 
terms').

\[%
\xymatrix{
\Set \ar@(ul,dl)_{\Tba} \ar@/^{.5cm}/[r]^\funaF & \Boole \ar@/^{.5cm}/[l]^U}
\]

\noindent In particular, for any $f:X \to \Tba Y$ there is $\wh{f}:
\Tba X \to \Tba Y$ which extends $f$ and can be defined as the
composition $\mu_{Y} \circ \Tba f$.
Logicians will recognise $\wh{f}$ as the \emph{substitution} induced by $f$.

\begin{definition}\label{d:ind_extension}
Given a set $X$ and a Boolean-type algebra $\bbB$, a map $f: X \to \funU\bbB$
is called an \emph{assignment}.
Because of the adjunction $\funaF \dashv \funU$, such an assignment has a unique extension to a $\Boole$-homomorphism, 
denoted by
\[ \ti{f}: \funaF X \to \bbB.
\] 
This map $\ti{f}$ is the \emph{meaning} function induced by $f$.
\end{definition}

\begin{definition}
\label{d:funaQ}
A Boole-type algebra $\bbB$ is a \emph{Boolean algebra} if it satisfies
the inequalities of Table~\ref{tb:clax}.

We let $\funaQ: \Set \to \BA^{\mathit{op}}$ denote the contravariant
power set algebra functor.
That is, given a set $X$, we let $\funaQ X$ denote the power set
algebra of $X$, and for a map $f: X \to Y$, the homomorphism $\funaQ f: 
\funaQ Y \to \funaQ X$ is provided by the map $f^{-1} = \funQ f$.
\end{definition}

\subsection{Presentations of Boolean algebras}
\label{ss:presentations}

It has become a standard tool in mathematics to define an algebraic structure
by means of a \emph{presentation} by generators and relations.
Usually, these definitions are given in the category-theoretic sense, and 
in particular do not distinguish isomorphic structures.
Our proof-theoretic analysis of the logic requires us to be very precise
here, and for this purpose we have developed a small piece of theory on
`concrete presentations'.
We want to stress the fact that whereas we only talk about Boolean algebras
here, the results in this section in fact apply to a wide universal
algebraic setting.

\begin{definition}
\label{d:funB1}
A \emdef{presentation} is a pair $\pGR$ consisting of a set $G$ of
\emdef{generators} and a set $R \sse \Tba(G) \times \Tba(G)$.  
Given such a relation $R$, let ${\equiv_{R}} \sse \Tba(G) \times \Tba(G)$ 
be the least congruence relation on the term algebra $\funaF G$ extending $R$
such that the quotient $\funaF G/_{\equiv_{R}}$ is a Boolean algebra.  
We say that this quotient is the Boolean algebra \emph{presented by $\pGR$},
and denote it as $\funB\pGR$.
Given a presentation $\pGR$, we let 
\begin{equation}
\label{eq:defunit}
\unitBC_{\pGR}: g \mapsto [g].
\end{equation}
define a map $\unitBC_{\pGR}: G \to \funU\funB\pGR$.
\end{definition}

It is straightforward to verify that $\ti{\unitBC}_{\pGR}$ is the quotient
morphism from $\funaF G$ to $\funB\pGR$, with kernel $\ker(\ti{\unitBC}_{\pGR})
= {\equiv_{R}}$.

Relating this definition of presentations to the more usual one, first
observe that a `relation' is nothing but an equation over the set of 
generators (but note that generators should not be seen as \emph{variables}).
Accordingly, given a presentation $\pGR$, a Boolean algebra $\bbB$, and an
assignment $f: G \to \funU\bbB$, we say that a relation $(s,t) \in R$ is
\emph{true} in $\bbB$ under $f$, notation: $\bbB,f \models s \is t$, if 
$\ti{f}(s) = \ti{f}(t)$.
$\bbB$ is a \emph{model} for $R$ under $f$ if $\bbB,f \models s \is t$ for 
all $(s,t) \in R$.
It is straightforward to verify that $\funB\pGR$ is a model for $R$ under
$\unitBC_{\pGR}$.
We can now formulate the following proposition, of which we omit the
(straightforward) proof.

\begin{prop}
Let $\pGR$ be a presentation, and let $\bbB$ be a model for $R$ under 
the assignment $f: G \to \funU\bbB$.
Then there is a unique homomorphism $f': \funB\pGR \to \bbB$ that
extends $f$ in the sense that $f'([g]) = f(g)$.
In a diagram:
\begin{equation*}
\xymatrix{
  G  \ar[drr]_{f}   \ar[rr]^{\unitBC_{\pGR}} 
&& \funU\funB\pGR     \ar[d]^{f'} 
\\
&& \funU\bbB
}
\end{equation*}
\end{prop}
The universal property of $\funB\pGR$ expressed by the above proposition
is usually taken as the definition of the Boolean algebra presented by a
presentation.

In order to turn the class of presentations into a category we need to define
a notion of morphism between two presentations.

\begin{definition}
A \emph{presentation morphism} from one presentation $\pGR$ to another
$\pGRp$ is a map $f: G \to \Tba(G')$ satisfying $\wh{f}(s) \equiv_{R'} 
\wh{f}(t)$ for all $s,t \in \Tba(G)$ such that $(s,t) \in R$.
Given two presentation morphisms $f: \pGR \to \pGRp$ and $g: \pGRp
\to \prs{G''}{R''}$, we define their \emph{composition} $g\circ f: G \to
\Tba(G'')$ as the map
given by
\[
g \circ f  (x) := \wh{g}(f(x)),
\]
and the \emph{identity presentation} on $\pGR$ as the function $\id_{\pGR}: G
\to \Tba(G)$ mapping a generator $x \in G$ to the term $x \in \Tba{G}$.
\end{definition}

The verification that the above defines a category is routine.
Category theorists will note that identity and composition are those
of the \emph{Kleisli category} associated with the monad $\Tba$.

\begin{definition}
\label{d:Prs}
We will let $\Prs$ denote the category with presentations as objects and 
presentation morphisms as arrows.
\end{definition}

We will now extend the construction $\funB$ of a Boolean algebra out of a 
presentation to a functor $\funB: \Prs \to \BA$, and define a functor $\funC:
\BA \to \Prs$ in the opposite direction.

\begin{definition}
\label{d:funB2}\label{d:funC}
Given a presentation morphism $f: \pGR \to \pGRp$, it is easy to
see that the map $\funB f: \funaF G/_{\equiv_{R}} \to \funaF G'/_{\equiv_{R'}}$
given by
\[
\funB f:\;
[s]_{\pGR} \;\mapsto\; [\wh{f}(s)]_{\pGRp}
\]
is well-defined.

Conversely, given a Boolean algebra $\bbB$, define its
\emdef{canonical presentation} as the pair $\funC\bbB :=
\prs{\funU\bbB}{\Delta_{\bbB}}$. 
Here $\funU\bbB$ is the underlying set of $\bbB$, and $\Delta_{\bbB}$ is the
\emdef{diagram} of $\bbB$, defined as follows:
\[\begin{array}{llll}
\Delta_{\bbB} &:=&&
\{ (a,\neg b) \mid a,b \in \funU\bbB \mbox{ with } a = \neg^{\bbB}b \}
\\ && \cup & \{ (a, \textstyle{\bw} \phi) \mid \{ a \} \cup \phi \sse_{\om} \funU\bbB 
    \mbox{ with } a = \textstyle{\bw}^{\bbB}\phi \}
\\ && \cup & \{ (a, \textstyle{\bv} \phi) \mid \{ a \} \cup \phi \sse_{\om} \funU\bbB 
    \mbox{ with } a = \textstyle{\bv}^{\bbB}\phi \}.
\end{array}\]
Given a homomorphism $f: \bbB \to \bbB'$ between two Boolean algebras, we
let
\[
\funC f:  b \mapsto f(b)
\]
define a map $\funC f: \funU\bbB \to \Tba(\funU\bbB')$.
\end{definition}

\begin{prop}
\label{p:BCfun}
$\funB: \Prs \to \BA$ and $\funC: \BA \to \Prs$ are functors.
\end{prop}

Further on we will make good use of the following definition.

\begin{definition}
A presentation morphism $f: \pGR \to \pGRp$ is a
\emph{pre-isomorphism} if there is a morphism $g: \pGRp \to \pGR$ such
that $\wh{g}\wh{f}(s) \equiv_{R} s$ and $\wh{f}\wh{g}(s') \equiv_{R'}
s'$, for all terms $s \in \Tba G$ and $s' \in \Tba G'$.  This $g$ is
called a \emph{pre-inverse} of $f$.
\end{definition}

\begin{prop}
\label{p:piBi}
Let $f: \pGR \to \pGRp$ be a presentation morphism.
Then $f$ is a pre-isomorphism iff $\funB f$ is an isomorphism.
\end{prop}

\begin{proof}
For the direction from left to right, let $f$ be a pre-isomorphism.
We confine ourselves to proving that $\funB f$ is injective.
For this purpose assume that $\funB f([s]_{\pGR}) = \funB f([t]_{\pGR})$.
Then by definition we have $[\wh{f} s]_{\pGRp} = 
[\wh{f} t]_{\pGRp}$, or equivalently, $\wh{f}s \equiv_{R'} \wh{f}t$.
From this it follows by the assumption that $s \equiv_{R} \wh{g}\wh{f}s
\equiv_{R} \wh{g}\wh{f}t \equiv_{R} t$, and so it is immediate that $[s]_{\pGR}
= [t]_{\pGR}$.

Conversely, assume that $\funB f$ is an isomorphism between $\funB \pGR$ and 
$\funB \pGRp$.
Let $g: G' \to \Tba G$ be such that $g(x') \in (\funB f)^{-1}[x']$ for every
generator $x' \in G'$.
We claim that $\funB g = (\funB f)^{-1}$.
To see this, note that it is straightforward to check that $g(s') \in 
(\funB f)^{-1}[s']$; from this it follows that $(\funB f)^{-1} ([s']_{\pGRp} )=
[\wh{g} s']_{\pGR}$.

In order to see that $g$ is a pre-inverse of $f$, consider an arbitrary term $s
\in \Tba G$.
Clearly we have $[s]_{\pGR} = (\funB f)^{-1}(\funB f) [s]_{\pGR}$, and so by
definition and the above observation, we find $[s]_{\pGR} = 
(\funB f)^{-1}[\wh{f} s]_{\pGRp} = [\wh{g}\wh{f} s]_{\pGR}$.
This means that $s \equiv_{R} \wh{g}\wh{f} s$, as required.
Conversely, let $s'$ be an arbitrary term in $\Tba G'$.
Then we have $[s']_{\pGRp} = (\funB f)(\funB f)^{-1}[s']_{\pGRp}
= \funB f [\wh{g}s']_{\pGR} = [\wh{f}\wh{g}s']_{\pGRp}$, or equivalently, 
$s' \equiv_{R'} \wh{f}\wh{g}s'$.
\end{proof}

The functors $\funB$ and $\funC$ are very close to forming an equivalence
between the categories $\Prs$ and $\BA$.
More precisely, we can formulate the following connections.
Given a presentation $\pGR$, it is not hard to verify that the insertion of
generators $\eta_{\pGR}: G \to \funU\funB\pGR$ defined in \eqref{eq:defunit}
is in fact a presentation morphism 
\[
\unitBC_{\pGR}: \pGR \to \funC\funB\pGR.
\]
Conversely, given a Boolean algebra $\bbB$, let $\id_{B}$ denote the identity
map on $B := \funU\bbB$, and recall that $\ti{\id}_{B}$ denotes the unique
homomorphism $\ti{\id}_{B}: \funaF\funU\bbB \to \bbB$ extending $\id_{B}$.
It is not difficult to show that $\ti{\id}_{B}(t(b_{1},\ldots,b_{n})) =
t^{\bbB}(b_{1},\ldots,b_{n})$, and so we may think of $\ti{\id}$ as an
\emph{evaluation map}.
We leave it for the reader to verify that for all $s,t \in \funaF\funU\bbB$, we 
have 
\begin{equation}
s \equiv_{\funC\bbB} t \mbox{ iff } \ti{\id}_{B}(s) = \ti{\id}_{B}(t).
\end{equation}
From this it follows that the map $\counitBC_{\bbB}: 
\funB\funC\bbB \to \bbB$ given by putting, for any
$t(b_{1},\ldots,b_{n}) \in \Tba(\funU\bbB)$:
\begin{equation}
\label{eq:defcounit}
\counitBC_{\bbB}: 
[t(b_{1},\ldots,b_{n})] \mapsto t^{\bbB}(b_{1},\ldots,b_{n})
\end{equation}
is a well-defined homomorphism from $\funB\funC\bbB$ to $\bbB$.

\begin{thm}
\label{t:BCadj}
The functors $\funB$ and $\funC$ form an adjoint pair $\funB \dashv
\funC$, with unit $\unitBC: \Id_{\Prs} \ntrto \funC\funB$ and counit
$\counitBC: \funB\funC \ntrto \Id_{\BA}$ given by \eqref{eq:defunit}
and \eqref{eq:defcounit}, respectively.  Furthermore, each arrow
$\unitBC_{\pGR}: \pGR \to \funC\funB\pGR$ is a pre-isomorphism, and
each arrow $\counitBC_{\bbB}: \funB\funC\bbB \to \bbB$ is an
isomorphism.
\end{thm}

\begin{proof}
Let us start with showing that $\unitBC: \Id_{\Prs} \ntrto \funC\funB$ is indeed
a natural transformation.
That is, given an presentation morphism $f: \pGR \to \pGRp$ we have to
show that the following diagram commutes.
\begin{equation*}
\xymatrix{
  \pGR  \ar[d]_{f}   \ar[r]^{\unitBC_{\pGR}} 
& \funC\funB\pGR     \ar[d]_{\funC\funB f} 
\\
  \pGRp \ar[r]^{\unitBC_{\pGRp}} 
& \funC\funB\pGRp  
}
\end{equation*}
For this purpose it suffices to check that the two compositions, $\funC\funB f
\circ \unitBC_{\pGR}(x)$ and $\unitBC_{\pGRp}\circ f (x)$ agree on an arbitrary
generator $x \in G$.
But this is immediate:
\[
\funC\funB f \circ \unitBC_{\pGR}(x)
= \funC\funB f [x] = [\wh{f} x] 
= \wh{\unitBC}_{\pGRp}(\wh{f}x)
= \wh{\unitBC}_{\pGRp}(f x)
= (\unitBC_{\pGRp}\circ f) (x).
\]
In order to prove that $\unitBC_{\pGR}$ is a pre-isomorphism, let $g: 
\funU\funB\pGR \to \Tba G$ be any map such that $g([s]) \in [s]$ for any element 
$[s] \in \funU\funB\pGR$.
It is easy to check that $g$ is a presentation morphism and that $\unitBC_{\pGR}$
and $g$ are pre-inverses of each other.
From this it is immediate that $\unitBC_{\pGR}$ is a pre-isomorphism.

Turning to the counit of the adjunction, let $f: \bbB \to \bbB'$ be a 
homomorphism between Boolean algebras.
Let $[t(b_{1},\ldots,b_{n})]$, with each $b_{i}$ in $\bbB$, be an arbitrary
element of $\funB\funC\bbB$. 
Then we compute
\begin{align*}
f\circ\counitBC_{\bbB}[t(b_{1},\ldots,b_{n})] &= f(t^{\bbB}(b_{1},\ldots,b_{n}))
  \tag*{(definition of $\counitBC$)}
\\ &= t^{\bbB'}(fb_{1},\ldots,fb_{n})
  \tag*{($f$ is a homomorphism)}
\\ &= \counitBC_{\bbB'}[t(fb_{1},\ldots,fb_{n})]
  \tag*{(definition of $\counitBC$)}
\\ &= \counitBC_{\bbB'}[\wh{f}(t(b_{1},\ldots,b_{n}))]
  \tag*{(definition of $\wh{f}$)}
\\ &= \counitBC_{\bbB'}(\funB\funC f)[t(b_{1},\ldots,b_{n})]
  \tag*{(definition of $\funB$ and $\funC$)}
\end{align*}
This shows that the following diagram commutes:
\begin{equation*}
\xymatrix{
  \funB\funC\bbB  \ar[d]_{\funB\funC f}   \ar[r]^{\counitBC_{\bbB}} 
& \bbB     \ar[d]_{f} 
\\
  \funB\funC\bbB' \ar[r]^{\counitBC_{\bbB'}} 
& \bbB'
}
\end{equation*}
and thus proves that $\counitBC$ is a natural transformation.

To show that $\counitBC_{\bbB}$ is an isomorphism, it suffices to check 
injectivity.
But by a straightforward term induction it is easy to prove that every term
$t(b_{1},\ldots,t_{k})$ in $\Tba\funU\bbB$ satisfies
\[
t(b_{1},\ldots,b_{n}) \equiv_{\funC\bbB} t^{\bbB}(b_{1},\ldots,b_{n}).
\]
Hence if $\counitBC_{\bbB}[s(a_{1},\ldots,a_{k})] = 
\counitBC_{\bbB}[t(b_{1},\ldots,b_{n})]$, then by
$s(a_{1},\ldots,a_{k}) \equiv_{\funC\bbB} s^{\bbB}(a_{1},\ldots,a_{k})
= t^{\bbB}(b_{1},\ldots,b_{n}) \equiv_{\funC\bbB}  t(b_{1},\ldots,b_{n})$, 
we immediately find that $[s(a_{1},\ldots,a_{k})] = [t(b_{1},\ldots,b_{n})]$,
as required.

Finally, in order to prove that $B \dashv C$,
by~\cite[Theorem IV.1.2]{macl:cate98} it 
suffices to prove that (i) for any Boolean algebra $\bbA$, the composition
\[
\funC\bbA \stackrel{\unitBC_{\funC\bbA}}{\longrightarrow}
\funC\funB\funC\bbA \stackrel{\funC\counitBC_{\bbA}}{\longrightarrow}
\funC\bbA
\]
is the identity on $\funC\bbA$, and that (ii) for any presentation $\pGR$, the
composition
\[
\funB\pGR \stackrel{\funB\unitBC_{\pGR}}{\longrightarrow}
\funB\funC\funB \pGR \stackrel{\counitBC_{\funB\pGR}}{\longrightarrow}
\funB\pGR
\]
is the identity on $\funB\pGR$.  Both of these facts can be checked by
a straightforward unravelling of the definitions, which we will leave
as an exercise for the reader.
\end{proof}

\begin{rem}
What keeps $\funB$ and $\funC$ from forming an \emph{equivalence} of
categories is that the unit $\unitBC$ is a `natural pre-isomorphism' rather
than a natural isomorphism.
We could remedy this by changing the notion of arrow in the category of
presentations but this would be disadvantageous in our completeness proof,
when we construct a stratification of our logic.
\end{rem}

\begin{rem}
  We indicate how the present section generalises beyond Boolean
  algebras, as suggested by a referee. We have been working with three
  categories, $\BA$, $\Boole$, and $\Prs$. Instead of $\Boole$
  consider a category $\cal B$ with forgetful functor $U:\cal
  B\to\Set$ and left-adjoint $F$ of $U$. Instead of $\BA$ consider a
  category $\cal A$ and a full inclusion $I:\cal A \to \cal B$ with a
  left-adjoint $L$ of $I$. Now, we can define a category
  $\Prs$. $\Prs$ has as as objects pairs $\langle G, R\rangle$ where
  $G$ is a set and $R$ is a relation given by a pair of arrows
  $R\rightrightarrows UFG$, or equivalently, by $FR\rightrightarrows
  FG$. A presentation morphism $f:\langle G, R\rangle \to \langle G',
  R'\rangle$ is then an algebra morphism $f:FG\to FG'$ such that for
  all $A\in\cal A$ and all $v:FG'\to IA$, if $v$ equalises
  $FR'\rightrightarrows FG'$ then $v\circ f$ equalises
  $FR\rightrightarrows FG$. The functors $B:\Prs\to\cal A$ and $C:\cal
  A\to \Prs$ can then be defined as above. Indeed, for $A\in\cal A$ we
  let the canonical presentation $CA$ be the kernel pair of the map
  $UFUIA\to UIA$, given by the counit of $F\dashv U$ at $IA$; and
  $B\langle G, R\rangle$ is given by the coequaliser of
  $LFR\rightrightarrows LFG$. As in Theorem~\ref{t:BCadj}, one can now
  show that $B\dashv C$ and that the counit $BC\to\Id$ is an
  iso. Moreover, the proofs do not depend on the base category $\Set$
  and only require rather general assumptions about kernel pairs and
  coequalisers (which are certainly fullfilled whenever $\cal A$
  and $\cal B$ are varities, that is, classes of algebras given by
  operations of finite arity and equations).
\end{rem}


%% file: sec-nabla.tex
%
\section{Moss' coalgebraic logic}
\label{s:moss}

In this section we will recall the definitions of Moss' coalgebraic logic 
and its semantics~\cite{moss:coal99}, or rather, the finitary version 
thereof developed by Venema~\cite{vene:auto06}.

\subsection{Syntax}

As mentioned in the introduction, the key idea underlying the syntax of
Moss' language for reasoning about $\T$-coalgebras is to include a modal
operator $\nb$ into the language whose `arity' is given by the functor 
$\T$ itself, in the same way that $\Pom$ is the `arity' of our conjunction
and disjunctions.
In the \emph{finitary} version of the language, the arity of $\nb$ is 
given by the finitary version $\Tom$ of $\T$.
In brief, the language $\Lmoss$ will be defined by the following
grammar:
\[
a \isbnf \neg a \divbnf \bwsmall\phi \divbnf \bvsmall\phi
  \divbnf \nb \alpha
\]
where $\phi \in \Pom \Lmoss$ and $\alpha \in \Tom\Lmoss$.
For the purpose of this paper we need some further syntactic definitions.

\begin{definition}
\label{d:syntax}
Let $\T:\Set \to \Set$ be a standard, weak pullback preserving set functor 
and let $\Tom$ be the finitary version of $\T$.
The language $\Lmoss$ of the finitary Moss language for $\T$ is defined 
inductively.
We first define $\Lmoss_{0}$ as the set $\Tba(\nada)$ of closed Boolean
formulas (see Definition~\ref{d:BAsyntax}).
For the inductive step, we start with introducing the set functor $\Tomnb$
defined by, for a given set $X$ and function $f: X \to Y$,
\begin{eqnarray*}
   \Tomnb X & \coloneqq & \{ \nb \alpha \mid \alpha \in \Tom X \},
\\ \Tomnb f (\nb\al) & \coloneqq & \nb \T f(\alpha).
\end{eqnarray*}
We continue the inductive definition by putting
\[
\Lmoss_{i+1} \isdef \Tba \Tomnb \Lmoss_{i}.
\]
Finally, we define $\Lmoss$ as the union $\Lmoss \isdef\bigcup_{i \in \om}
\Lmoss_{i}$, and fix the \emph{rank} or \emph{depth} of a formula $a \in 
\Lmoss$ is the smallest natural number $n$ such that $a \in \Lmoss_{n}$.
\end{definition}

Using BNF notation, we can recast the above definition as
\begin{align*}
\Lmoss_0   \ni a \isbnf 
  & \neg a \divbnf \bwsmall\phi \divbnf \bvsmall\phi
\intertext{where $\phi \sse_{\om} \Lmoss_{0}$, and}
\Lmoss_{i+1} \ni a \isbnf 
  & \nb \alpha \divbnf \neg a \divbnf \bwsmall\phi \divbnf \bvsmall\phi
\end{align*}
where 
$\alpha \in \Tom\Lmoss_i$ and 
$\phi \in \Pom \Lmoss_{i+1}$.

Despite its unconventional appearance, the language $\Lmoss$ admits
fairly standard definitions of most syntactical notions. 
As an example we mention the notion of a subformula.

\begin{definition}
\label{d:sfor}
We define the set $\Sfor(a)$ of \emph{subformulas} of $a$ by the following
induction:
\begin{eqnarray*}
   \Sfor(\neg a)       & \isdef & \{ \neg a \} \cup \Sfor(a)
\\ \Sfor(\bwsmall\phi) & \isdef & \{ \bwsmall\phi \} \cup
      \bcsmall_{a\in\phi}\Sfor(a)
\\ \Sfor(\bvsmall\phi) & \isdef & \{ \bvsmall\phi \} \cup
      \bcsmall_{a\in\phi}\Sfor(a)
\\ \Sfor(\nb\al)       & \isdef & \{ \nb\al \} \cup
      \bcsmall_{a\in\Base(\al)}\Sfor(a).
\end{eqnarray*}
The elements of $\Base(\al)\sse\Sfor(\nb\al)$ will be called the
\emph{immediate} subformulas of $\nb\al$.
\end{definition}

On the basis of this definition it is not difficult to prove that every
formula in $\Lmoss$ has only \emph{finitely} many subformulas.
This is in fact the reason why we call our language the \emph{finitary}
version of Moss'.

\begin{rem}
\label{r:Tsynt}
In order to formulate and understand the interaction principles between
nabla and the Boolean operations, we need to think of the propositional 
connectives as \emph{functions} on formulas.
Taking disjunction as an example, observe that we may think of it as a map
$\bv: \Pom\Lmoss \to \Lmoss$.
Thus we may apply the functor $\Tom$ to this map, obtaining $\T\bv:
\Tom\Pom\Lmoss \to \Tom\Lmoss$.
(Recall from our discussion on the finitary version of a functor that to 
simplify notation we will write $\T\bv$ rather than $\Tom\bv$ .)
Hence, for $\Phi \in \Tom\Pom\Lmoss$, we find $(\T\bv)\Phi \in \Tom\Lmoss$,
which means that $\nb(\T\bv)\Phi$ is a well-formed formula.
The same applies to the formula $\nb(\T\bw)\Phi$, and
similarly, we may think of negation as a map $\neg: \Lmoss \to \Lmoss$, and
obtain $\T\neg: \T\Lmoss \to \T\Lmoss$; thus for any formula $\nb\al$, we 
may also consider the formula $\nb(\T\neg)\al$.
\end{rem}

\begin{rem}
\label{r:prop}
The reader may be surprised that we did not include propositional variables
in our language.
The reason for this is that we may \emph{encode} these into the functor.
More precisely, given a functor $\T$ and a set $\Prop$ of proposition letters,
recall from Example~\ref{ex:1}(5) that the $\T$-models over $\Prop$ can be 
identified with the coalgebras for the functor $\T_{\Prop} = \funP(\Prop)
\times T$.
Hence we may use the language $\Lmoss$ associated with $\T_{\Prop}$ to
describe the $\Prop$-models based on $\T$-coalgebras, see
Example~\ref{ex:sem}(3).
\end{rem}

\begin{convention}
Since in this paper we will not only be dealing with formulas and sets of 
formulas, but also with elements of the sets $\Tom\Lmoss$, $\Pom\Tom\Lmoss$
and $\Tom\Pom\Lmoss$, it will be convenient to use some kind of \emph{naming
convention}, see Table~\ref{tb:naming} below.

\begin{table}[thb]
\begin{center}
\begin{tabular}{|r|l|}
 \hline
      Set & Elements\\
      \hline
      $\Lmoss$& $a,b,\dotsc$\\
      $\Tom \Lmoss$ & $\alpha,\beta, \dotsc$\\
      $\Pom \Lmoss$ & $\phi,\psi, \dotsc$\\
      $\Pom \Tom \Lmoss$& $A, B, \dotsc$\\
      $\Tom \Pom \Lmoss$ & $\Phi, \Psi,\dotsc$\\ \hline
    \end{tabular}
  \end{center}
\caption{Naming convention 
}
\label{tb:naming}
\end{table}
\end{convention}

It will be useful later on to have a more categorical description of the
finitary Moss language for a functor $\T$.
For this purpose we need the following definition.

\begin{definition}
\label{d:Mossfun}
We define the category $\Boolenb$ of \emph{Moss algebras} as the algebras 
for the Moss functor $\Mossfun: \Set\to\Set$, given as: 
\[
\Mossfun \isdef \Id + \Pom  + \Pom  + \Tom ,
\]
That is, for a set $S$, $\Mossfun S$ is the disjoint union of $S$, two 
(disjoint copies) of $\Pom S$, and $\Tom S$; for a map $f$, $\Mossfun
f$ is defined accordingly.

A Moss algebra will usually be introduced as a quadruple $\bbB =
\struc{B,\neg^{\bbB},\bw^{\bbB},\bv^{\bbB},\nb^{\bbB}}$, where 
$\struc{B,\neg^{\bbB},\bw^{\bbB},\bv^{\bbB}}$ is a $\Boole$-type algebra,
called the \emph{Boolean reduct} of $\bbB$, and $\nb^{\bbB}: \Tom B
\to B$ is the \emph{nabla operator} of $\bbB$.
\end{definition}

Given a Moss algebra $\bbB$, there is a unique, natural way to interpret
$\Lmoss$-terms as elements of the carrier $B$ of $\bbB$.
This \emph{meaning function} $\mng_{\bbB}: \Lmoss \to \funU\bbB$ can be defined by 
a straightforward induction on the complexity of formulas.
For instance, the clauses for $\bw$ and $\nb$ are
\begin{eqnarray*}
\mng_{\bbB}(\bwsmall\phi) &\isdef& \bwsmall^{\bbB}(\funP\mng_{\bbB})(\phi)
\\
\mng_{\bbB}(\nb\al)       &\isdef& \nb^{\bbB}(\T\mng_{\bbB})(\al)
\end{eqnarray*}

Categorically speaking, this means the following.
We may view Moss' language itself as a Moss algebra, by interpreting the 
function symbols as the corresponding syntactic operation, as usual in
universal algebra.
Note that in order to prove that $\nb^{\Lmoss}\al$ belongs to $\Lmoss$,
it is crucial that $\nb$ is a finitary operation: from $\al \in \Tom\Lmoss$
it follows that $\al \in \Tom\Lmoss_{n}$ for some finite $n$, and then
we may proceed with $\nb\al \in \Lmoss_{n+1}\sse\Lmoss$.
The arising algebra, that we will also denote as $\Lmoss$, is a rather
special Moss algebra, namely, the \emph{initial} one.
Apart from the fact that the syntax of $\Lmoss$ is slightly unusual, the
proof of the proposition below is standard universal algebra, and so we
omit it.

\begin{prop}
\label{p:Lmoss-init}
$\Lmoss$ is the initial Moss algebra: given an arbitrary Moss
algebra $\bbB$, the meaning function $\mng_{\bbB}$ is the unique
homomorphism from $\Lmoss$ to $\bbB$.
\end{prop}

Before moving on to the coalgebraic semantics of $\Lmoss$, we finish our
discussion of its syntax with the following definition, for future
reference.

\begin{definition}
\label{d:Tomnb}\label{d:Tnb}
Let $\T:\Set \to \Set$ be a set functor and let $\Tom$ be the finitary
version of $\T$.  We define the functor $\Tnb: \Set \to \Boole$ by
putting
\[
\Tnb \isdef \akk{ \Lmoss_0}\cof \Tomnb \cof \Lmoss_0.
\]
\end{definition}

\akk{On occasion, we will consider $\Tnb$ also as a $\Boole$ valued
  functor allowing us to write $\Tnb=\funaF \Tomnb \Tba$.  The
notation $\Tnb$} is in accordance with the definition of $\Lmoss_{1}$
as the fragment of rank one formulas in $\Lmoss$, by the observation
that $\Lmoss_{1} = (\Tba\cof\Tomnb)(\Lmoss_{0}) =
\Tba\Tomnb\Tba(\nada)$.

\subsection{Semantics}

Given all the preparations we have made in the previous sections, the 
definition of the semantics of the language is completely straightforward.

\begin{definition}\label{def:moss_sem}
Let $\T:\Set \to \Set$ be a standard, weak pullback preserving functor,
and let $\bbX = \struc{X,\xi}$ be a $\T$-coalgebra.
The satisfaction relation ${\forces_{\bbX}} \subseteq X \times \Lmoss$ is defined 
by the following induction on the complexity of formulas:
\[\begin{array}{lcl}
   x \forces_{\bbX} \neg a  & \mbox{if} & 
   x \not\forces_{\bbX} a,
\\ x \forces_{\bbX} \bw\phi & \mbox{if} & 
   x \forces_{\bbX} a \text{ for all } a \in \phi,
\\ x \forces_{\bbX} \bv\phi & \mbox{if} & 
   x \forces_{\bbX} a \text{ for some } a \in \phi,
\\ x \forces_{\bbX} \nb \al & \mbox{if} & 
   \xi(x) \rel{\Tb{\forces_{\bbX}}} \al.
\end{array}
\]
If $x \forces_{\bbX} a$ we say that $a$ is \emph{true}, or \emph{holds} at $x$
in $\bbX$.
We may omit the superscript when no confusion is likely, writing $\forces$
instead of $\forces_{\bbX}$.

In case $a$ \emph{holds throughout $\bbX$}, that is, at every state of $\bbX$,
we write $\bbX \forces a$.
\end{definition}

\noindent Before we turn to look at some examples, we should argue for the 
\emph{well-definedness} of the relation $\forces$.
In particular, when looking at the clause for the nabla modality, the reader
might be worried whether this is an inductive definition at all, since the
defining clause, `$\xi(x) \rel{\Tb{\forces}} \al$', refers to the \emph{full}
forcing relation.
The point is that because of our assumptions, $\Tb$ commutes with
restrictions, and so we have
\begin{equation}
\label{eq:sem}
(\xi(x),\al) \in \Tb({\forces}) \iff 
(\xi(x),\al) \in \Tb({\forces\rst{X\times\Base(\al) }}).
\end{equation}
Thus, in order to determine whether $\nb\al$ holds at $x$ or not, we only
have to know the interpretation of the \emph{immediate subformulas of $\al$}
(that is, the elements of $\Base(\al)$).
In other words, if using the right hand side of \eqref{eq:sem} rather than
the left hand side, we would have an equivalent, inductive, definition of 
the semantics.

\begin{exa}
\label{ex:sem}\hfill
\begin{enumerate}[(1)]
\item
Let $\T$ be the $C$-stream functor given by $\T X = C \times X$ for some set 
$C$.
Then $\nb_{\T}$ takes as its argument a pair $(c,a)$ where $c \in C$ and $a$ is 
a formula in $\Lmoss$.
The formula $\nb(c,a)$ is true in a $\T$-coalgebra $(X,\xi)$ at a state $x$
if $\xi(x) = (c',y)$ with $c=c'$ and $y \forces a$.
\item 
The nabla operator $\nabla_{\funP}$ associated with the power set functor
$\funP$ is the \emph{cover modality} discussed in the introduction.
\item
If $\T_{\Prop}$ is the $\T$-model functor of Example~\ref{ex:2}(5), associated
with a functor $\T$ and a set $\Prop$ of proposition letters, then
$\nb_{\T_{\Prop}}$ takes as its
argument a pair $(\pi,\al)$ consisting of a set $\pi \sse \Prop$ and a set
$\al \sse_{\om} \Lmoss^{\T}$.
The meaning of the formula $\nb_{K}(\pi,\al)$ can be expressed as
\[
\nb_{\T_{\Prop}}(\pi,\al) \equiv (\bw_{p\in\pi}p \land \bw_{p\not\in\pi}\neg p)
\land \nb_{\T}\al.
\]
\item
Finally, let $\T=D_{\om}$ be the finitary distribution functor, 
In this case, $\nb_{D_{\om}}$ takes as argument a distribution $\mu: \Lmoss \to
[0,1]$ of finite support.
Given a $\T$-coalgebra $\bbX = (X,\xi)$ and some $x \in X$ we have 
$x \forces_{\bbX} \nb_{D_{\om}} \mu$ if for all $y \in X$ and all 
$a \in \Lmoss$ there are real numbers 
$\rho_{y,a} \in [0,1]$ such that
\begin{align*}
    \rho_{y,a} \not= 0 \quad \mbox{implies} \quad & y \forces a, \xi(x)(y) \not= 0, 
	        \mu(a) \not= 0 \quad  &\mbox{and}\\
              &	\sum_{a' \in \Lmoss} \rho_{y,a'} = \xi(x)(y)  \quad \mbox{for all } y \in X  \quad &\mbox{and}  \\
            &   \sum_{y \in X} \rho_{y,a} = \mu (a) \quad \mbox{for all } a \in \Lmoss. \quad
\end{align*}
     \end{enumerate}
\end{exa}

\noindent The state-based semantics of the logics as presented in
Definition~\ref{def:moss_sem} can be brought in accordance with the earlier
algebraic perspective by the observation that every $\T$-coalgebra naturally
induces a Moss algebra, namely its \emph{complex algebra}.

\begin{definition}
\label{d:cplxalg1}
Let $\T:\Set \to \Set$ be a standard, weak pullback preserving functor,
and let $\bbX = \struc{X,\xi}$ be a $\T$-coalgebra.
The \emph{complex algebra} $\bbX^{+}$ of $\bbX$ is defined as the Moss
algebra $\bbB$ which has the power set algebra $\funaQ(X)$ as its Boolean
reduct, while 
\[
\nb^{\bbX^{+}} \isdef \funQ\xi \cof \nbsem_{X}
\]
defines the nabla operation of $\bbX^{+}$.
\end{definition}

In words: the Boolean function symbols $\neg,\bv$ and $\bw$ are 
interpreted as the complementation, union and intersection operations on
the power set of $X$.
To understand the definition of the nabla operation, observe that applying
the contravariant power set functor to the coalgebra map $\xi$, we obtain
a function $\funQ\xi: \funQ\T X \to \funQ X$, so if we compose this
map with the $\T$-transformation $\nbsem_{X}: \T\funQ X \to \funQ\T X$,
we obtain a map $\funQ\xi \cof \nbsem_{X}: \T\funQ X \to \funQ X$ of the
right shape.

It follows by Proposition~\ref{p:Lmoss-init} that every $\Lmoss$-formula $a$
can uniquely be assigned a meaning $\mng_{\bbX^{+}}(a) \in \funP X$ in the
complex algebra of a $\T$-coalgebra $\bbX$ --- in the sequel we will write
$\mng_{\bbX}$ rather than $\mng_{\bbX^{+}}$.
The Proposition below states that the two approaches to the coalgebraic 
semantics of $\Lmoss$ coincide, so that we can speak without hesitation of 
`the' meaning of a formula in a $\T$-coalgebra.

\begin{prop}
\label{p:onesemantics}
Let $\T:\Set \to \Set$ be a standard, weak pullback preserving functor,
and let $\bbX = \struc{X,\xi}$ be a $\T$-coalgebra.
Then we have
\begin{equation*}
\mng_{\bbX}(a) = \{ x \in X \mid x \forces a \},
\end{equation*}
for every formula $a \in \Lmoss$.
\end{prop}
\begin{proof}
The proof of this proposition proceeds by a routine formula induction.
\end{proof}

\subsection{First observations}

In this subsection we gather first observations on $\Lmoss$.
First we show that Moss' logic is \emph{adequate}; that is, it cannot 
distinguish behaviorally equivalent states.

\begin{thm}[Adequacy]
\label{t:adequacy}
Let $\T:\Set \to \Set$ be a standard, weak pullback preserving functor,
and let $f: X \to Z$ be a coalgebra morphism between the $\T$-coalgebras
$(X,\xi)$ and $(Z,\zeta)$.
For all formulas $a \in \Lmoss$ and all states $x \in X$ we have
\begin{equation}
\label{eq:adeq1}
x \forces_{\bbX} a \text{ iff } f(x) \forces_{\bbZ} a.
\end{equation}
\end{thm}

We leave it as an exercise for the reader to give a \emph{direct} proof
of Theorem~\ref{t:adequacy} --- a straightforward induction will suffice,
using the fact that $\Tb$ distributes over relation composition in the
case of a formula $a = \nb\al$.
We will give a proof based on the algebraic approach, involving the 
initiality of $\Lmoss$ (Proposition~\ref{p:Lmoss-init}), and the following
result.

\begin{prop}
\label{p:nbhom}
Let $\T:\Set \to \Set$ be a standard, weak pullback preserving functor,
and let $f: X \to Z$ be a coalgebra morphism between the $\T$-coalgebras
$\bbX = (X,\xi)$ and $\bbZ = (Z,\zeta)$.
Then $\funQ f$ is an algebraic homomorphism from $\bbZ^{+}$ to $\bbX^{+}$.
\end{prop}

\begin{proof}
It is well-known that $\funQ f$ is a homomorphism from the power set 
algebra $\funaQ(Z)$ to $\funaQ(X)$.
Thus it is left to show that $\funQ f$ also is a homomorphism with respect
to the nabla operators.
For that purpose, consider the following diagram:
\[\xymatrix{
    \T\funQ Z \ar[d]_{\T\funQ f} \ar[r]^{\nbsem_Z} 
  & \funQ\T Z \ar[d]^{\funQ\T f} \ar[r]^{\funQ\zeta} 
  & \funQ Z   \ar[d]^{\funQ f}
\\  \T\funQ X \ar[r]_{\nbsem_X} 
  & \funQ\T X \ar[r]^{\funQ\zeta} 
  & \funQ X
}\]
The left rectangle commutes since $\nbsem$ is a distributive law of $\T$
over $\funQ$ (see Proposition~\ref{p:nbdlfunQ}), and the right rectangle
commutes by functoriality of $\funQ$ and the assumption that $f$ is a
coalgebra morphism.
As a corollary, the outer diagram commutes, but by definition of
$\nb^{\bbX^{+}}$ and $\nb^{\bbZ^{+}}$ this just means that $\funQ f$
is a homomorphism for $\nb$.
\end{proof}

On the basis of the previous proposition, the proof of the Theorem is 
almost immediate.

\begin{proofof}{Theorem~\ref{t:adequacy}}
By initiality of $\Lmoss$ as a Moss algebra, $\mng_{\bbX}$ is the 
unique homomorphism $\mng_{\bbX}: \Lmoss \to \bbX^{+}$.
But it follows from Proposition~\ref{p:nbhom} that $\funQ f \cof 
\mng_{\bbZ}$ is also a homomorphism from $\Lmoss$ to $\bbX^{+}$, so
that we may conclude that
\begin{equation}
\label{eq:adeq2} 
\mng_{\bbX} = \funQ f \cof \mng_{\bbZ}.
\end{equation}
Now let $x$ and $a$ be as in the statement of the theorem, then we have
\begin{align*}
x \forces_{\bbX} a 
   & \text{ iff } x \in \mng_{\bbX}(a) 
   & \text{(Proposition~\ref{p:onesemantics})}
\\ & \text{ iff } x \in \funQ f (\mng_{\bbZ}(a))
   & \text{(\ref{eq:adeq2})}
\\ & \text{ iff } fx \in \mng_{\bbZ}(a)
   & \text{(definition of $\funQ f$)}
\\ & \text{ iff } fx \forces_{\bbZ} a 
   & \text{(Proposition~\ref{p:onesemantics})}
\end{align*}
From this the theorem is immediate.
\end{proofof}

\subsection{Logic}

The purpose of this paper is to provide a sound and complete axiomatization 
of the set of \emph{coalgebraically valid} formulas in this language, that 
is, the set of $\Lmoss$-formulas that are true in every state of every 
coalgebra.
Since our completeness proof will be algebraic in nature, for our purposes it
will be convenient to formulate our results in terms of \emph{equations}, or
rather, \emph{inequalities}.

\begin{definition}
An \emph{inequality} is an expression of the form $a \isleq b$, where $a$ and
$b$ are formulas in $\Lmoss$.
Similarly, an \emph{equation} is an expression of the form $a \is b$.
\end{definition}

One may think of the inequality $a \isleq b$ as abbreviating the equation $a
\land b \is a$, and we will see the equation $a \is b$ as representing the set
$\{ a \isleq b, b \isleq a \}$ of inequations. 
(In fact, in our Boolean setting, we could even represent the equation $a \is b$ 
by the single inequality $(a \land \neg b) \lor (\neg a\land b) \isleq \bot$.)
Thus it does not really matter whether we base our logic on equations or on
inequalities, and in the sequel we will move from one perspective to the other
if we deem it useful.

\begin{definition}
An inequality $a \isleq b$ \emph{holds} in a Moss algebra $\bbA$, notation:
$\bbA \models a \isleq b$, if $\mng_{\bbA} (a) \leq_{\bbA} \mng_{\bbA} (b) $.
\end{definition}

Given the Boolean basis of our logics, we can express coalgebraic validity 
in terms of equational validity, and vice versa.
More precisely, given a $\T$-coalgebra $\bbX = \struc{X,\xi}$, it is easy to
see that 
\begin{align*}
\bbX \forces a &\iff \bbX^{+} \models \top \isleq a 
\intertext{and, conversely,}
\bbX^{+} \models a \isleq b &\iff \bbX \forces \neg a \lor b.
\end{align*}
As a consequence, in order to axiomatize the coalgebraically valid
formulas, we may just as well find a derivation system for the
inequalities that are valid in all complex algebras.

\begin{definition}
An inequality $a \isleq b$ is \emph{($\T$-coalgebraically) valid}, notation: 
$a \models_{\T} b$, if it holds in every complex algebra $\bbX^{+}$.
\end{definition}

As an example of a validity, we mention the following, for an arbitrary
$\Phi \in \Tom\Pom\Lmoss$:
\[
\tag{$\nb 3_{f}$}
\nb(\T\bvsmall)\Phi \isleq 
\bv \Big\{ \nb \beta \mid \beta \Tin \Phi \Big\}
\]
(see Remark~\ref{r:Tsynt} for an explanation of the syntax).
Note that the right hand side of ($\nb3_{f}$) is a well-defined formula
only if the disjunction is finite; we can guarantee this by requiring $\T$
to map finite sets to finite sets.
(We will come back to this issue in the next section.)

\begin{prop}
\label{p:nb3s}
If $\T$ is a weak pullback preserving, standard set functor that maps finite
sets to finite sets, then the formula $(\nb3_{f})$ is valid for every $\Phi
\in \Tom\Pom\Lmoss$.
\end{prop}

\begin{proof}
In order to understand the validity of ($\nb3_{f}$), fix some $\T$-coalgebra
$\bbX = \struc{X,\xi}$.

First observe that for any $\phi \sse_{\om} \Lmoss$ we have $\bbX,x \forces
\bv\phi$ iff $\bbX,x \forces a$, for some $a \in \phi$.
Putting it differently, the relations ${\forces} \corel {\in}$ and  ${\forces}
\corel \cv{\bv}$ coincide.
From this it follows that 
\begin{equation}
\label{eq:soundbv}
\Tb({\forces} \corel {\in}) \;=\;
\Tb({\forces}\corel \cv{\bvsmall}).
\end{equation}

\noindent Now fix some object $\Phi \in \Tom\Pom\Lmoss$, and suppose
that $x$ is a state in $\bbX$ such that $x \forces \nb(\T\bv)\Phi$.
From this it follows that the pair $(\xi(x),(\T\bv)(\Phi))$ belongs to
the relation $\Tb\forces$, and so $(\xi(x),\Phi)$ belongs to
$(\Tb{\forces}) \corel \cv{(\T\bv)} = \Tb({\forces}\corel \cv{\bv})$.
But then by (\ref{eq:soundbv}), we find $(\xi(x),\Phi) \in
\Tb({\forces} \corel {\in}) = \Tb{\forces} \corel \Tb{\in}$.  In other
words, there is some object $\be$ such that $\xi(x) \rel{\Tb{\forces}}
\be$ and $\be \Tin \Phi$.  Clearly then $x \forces \nb\be$, and so we
have $x \forces \bv \{ \nb \beta \mid \beta \Tin \Phi \}$, as
required.
\end{proof}


%% file: sec-derivation.tex
\section{The derivation system}
\label{s:derivation}

\subsection{Introduction}

In this section we introduce our derivation system $\nax$ for the finitary
version of Moss' logic, as given in the previous section.
First we fix some general notation and terminology concerning derivations.

\begin{definition}
\label{d:dy}
Given a derivation system $\mathbf{D}$, we let each of $\vdash_{\mathbf{D}} a 
\isleq b$, $a \sqleq_{\mathbf{D}} b$ and $b \sqgeq_{\mathbf{D}} a$ denote the
fact that the inequality $a \isleq b$ is derivable in $\mathbf{D}$, and we write
$a \equiv_{\mathbf{D}} b$ if both $a \sqleq_{\mathbf{D}} b$ and $b
\sqleq_{\mathbf{D}} a$.
\end{definition}

In other words, where $a \isleq b$ and $a \is b$ are syntactic expressions 
in an object language, the expressions $a \sqleq_{\mathbf{D}} b$ and 
$a \equiv_{\mathbf{D}} b$ denote statements, in the metalanguage, \emph{about}
the derivability of such expressions $a \isleq b$ and $b \isleq a$.
In case no confusion is likely concerning the derivation system at hand, we will
drop subscripts, simply writing $a \equiv b$ and $a \sqleq b$.
\bigskip

In principle, the derivation system that we are looking for, should have 
axioms and rules of three kinds.
First of all, it will have a propositional core taking care of the Boolean
basis of our setting. 
For this purpose, any sound and complete set of axioms and derivation rules 
would do; for concreteness, we propose the set given in Table~\ref{tb:clax}.
Recall that our language has $\bv$ and $\bw$ as primitive connectives.

\begin{table}[bht]
\begin{center}
\begin{tabular}{|cc|}
\hline  & \\

\AXC{}
\UIC{$a \isleq a$}
\DisplayProof 
&
\AXC{$a \isleq b$}
\AXC{$b \isleq c$}
\BIC{$a \isleq c$}
\DisplayProof 
\\ & \\

\AXC{$\{ a \isleq b \mid a \in \phi \}$}
\UIC{$\bv\phi \isleq b$}
\DisplayProof 
&
\AXC{$a \isleq b$}
\RL{$b \in \psi$}
\UIC{$a \isleq\bv\psi$}
\DisplayProof 
\\ & \\

\AXC{$\{ a \isleq b \mid b \in \psi \}$}
\UIC{$a \isleq \bw\psi$}
\DisplayProof 
&
\AXC{$a \isleq b$}
\RL{$a \in \phi$}
\UIC{$\bw\phi \isleq b$}
\DisplayProof 
\\ & \\

\multicolumn{2}{|c|}{%
\AXC{}
\UIC{$\bw \{ \bv\phi \mid \phi \in X\} \isleq 
   \bv \{ \bw\ga[X] \mid \ga \in \mathit{Choice}(X) \}$}
\DisplayProof 
}
\\ & \\

\AXC{$\bw (X \cup \{ \neg a\}) \isleq \bv Y$}
\UIC{$\bw X \isleq \bv (Y \cup \{ a\})$}
\DisplayProof 
&
\AXC{$\bw (X \cup \{ a\}) \isleq \bv Y$}
\UIC{$\bw X \isleq \bv (Y \cup \{ \neg a\})$}
\DisplayProof

\\ & \\ \hline
\end{tabular}
\end{center}
\caption{Axioms and rules for classical propositional logic}
\label{tb:clax}
\end{table}

Second, our system will need some kind of \emph{congruence rule} for the nabla
modality.
Since $\nb$ has a rather unusual form, perhaps it is not a priori clear what
such a rule would look like.
The naive way to formulate a congruence rule for $\nb$ would be as
\begin{equation}
\label{eq:cgr}
\mbox{from } \al \rel{\Tb{\equiv}} \be \mbox{ infer } \nb\al \equiv \nb\be
\end{equation}
Problem is that the premiss of \eqref{eq:cgr} is not itself an equation, or a
set of equations.
This problem can be remedied by invoking some properties of relation lifting.
More precisely, note that from Proposition~\ref{p:st-rl} we may derive the
equivalence $\al \rel{\Tb{\equiv}} \be \iff \al \rel{\Tb Z} \be$, for some
$Z \sse \Base(\al) \times \Base(\be)$.
This would lead to the following formulation of a congruence rule:
\[
\AXC{$\{ a \is b \mid (a,b) \in Z \}$}
\RL{$(\alpha,\beta) \in \Tb Z$}
\UIC{$\nb\alpha \is \nb\beta$}
\DisplayProof \\ \\
\]
The above rule is supposed to have a set of premisses:  $\{ a \is b \mid (a,b) 
\in Z \}$, where $Z \sse \Base(\al) \times \Base(\be)$ is a relation such that 
$(\alpha, \beta) \in \Tb Z$ --- the latter condition is formulated as a
\emph{side condition} of the rule.

As it turns out, however, we also want $\nb$ to be \emph{order-preserving}, and
the most straightforward way to formulate that would be by strengthening
\eqref{eq:cgr} to 
\begin{equation}
\label{eq:cgrmon}
\mbox{from } \al \rel{\Tb {\sqleq}} \be \mbox{ infer } \nb\al \sqleq \nb\be.
\end{equation}
If we want to turn this into a syntactically well-formed derivation rule again,
we obtain our first derivation rule ($\nb1$):
\[
\tag{$\nb 1$}
\AXC{$\{ a \isleq b \mid (a,b) \in Z \}$}
\RL{$(\alpha,\beta) \in \Tb Z $}
\UIC{$\nb\alpha \isleq \nb\beta$}
\DisplayProof \\ \\
\]
which can be read as a congruence and monotonicity rule in one.
It has the additional advantage of being formulated in terms of our primitive
symbol, $\isleq$.

\begin{exa}
First, consider the $C$-labelled binary tree functor $\Btree = C \times \Id 
\times \Id$ of Example~\ref{ex:2}.
Here, an application of rule ($\nb 1$) looks as follows:
\[
\AXC{$\{ a_1  \isleq b_1, a_2 \isleq b_2\}$}
\UIC{$\nb(c,a_1,a_2) \isleq \nb(c,b_1,b_2)$}
\DisplayProof \\ \\
\]
where $c$ is an arbitrary element of $C$.
Note that no inequality of the form $\nb(c,a_1,a_2) \isleq \nb(d,b_1,b_2)$
with $c \not= d$ can be derived using ($\nb 1$) because 
$(\nb(c,a_1,a_2),\nb(d,b_1,b_2)) \not\in \Tb (Z)$ for any relation $Z$.

In the case of the power set functor $\funP$, an application of the rule
($\nb 1$) looks as follows:
\[
\AXC{$\{ a  \isleq b \mid (a,b) \in Z\}$}
\RL{$(\alpha,\beta) \in \rl{\Pow}Z$}
\UIC{$\nb \alpha \isleq \nb \beta$}
\DisplayProof \\ \\
\]
where $\alpha,\beta \in \Pom \Lmoss$ are finite sets of formulas.
It can be easily seen that the premiss of the rule can be satisfied iff for
all $a \in \alpha$ there is a $b \in \beta$ such that $a \isleq b$, and
vice versa.
\end{exa}

In addition, any complete derivation system for Moss' language will need some 
\emph{interaction principles} describing the interaction between the nabla
modality and the Boolean connectives.
As we will see, the interaction principles between $\nb$ and the Boolean
connectives $\bv$ and $\bw$ will take the form of two \emph{distributive
laws} (in the logical meaning of the word).
We postpone discussing the role of negation in our system until
subsection~\ref{ss:neg}, and before giving the general formulation of the 
laws for $\bw$ and $\bv$, we first discuss a simple, special, case.

\subsection{Functors restricting to finite sets}

For a gentle introduction of our derivation system we first consider the special
case where the functor restricts to finite sets.

Turning to the interaction principles, we first consider the interaction between 
the coalgebraic modality and \emph{conjunctions}.
More specifically, the purpose of axiom ($\nb2$) will be to rewrite a conjunction
of nabla formulas as an equivalent `disjunction of nablas of conjunctions', and
we think of this axiom as a distributive law (in the logical sense).
Formally, recall from Definition~\ref{d:srd} that given a finite set $A \in 
\Pom\Tom\Lmoss$, the set $\SRD(A) \sse \Tom\Pom\Lmoss$ denotes the set of 
\emph{slim redistributions} of $A$.
Also recall that given an object $\Phi \in \Tom\Pom\Lmoss$, we find 
$(\T\bw)\Phi \in \Tom\Lmoss$, which means that $\nb(\T\bw)\Phi$ is a
well-formed formula.
We can now formulate the axiom ($\nb 2$) as the following inequality:
\begin{equation}
\tag{$\nb 2_{f}$}
\bw \Big\{ \nb\al \mid  \al \in A \Big\} \isleq 
\bv \Big\{ \nb (\T\bwsmall)\Phi \mid \Phi \in \SRD(A) \Big\}
\end{equation}

\begin{exa}
First consider the case of the $C$-labelled binary tree functor $\Btree$ of 
Example~\ref{ex:2}.
In Example~\ref{ex:srd} we discussed the shape of the collection of slim
redistributions of a collection $A \sse_{\om} \Tom\Lmoss$.
From this it should be clear that we obtain the following three instances 
of ($\nb2_{f}$).

\begin{enumerate}[(1)]
\item 
If $A=\nada$, we obtain
\[
\top \isleq \bv \{ \nb (c,\top,\top) \mid c \in C \}
\]
\item
If $A$ contains two elements $(c,a_{1},a_{2})$ and $(c',a'_{1},a'_{2})$
with $c \neq c'$, then we obtain
\[
\bw \{\nb\al \mid \al \in A \} \isleq \bot.
\]
\item
If $\pi_{C}[A]$ contains a unique element $c_{A}$, then we obtain
\[
\bw \{\nb\al \mid \al \in A \} \isleq
\nb(c_{A},\pi_1[A],\pi_2[A])  
\]
where $\pi_{C}$, $\pi_{1}$ and $\pi_{2}$ are the projection functions, as 
in Example~\ref{ex:srd} and where we used the optimization outlined in Remark~\ref{rem:superslim}.
\end{enumerate}

\noindent
Second, in the case of the power set functor in Example~\ref{ex:srd2}, $\T =\Pow$, an instance of
($\nb 2_f$) looks as follows
\begin{equation}
\label{eq:dl1}
\bw_{\al\in A} \nb \al \isleq
\bv \Big\{ \nb \{ \bwsmall\be \mid \be \in \Phi \}
\mid \bcsmall A = \bcsmall \Phi \text{ and }
   \al\cap\be\neq\nada \text{ for all } \al\in A, \be \in \Phi
\Big\}
\end{equation}
\end{exa}

\begin{rem}
In fact, we could have formulated this principle as an \emph{equation} rather 
than as an inequality, since the opposite inequality of ($\nb 2_{f}$) can be
derived on the basis of $(\nb 1)$.
To see this, observe that for any formula $a \in \Lmoss$ and any set $\phi \in
\Pom\Lmoss$ it holds that $a \in \phi$ implies that $a \sqgeq \bw\phi$.
Reformulating this as $({\in};\bw)\; \sse \; {\sqgeq}$, and using the properties of 
relation lifting we find that $\Tb{\in};\T{\bwsmall} \sse \Tb{\sqgeq}$.
From this it follows that, whenever $\al \in \Tom\Lmoss$ is a lifted member of
$\Phi \in \Tom\Pom\Lmoss$, we find that $(\T\bwsmall)\Phi \rl{T}(\sqleq)\al$.
From this, one application of ($\nb 1$) yields the existence of a derivation for
the inequality $\nb(\T\bwsmall)\Phi \isleq \nb\al$.
Since this holds for any $\al$ and $\Phi$ with $\al \rl{T}{\in} \Phi$, we 
may conclude that 
\[
\bv \Big\{ \nb (\T\bwsmall)\Phi \mid \Phi \in \SRD(A) \Big\}
\sqleq
\bw \Big\{ \nb\al \mid  \al \in A \Big\}.
\]
That is, the opposite inequality of ($\nb 2_{f}$) is indeed derivable.
\end{rem}

Our second interaction principle, ($\nb3$), involves the interaction between $\nb$
and the \emph{disjunction} operation.
And again, we think of this axiom as a distributive law (in the logical sense),
stating that the coalgebraic modality distributes over disjunctions.
More precisely, the rule reads as follows:
\[
\tag{$\nb 3_{f}$}
\nb(\T\bvsmall)\Phi \isleq 
\bv \Big\{ \nb \beta \mid \beta \rl{T}(\in) \Phi \Big\}
\]

\begin{exa}
In the case of the functor $\Btree=C \times \Id \times \Id$, axiom ($\nb 3_f$)
is of the following shape:
\[ \nb (c,\bvsmall A,\bvsmall B) \isleq 
  \bigvee \{ \nb (c,a,b) \mid a \in A, b \in B \}.
\]

For the power set functor $\funP$, an instance of axiom ($\nb 3_f$) looks as
follows
\[
\nb \{ \bvsmall\be \mid \be\in \Phi\} \isleq 
   \bigvee \{ \nb \alpha \mid \alpha \subseteq \bcsmall \Phi \;
   \mbox{and} \; \alpha \cap \beta \not= \emptyset \; \mbox{for all} \;
   \beta \in \Phi \;  \} .\] 
\end{exa}

\begin{rem}
In this case the opposite inequality can be derived on the basis of ($\nb1$)
as well.
Here we use the fact that $a \in \phi$ implies $a \sqleq \bv\phi$, or in other
words, that ${\in};{\bv} \sse {\sqleq}$.
This implies that $\Tb{\in};\T{\bv} \sse \Tb{\sqleq}$, and hence, 
whenever $\be$ is a lifted member of $\Phi$, we find that $\be \Tb{\sqleq}
(\T\bv)\Phi$.
Thus an application of ($\nb 1$) shows the derivability of the inequality
$\nb\be \isleq \nb(\T\bv)\Phi$.
And since this applies to every lifted member of $\Phi$, we may conclude that 
\[
\bv \Big\{ \nb \beta \mid \beta \rl{T}(\in) \Phi \Big\}
\sqleq
\nb(\T\bvsmall)\Phi,
\]
meaning that, indeed, the opposite inequality of ($\nb 3_{f}$) is derivable.
\end{rem}

Summarizing, in the case of a set functor $\T$ that preserves finite sets, 
our derivation system
$\nax_{f}$ extends that of classical proposition logic (Table~\ref{tb:clax}) 
with one congruence/monotonicity rule, and two axioms that take the form of 
distributive laws, see Table~\ref{tb:naxfin}.
The point of restricting to this case is to ensure that the axioms
($\nb2_{f}$) and ($\nb3_{f}$) are well-formed pieces of syntax, in the sense that the
disjunctions on the right hand side are \emph{finite}.

\begin{rem}
	The requirement on the given set functor $\T$ to preserve finite sets is obviously sufficient
	in order to ensure that the axioms ($\nb2_{f}$) and ($\nb3_{f}$) are well-formed.
	Note, however, that there are set functors that do not restrict to finite sets and
	for which the axioms ($\nb2_f$) and ($\nb3_f$) are nevertheless syntactically
	well-formed.
	
	Consider for example the bag functor $\Bag_\om$ from Example~\ref{ex:1}. In order to show that ($\nb2_f$) and ($\nb3_f$) are well-formed we have to prove that the sets
	\begin{align}
		\{ \Phi \in \Bag_\om \Pom X \mid \Phi \in \SRD(A) \}  &\quad \mbox{for } \; 
		A \in \Pom \Bag_\om X \; \mbox{and} \label{finite_number1}\\
		\{ \beta \in \Bag_\om X \mid \beta (\overline{\Bag_\om} \in ) \Phi \} &\quad \mbox{for } \; 
		\Phi \in \Bag_\om \Pom X \label{finite_number2}
	\end{align}
	are finite. Using the characterisation of the relation lifting for $\Bag_\om$ in
	Example~\ref{ex:rellift} this is not diffcult to see: Let us consider first the set
	in (\ref{finite_number1}), ie., we consider some $A \in \Pom \Bag_\om X$
	and we want to prove that the set $\{ \Phi \in \Bag_\om \Pom X \mid \Phi \in \SRD(A) \}$
	is finite. If $\Phi \in \SRD(A)$ then by the definition of slim redistributions we have
	$(\alpha,\Phi) \in (\overline{\Bag_\om} \in)$ for all $\alpha \in A$ and
	$\Phi \in \Bag_\om \Pom (\bigcup_{\alpha' \in A} \Base(\alpha'))$. Therefore, using
	Proposition~\ref{p:st-rl}, we get that
	\[ (\alpha,\Phi) \in \overline{\Bag_\om} \left( \in \rst{\Base(\alpha) \times
	\Pom (\bigcup_{\alpha' \in A} \Base(\alpha'))} \right) \quad \mbox{for all } \alpha \in A.\]
	This implies, by the definition of $\overline{\Bag_\om}$ from Example~\ref{ex:rellift},
	that there exists a function
	\[ \rho: \in \rst{\Base(\alpha) \times \Pom (\bigcup_{\alpha' \in A} \Base(\alpha'))} \to \bbN \]
	such that for all $\alpha  \in A$, all $x \in \Base(\alpha)$ and all $U \in
	 \Pom (\bigcup_{\alpha' \in A} \Base(\alpha'))$ we have
	\[
		  \Phi(U) = \sum_{x' \in \Base(\alpha), x' \in U} \rho(x',U)
		  \quad \mbox{and} \quad
		\rho(x,U)  \leq \alpha(x)
	\]
	Therefore we have $\Phi(U) \leq  \sum_{x \in U} \alpha(x)$. This shows that the range
	of $\Phi$ has an upper bound an thus, as $\Phi$ is determined by its values on the
	finite set $ \Pom (\bigcup_{\alpha' \in A} \Base(\alpha'))$,  
	there can only finitely many $\Phi$'s that satisfy the requirement of
	a slim redistribution for the set $A$.
	In a similar way one can show that the set $\{\beta \in \Bag_\om X \mid \beta (\overline{\Bag_\om} \in ) \Phi  \}$ in (\ref{finite_number2}) is finite
	for all $\Phi \in \Bag_\om \Pom X$. We leave the details of the argument as an exercise
	to the reader.
	
	One example for a set functor for which the finitary axioms ($\nb2_f$) and ($\nb3_f$) are not
	well-formed is provided by the finitary probability functor $D_\om$ in Example~\ref{ex:1}.
\end{rem}

\begin{table}
\begin{center}
\begin{tabular}{|lc|}
\hline  & \\
($\nb 1$) & \AXC{$\Big\{ a \isleq b \mid (a,b) \in Z \Big\}$}
\RL{$(\alpha,\beta) \in \Tb Z $}
\UIC{$\nb\alpha \isleq \nb\beta$}
\DisplayProof 
\\ & \\
($\nb 2_{f}$) & 
$\bwlarge \Big\{ \nb\al \mid  \al \in A \Big\} \isleq 
\bvlarge \Big\{ \nb (\T\bwsmall)\Phi \mid \Phi \in \SRD(A) \Big\}$
\\ & \\
($\nb 3_{f}$) & 
$\nb(\T\bvsmall)\Phi \isleq 
\bvlarge \Big\{ \nb \beta \mid \beta \rl{T}(\in) \Phi \Big\}$
\\ & \\
\hline
\end{tabular}
\end{center}
\caption{Rules and axioms of the system $\nax$ (in case $\T$ preserves
finite sets)}
\label{tb:naxfin}
\end{table}

\subsection{The derivation system $\nax$}

In the case that we are dealing with an arbitrary set functor $\T$ (not
necessarily preserving finite sets), we would like to use the same derivation
system as given in Table~\ref{tb:naxfin}.
Unfortunately however, in this case the axioms ($\nb 2_{f}$) and ($\nb 3_{f}$)
are no longer well-formed syntactic expressions, since we cannot guarantee that
the disjunctions on the right hand sides are taken over a \emph{finite} set.
In order to deal with this problem, we use the following trick: we replace an
axiom of the form 
\[
a \isleq \bv \{ a_{i} \mid i \in I \}
\]
with the derivation rule
\[
\AXC{$\{ a_{i} \isleq b \mid i \in I \}$}
\UIC{$a \isleq b$}
\DisplayProof 
\]
The price that we have to pay for this transformation is that our derivation
system will be \emph{infinitary}.

\begin{definition}\label{def:nax}
The derivation system $\nax$ is given by the axioms and derivation rules 
of Table~\ref{tb:nax}, together with the complete set of axioms and rules
for classical propositional logic given in Table~\ref{tb:clax}.
\begin{table}
\begin{center}
\begin{tabular}{|lc|}
\hline & \\
($\nb 1$) &
\AXC{$\{ a \isleq b \mid (a,b) \in Z \}$}
\RL{$(\alpha,\beta) \in \Tb Z$}
\UIC{$\nb\alpha \isleq \nb\beta$}
\DisplayProof 
\\ & \\
($\nb 2$) &
\AXC{$\{ \nb (\T\bw)(\Phi) \isleq b \mid \Phi\in \SRD(A)\}$}
\UIC{$\bw\{\nb\alpha \mid \alpha\in A\} \isleq b$}
\DisplayProof 
\\ & \\
($\nb 3$) &
\AXC{$ \{ \nb\alpha \isleq b \mid \alpha \Tin \Phi \}$}
\UIC{$\nb(\T\bv)(\Phi) \isleq b$}
\DisplayProof 
\\ & \\
\hline
\end{tabular}
\end{center}
\caption{Rules of the system $\nax$}
\label{tb:nax}
\end{table}
\end{definition}
Our notions of derivation and derivability are completely standard.

\begin{definition}\label{def:deriv}
A \emph{derivation} is a well-founded tree, labelled with inequalities, such 
that the leaves of the tree are labelled with axioms of $\nax$, whereas with
each parent node we may associate a derivation rule of which the conclusion
labels the parent node itself, and the premisses label its children.
If $\D$ is a derivation of the inequality $a\isleq b$, we write 
\AXC{$\D$}
\UIC{$a \isleq b$}
\DisplayProof 
or $\D: a \sqleq b$.
If we want to suppress the actual derivation, we write $\vdash_{\nax} a \isleq b$
or (in accordance with Definition~\ref{d:dy}) $a \sqleq_{\nax} b$.
\end{definition}

Note that $\nax$ is not a Gentzen-style derivation system; in particular, 
we do not have left- and right introduction- and elimination rules for
$\nb$.
Readers who are interested to see a detailed development of the \emph{proof
theory} of nabla-style coalgebraic logic, are referred to B\'{\i}lkov\'a,
Palmigiano \& Venema~\cite{bilk:proo08} (for the power set case).

\subsection{Soundness and completeness}
\label{ss:main}

We can now very concisely formulate the main result of this paper as the
following soundness and completeness result:

\begin{thm}
\label{t:main}
Let $\T$ be a standard set functor that preserves weak pullbacks.
For all formulas $a,b \in \Lmoss$ we have
\begin{equation}
\label{eq:main}
\vdash_{\nax}  a \isleq b 
\qquad \mbox{iff} \qquad
a \models_{\T} b.
\end{equation}
\end{thm}

In words, Theorem~\ref{t:main} states that for any two $\Lmoss$-formulas
$a$ and $b$, the inequality $a\isleq b$ is derivable in our derivation
system $\nax$ iff it is valid in all $\T$-coalgebras.
Our proof of this result will be based on many auxiliary results, which we
will discuss in the next two sections.
The final proof will be given at the end of section~\ref{s:completeness}.

\subsection{The role of negation}
\label{ss:neg}

At this point, the reader may be surprised or even worried that we have 
formulated our derivation system for a Boolean-based coalgebraic modal 
logic, without mentioning the negation connective (or the implication, for 
that matter) in relation to the nabla modality at all.
Surely there must be some validities involving both $\nb$ and $\neg$?
The point is that indeed there are such interaction principles, but we do not
need to formulate them explicitly as axioms or derivation rules since they are 
already \emph{derivable} in the system $\nax$.
The intuition underlying this fact is that in a bounded distributive lattice, all
existing complementations are completely determined by the lattice operations:
the complement $\neg a$ of an element $a$, if existing, is the unique element $b$
such that $a \land b = \bot$ and $a \lor b = \top$.

Nevertheless, the key principle relating $\nb$ to $\neg$ will be needed in
our proofs below, and so we discuss it in some detail.
For a smooth formulation we need the following definition.

\begin{definition}
Given an element $\al \in \Tom\Lmoss$, let $Q(\al) \sse \Tom\Lmoss$ be the
set defined by
\[
Q(\al) :=
\Big\{ T (\bwsmall \circ \funP\neg) \Psi \mid
\Psi \in \Tom\Pom \Base(\al) \mbox{ and } 
(\al,\Psi) \not\in \Tb{\not\in} 
\Big\}.\vspace{-18 pt}
\]
\end{definition}

\noindent To unravel this definition, observe that $\funP\neg: \Pom\Lmoss \to \Pom\Lmoss$,
and so we have $\bw\circ\funP\neg: \Pom\Lmoss \to \Lmoss$.
Thus we find that for $\Psi \in \Tom\Pom \Base(\al) \sse \Tom\Pom\Lmoss$
we have $(\T(\bw\circ\funP\neg))\Psi \in \Tom\Lmoss$ indeed.

In case $\T$ preserves finite sets, $Q(\al)$ is a finite set, and we can
express the principle relating $\nb$ and $\neg$ as follows:
\[
\tag{$\nb 4_{f}$}
\neg\nb\al \is 
\bv \Big\{ \nb \be \mid \be \in Q(\al) \Big\}.
\]
In other words: the negation of a nabla is equivalent to a disjunction of 
nablas of conjunctions of negations of the base formulas.
Putting it yet differently, in the case of $\T$ preserving finite sets,
we can define the \emph{Boolean dual} $\Delta$ of $\nabla$, just in terms of 
$\nb$ and $\bv$.
For more information on this dual modality $\Delta$ the reader is referred
to Kissig \& Venema~\cite{kiss:comp09}.

In the general case, that is, if the functor $\T$ does not necessarily take
finite sets to finite sets, we can express the interaction between $\nb$ and
$\neg$ in the form of a derivation rule,
\[
\tag{$\nb 4_{L}$}
\AXC{$\{ \nb \be \isleq b
\mid 
\be \in Q(\al) \}$}
\UIC{$\neg\nb\al \isleq b$}
\DisplayProof
\]
and a collection of axioms:
\[
\tag{$\nb 4_{R}$}
\{ \nb\be \isleq \neg\nb\al \mid \be \in Q(\al) \},
\]
corresponding to the directions $\isleq$ and $\succcurlyeq$ of ($\nb 4_{f}$),
respectively.
The point to make is that \emph{both} ($\nb 4_{L}$) and ($\nb 4_{R}$) are
\emph{derivable} in $\nax$.
We will prove this in detail for ($\nb 4_{L}$).
Given our completeness result, the derivability of $(\nb 4_{R})$ is an immediate
consequence of its validity~\cite{kiss:comp09}.
The actual derivation of $\nb\be \isleq \neg\nb\al$ for $\be \in Q(\al)$ is
rather involved, so we refrain from giving the details here.

In any case, the key instruments in the derivability of both ($\nb 4_{L}$) and
($\nb 4_{R}$) are the following two rules.

\begin{prop}
\label{p:negder}
For any finite set $\phi$ of formulas, the following rules are $\nax$-derivable:

\noindent
\begin{tabular}{ll}
\\

$(\nb 4a)$ &
\AXC{$\top \isleq \bv\phi$}
\AXC{$\Big\{\nb\al \isleq b \mid \al \in \T\phi \Big\}$}
\BIC{$\top \isleq b$}
\DisplayProof

\\ \\

$(\nb 4b)$ &
\AXC{$\Big\{ a \land a' \isleq \bot \mid a \neq a' \in \phi \Big\}$}
\RL{\hspace{1mm} $\al\neq\al' \in \T\phi$}
\UIC{$\nb\al \land \nb\al' \isleq \bot$}
\DisplayProof

\end{tabular}
\end{prop}

\begin{proof}
In the proof below, the following principle will be used a few times:
\begin{equation}
\label{eq:pr1}
\mbox{Given $f: S \to S'$, for $s \in S$, $\T f$ restricts
to a bijection $\T f: \T\{s\} \to \T\{f(s)\}$}
\end{equation}

We first show the derivability of ($\nb 4a$).
Assume that we have a derivation $\D_{\top}$ of $\top \isleq \bv\phi$, and a
derivation $\D_{\al}$ of $\nb\al \isleq b$, for each $\al\in\T\phi$.

Consider an arbitrary element $\Phi \in \T\{\phi\}$.
By Proposition~\ref{p:nbsem}(\ref{item:memberofdistri}), each lifted 
member $\al$ of $\Phi$ belongs to $\T\phi$.
If we apply ($\nb 3$) to the set $\{ \D_{\al} \mid \al \Tin \Phi \}$,
we obtain a derivation
\[
\D_{\Phi}:
\AXC{$\{ \D_{\al}: \nb\al \isleq b \mid \al \Tin \Phi \}$}
\UIC{$\nb(\T\bvsmall)(\Phi) \isleq b$}
\DisplayProof
\]
for each $\Phi\in\T(\{\phi\})$.

Applying our principle \eqref{eq:pr1} to the map $\bv: \Pom\Lmoss \to \Lmoss$,
we find that each $\be\in \T(\{ \bv\phi \})$ is of the form $\be = 
(\T\bv)(\Phi_{\be})$ for some $\Phi_{\be} \in \T(\{\phi\})$.
Thus in fact for each such $\be$ we have a derivation
\[
\D_{\be}: \nb\be \isleq b
\]

On the other hand, we may continue the derivation $\D_{\top}$ as follows.
Consider the bijection $f: \{ \top \} \to \{ \bv\phi \}$, which induces a
bijection $\T f: \T\{ \top \} \to \T\{ \bv\phi \}$.
Clearly we find that $f \sse \{\sqleq\}$, so that $\T f \sse \Tb{\sqleq}$.
From this it follows that we may apply the rule ($\nb 1$) to the inequality
$\top \isleq \bv\phi$ and obtain, for each $\ga\in\T\{\top\}$, the derivation
\[
\AXC{$\D_{\top}$}
\UIC{$\top \isleq \bv\phi$}
\LL{$\nb 1$}
\UIC{$\nb\ga\isleq\nb (\T f)\ga$}
\DisplayProof
\]

Combining the observations until now, we obtain the following derivation
$\D_{\ga}$ for each $\ga \in \T\{\top\}$:
\[
\D_{\ga}:\hspace{10mm}
\AXC{$\D_{\top}$}
\UIC{$\top \isleq \bv\phi$}
\LL{$\nb 1$}\UIC{$\nb\ga\isleq\nb (\T f)\ga$}
\AXC{$\D_{(\T f)\ga}$}
\UIC{$\nb (\T f)\ga \isleq b$}
\LL{cut}
\BIC{$\nb\ga \isleq b$}
\DisplayProof
\]
Since $(\T\bw)(\Psi) \in \T\{\top\}$ for each $\Psi \in \T \{ \nada \}$, this
means that above we have obtained a derivation
\[
\D_{\Psi}: \nb(\T\bwsmall)(\Psi) \isleq b
\]
for each $\Psi\in \T\{\nada\}$.

Finally, consider the instantiation of ($\nb 2$) with $A = \nada$.
By Proposition~\ref{item:redistriofempty} we have 
$\SRD(\nada) = \T\{\nada\}$, so that the set $\left\{ \nb(\T\bw)(\Psi) \isleq b
\mid \Psi \in \T\{\nada\} \right\}$ is exactly the set 
of premises of this instantiation of ($\nb 2$).
Hence we may simply take the set of all derivations $\D_{\Psi}$, with 
$\Psi \in \T\{\nada\}$, and continue as follows:
\[
\AXC{$\Big\{ \D_{\Psi} \mid \Psi \in \T\{\nada\} \Big\}$}
\LL{$\nb2$}
\UIC{$\top \isleq b$}
\DisplayProof
\]
This finishes the proof of the derivability of ($\nb 4a$).
\medskip

In the case of ($\nb 4b$) we will proceed a bit faster, leaving the details
as to why our argumentation yields derivability rather than admissibility, 
as an exercise for the reader.
Let $\phi$ be a finite set of formulas such that
$a \land a' \equiv \bot$ for all distinct $a,a' \in \phi$, and let $\al$ and
$\al'$ be two distinct elements of $\T\phi$.
We will derive the inequality $\nb\al \land \nb\al' \isleq \bot$.
By ($\nb 2$) it suffices to show that
\[
\vdnax \nb (\T\bwsmall)(\Phi) \isleq \bot,
\]
where $\Phi$ is an arbitrary slim redistribution of the set $\{ \al, \al'\}$.

But if $\Phi \in \SRD(\{\al,\al'\})$, and both $\al$ and $\al'$ belong to 
$\T\phi$, then first of all we have $\Base(\Phi) \sse \funP \phi$,
because $\Phi \in \Tom \Pom (\Base(\alpha) \cup \Base(\alpha'))$ by
the definition of a slim redistribution and thus
$\Base(\Phi) \subseteq \funP (\Base(\alpha) \cup \Base(\alpha')) \subseteq
\funP \varphi$.
In addition, it follows by Proposition~\ref{p:nbsem}(1) that $\nada 
\not\in\Base(\Phi)$, and then by Proposition~\ref{p:nbsem}(\ref{item:singletonredistri})
that $\Base(\Phi)$ contains some set $\psi \sse \phi$ with $\size{\psi} > 1$.
Define the following function $d: \Base(\Phi) \to \funP(\phi) \cup \Big\{\{ \top \}
\Big\}$:
\[	 
d(\chi)  \coloneqq  
\left\{\begin{array}{lcl}
  \emptyset   & \mbox{if} & \size{\chi} > 1 
\\ \chi       & \mbox{if} & \size{\chi} = 1 
\\ \{ \top \} & \mbox{if} & \size{\chi} = 0
\end{array} \right.
\]
On the basis of our set of premises $\{ a \land a' \isleq \bot \mid a \neq a'\in
\T\phi \}$, for each $\chi \in \Base(\Phi) \sse \funP\phi$ we can find a
derivation for the  inequality $\bw\chi\isleq \bv d(\chi)$.
Putting these derivations together, and applying ($\nb 1$) with 
$Z= \{ (\bw \chi, \bv d(\chi)) \mid \chi \in \Base(\Phi) \}$, 
we obtain a derivation $\D_{\Phi}$ for the inequality $\nb (\T
\bw)(\Phi) \isleq \nb (\T \bv)(\T d (\Phi))$.

We also claim that we can derive the inequality $\nb (\T\bv)(\T d (\Phi)) \isleq
\bot$.
Since $\Base: \Tom \to \Pom$ is a natural transformation, we have that 
$\Base(\T d(\Phi)) = (\funP d) (\Base(\Phi)) = d[\Base(\Phi)]$.
Now recall that above we found a $\psi \in \Base(\Phi)$ with $\size{\psi} > 1$;
it follows that $\nada = d(\psi) \in \Base(\T d(\Phi))$, so that on the basis of 
Proposition~\ref{p:nbsem}(1) we may conclude that $\T d(\Phi)$ has \emph{no}
lifted members.
But then one single application of ($\nb 3$), with the \emph{empty} set of 
premisses, provides the desired derivation for $\nb (\T\bv)(\T d (\Phi)) \isleq
\bot$.

Finally then, an application of the cut rule gives $\nb (\T\bw)(\Phi) \isleq
\bot$, as required.
\end{proof}

As a corollary to this we can now prove the derivability of $(\nb 4_{L})$.

\begin{prop}
The rule $(\nb 4_{L})$ is derivable in $\nax$.
\end{prop}

\begin{proof}
Let $\al \in \Tom\Lmoss$ and $b \in \Lmoss$ be arbitrary, and assume that for 
all $\be \in Q(\al)$ we have $\nb\be \sqleq b$.
We will show that $\neg\nb\al \sqleq b$.

Consider the map $t: \Pom\Base(\al) \to \Lmoss$ given by
\[
t:\psi \mapsto \bw\{ a \in \Base(\al) \mid a \not\in\psi \} \land 
  \bw \{\neg b \mid b \in \psi \}.
\]
Then for all $\psi \sse \Base(\al)$ it is straightforward to verify that (i)
$t(\psi) \sqleq (\bw\circ\funP\neg)\psi$, and (ii) if $a \not\in \psi$ then
$t(\psi) \sqleq a$.

Define $\phi$ to be the \emph{range} of $t$.
Intuitively, think of $\phi$ as the set of atoms of a Boolean algebra; then it 
is not hard to see that 
\begin{equation}
\label{eq:nb4a1}
\top \sqleq \bv\phi.
\end{equation}
We claim that 
\begin{equation}
\label{eq:nb4a2}
\mbox{ for all } \ga\in\T\phi: \nb\ga \sqleq b \lor \nb\al.
\end{equation}
For the proof of \eqref{eq:nb4a2}, take an arbitrary $\ga \in \T\phi$.
By definition of $\phi$, the map $\T t$ is surjective when seen as $\T t: 
\Tom\Pom\Base(\al) \to \Tom\phi$, and so we may fix an element $\Psi \in
\Tom\Pom\Base(\al)$ such that $\ga = (\T t)\Psi$.
Now distinguish cases.

First assume that $(\al,\Psi) \not\in \Tb{\not\in}$.
It follows from (i) that $\ga = (\T t)\Psi \,\rel{\Tb{\sqleq}}\, 
(\T(\bw\circ\funP\neg))\Psi$, and so an application of ($\nb 1$) shows that
$\nb\ga \sqleq \nb (\T(\bw\circ\funP\neg))\Psi$.
Now by assumption we have $(\T(\bw\circ\funP\neg))\Psi \in Q(\al)$, and so
there is a derivation of the inequality $\nb(\T(\bw\circ\funP\neg))\Psi \isleq b$.
Then an application of the cut rule shows that $\nb\ga \sqleq b$.

If, on the other hand, the pair $(\al,\Psi)$ \emph{does} belong to the relation
$\rl{T}{\not\in}$, then by (ii) we obtain that $\ga = (\T t)\Psi \,\rl{T}{\sqleq}\,
\al$.
Now an application of ($\nb 1$) yields a derivation for $\nb\ga \isleq \nb\al$.

In either case, a simple propositional continuation of the derivation shows that
$\nb\ga \sqleq b \lor \nb\al$, which proves \eqref{eq:nb4a2}.

Finally, applying the derived rule ($\nb 4a$) to the premisses given by 
\eqref{eq:nb4a1} and \eqref{eq:nb4a2}, we obtain a derivation of the inequality
$\top \isleq b \lor \nb\al$.
But from this it follows by some straightforward classical propositional
manipulations that $\neg\nb\al \sqleq b$, as required.
\end{proof}

%% file: sec-onestep.tex
\section{One-step soundness and completeness}
\label{s:onestep}

As mentioned in the introduction, our completeness proof is based on
Pattinson's stratification method~\cite{patt:coal03}, which consists
of stratifying the logic in $\om$ many layers which are nicely glued
together by means of a so-called one-step version of the derivation
system.  The main technical hurdle in this method consists of showing
that this one-step derivation system is sound and complete with
respect to a natural one-step semantics.  In this section we will
first properly introduce our version of these notions, and then prove
the one-step soundness and completeness result.

\subsection{One-step semantics and one-step axiomatics}

Starting with the one-step semantics, fix a set $X$ and think of
$\funQ X$ as a set of formal objects or \emph{propositions}.  Recall
from Section~\ref{s:moss} that $\Tba \funQ X$ and $\Tnb \funQ X$ are
the sets of formulas of depth zero and depth one over this language,
respectively.  The point underlying the one-step semantics is that
there is a natural interpretation of the formulas in $\Tnb\funQ X$ as
sets of elements of $\T X$, or, expressed more accurately, as elements
of the Boolean algebra $\funaQ\T X$.  To explain this, first note that
we may see the identity map
\[
\iota: \funQ X \to \funQ X
\]
as a natural \emph{valuation} interpreting variables of $\funQ X$ as
subsets of $X$, and then extend this valuation to a unique
homomorphism
\[
\semzero{\cdot}^{\akk{X}} \coloneqq \ti{\iota}: \akk{\funaF\funQ X \to \funaQ X}.
\]
\akk{We find it convenient to denote $U\ti{\iota}:\Tba\funQ X \to
  \funQ X$ by the same symbol $\semzero{\cdot}^X$ and also to
  occasionally drop the superscript $^{\akk{X}}$.}  We may associate a
relation $\forces_{X}^{0} \sse X \times \Tba \funQ X$ with this map,
which we define inductively by putting
\[\begin{array}{lll}
x \forces_{X}^{0} p & \mbox{if} & x \in p, \; \mbox{ where } \; p \in \funQ X,
\\ x \forces_{X}^{0} \bv\phi & \mbox{if} & x \forces_{X}^{0} a \mbox{ for some }
   a \in \phi,
\\ x \forces_{X}^{0} \bw\phi & \mbox{if} & x \forces_{X}^{0} a \mbox{ for all }
   a \in \phi.
 \end{array}\]
 Clearly the relation between $\semzero{\cdot}$ and $\forces_{X}^{0}$ is given by
\[
x \in \semzero{a}  \ \mbox{ iff } \ x \forces_{X}^{0} a,
\]
for all $x \in X$ and all $a \in \Tba \funQ X$.

We note for future reference that $\semzero{\cdot}$ gives rise to a
natural transformation.
\begin{prop}
\label{p:semzeronat}
The family of homomorphisms $\{ \semzero{\cdot}^X\}_{X\in\Set}$ \akk{is a natural
  transformation $\funaF\funQ \ntrto
\funaQ $ and, therefore, also a natural transformation $\semzero{\cdot}: \Tba\funQ \ntrto \funQ$.}
\end{prop}

\begin{proof}
  Naturality of $\semzero{\cdot}$ is a matter of routine checking. The
  key for the proof is that for any function $f:X \to Y$, $\funQ f:
  \funQ Y \to \funQ X$ is a Boolean homomorphism.
\end{proof}

Turning our attention to depth-one formulas, perhaps the easiest way to explain
their one-step semantics is to introduce a similar relation ${\forces_{X}^{1}}
\sse \T X \times \Tnb \funQ X$:
\[\begin{array}{lll}
\T X, \xi \forces_{X}^{1} \nb\al & \mbox{if} & 
   (\xi,\al) \in \Tb(\forces_{X}^{0}),
\\ \T X, \xi \forces_{X}^{1} \bv\phi & \mbox{if} & \T X, \xi \forces_{X}^{1} c
   \mbox{ for some } c \in \phi,
\\ \T X, \xi \forces_{X}^{1} \bw\phi & \mbox{if} & \T X, \xi \forces_{X}^{1} c
   \mbox{ for all } c \in \phi.
\end{array}\]

\begin{rem}
\label{r:semsem}
It is instructive to have a look at the relationship between the
one-step semantics of depth-one formulas and the coalgebraic semantics
for arbitrary formulas from Definition~\ref{def:moss_sem}.  Roughly,
the definition of the one-step semantics of a formula captures
precisely what is needed to inductively define the semantics of the
logic.

More precisely, let $(X,\xi)$ be some $\T$-coalgebra and let, for $i<\om$,
$\mng_{i}: \Lmoss_{i} \to \funQ X$ be the map, that maps any formula $a \in
\Lmoss_{i}$ of modal rank $i$ to its coalgebraic meaning, that is, for all
$a \in \Lmoss_{i}$ and all $x \in X$ we let $x \in \mng_i(a)$ if $x \forces a$. 
Now we claim that for any $k<\om$ and any $\nb \alpha \in \Lmoss_{k+1}$ we
have
\begin{equation}
\label{eq:semsem0}
x \forces_{\bbX} \nb \alpha \ \mbox{ iff } \ 
\T X, \xi(x)  \forces_X^1 \nb (\T\mng_{k})\alpha.
\end{equation}
To see this, first observe that by induction on the Boolean structure
of $\Lmoss_{k}$-formulas, we may show that for any $a \in \Lmoss_{k}$
and any $x \in X$, we have $x \forces_{\bbX} a$ iff $x \forces_{X}^{0}
\mng_{k}(a)$.  In other words, we have
\begin{equation}
\label{eq:semsem1}
({\forces_{\bbX}) \rst{X \times \Lmoss_{k}}} = 
{\forces_{X}^{0}} \corel \cv{\mng_{k}}.
\end{equation}
Based on this, we may reason as follows:
\begin{align*}
x \forces_{\bbX} \nb \alpha
  &\iff \xi(x) \rel{\Tb{\forces_{\bbX}}} \al
  &\text{(definition of $\forces$)}
\\&\iff \xi(x) \rel{\Tb ({({\forces_{\bbX})}\rst{X\times\Lmoss_{k}}})} \al
\\&\iff \xi(x) \rel{\Tb {{\forces_{X}^{0}} \corel \cv{\mng_{k}}}} \al
  &\text{(equation \eqref{eq:semsem1})}
\\&\iff \xi(x) \rel{\Tb {{\forces_{X}^{0}}}} (\T\mng_{k}) \al
  &\text{(properties of relation lifting)}
\\&\iff \T X, \xi(x)  \forces_X^1 \nb (\T\mng_{k})\alpha.
  &\text{(definition of $\forces^{1}$)}
\end{align*}

\noindent In words: if we assume that we have already defined the 
interpretation of all formulas of modal rank $k$ then the one-step 
semantics allows us to extend this interpretation to formulas of rank $k+1$.
\end{rem}

The relation $\forces_{X}^{1}$ provides a natural semantics for terms of depth
one, and induces a natural semantic equivalence relation.

\begin{definition}
Given a set $X$, we define the one-step semantics $\semone{a'}$ of a formula
$a' \in \Tnb(\funQ X)$ as
\[
\semone{a'} := \{ \xi \in \T X \mid \T X, \xi \forces_{X}^{1} a' \}.
\]
We say that two formulas $a',b' \in \Tnb(\funQ X)$ are
\emph{semantically one-step equivalent}, notation: $a' \eqsem b'$, if
$\semone{a'} = \semone{b'}$.
\end{definition}

\begin{rem}
\label{r:onestep2}
Alternatively but equivalently, we can define the $\semone{\cdot}$
as follows.  Apply $\T$ to the map $\semzero{\cdot}$, and
compose with the function $\nbsem_{X}$ to obtain
\[
\nbsem_{X}\circ\T\semzero{\cdot}: \Tom\Tba\funQ X \to \funQ\T X.
\]
This map then provides us with an interpretation of the basic formulas in 
$\Tnb\funQ X = \Tba\Tomnb\Tba\funQ X$, namely the ones of the form $\nb\al \in
\Tomnb\Tba\funQ X$:
\[
\mu_{X}(\nb\al) := (\nbsem_{X}\circ\T\semzero{\cdot})(\al).
\]
Now $\semone{\cdot}$ may be identified with $\akk{U}\ti{\mu}_{X}:
\Tba\Tomnb\Tba\funQ X \to \akk{\funQ}\T X$. \akk{Occasionally, we will
  write $\semone{\cdot}$ also for the $\Boole$-morphism $\ti{\mu}_{X}:
  \funaF\Tomnb\Tba\funQ X \to \funaQ\T X$.}
\end{rem}

To match the semantic notions of equivalence between $\Tnb\funQ
X$-formulas, we introduce a one-step version of the derivation system
$\nax$, associated with the presentation $\funC\funaQ X$ of the power
set Boolean algebra $\funaQ X$.  Formal definitions will be given
below, but the basic idea is straightforward: modify $\nax$ by (i)
restricting attention to the depth-0 and depth-1 formulas over the set
$\funP X$ of (formal) variables, and (ii) adding the `true facts about
$\funaP X$' as additional axioms.  The resulting derivation system
naturally induces an interderivability relation on $\Tnb\funQ
X$-formulas that we shall denote as $\equiv_{\nax \funC\funaP X}$ for
reasons that we will clarify in Remark~\ref{r:sc1} further on.  This
then raises the question whether the two equivalence relations are the
same or not, and the main aim of this section is to provide an
affirmative answer to this question.

\begin{thm}[1-step soundness and completeness]
\label{t:sc1}
For any set $X$, and for any pair of formulas $c,d \in \Tnb\funP X$ we
have
\begin{equation}
\label{eq:1stsc}
c \eqsem d \ \mbox{ iff } \ c \equiv_{\nax\funC\funaP X} d.
\end{equation}
\end{thm}

Our proof of this result will be algebraic, and before we can move to
the details of the proof, we need to set up the appropriate framework
for this.

We now define the one-step derivation system $\nax\pGR$ associated
with a presentation $\pGR$.  Recall that $\Tba G$ and $\Tnb G =
\Tba\Tomnb\Tba(G)$ are the set of depth zero and depth one formulas in
$G$, respectively.  In this section if we want to stress the
difference between the two kinds of formulas, we shall use
$a,b,\ldots$ for formulas in $\Tba(G)$, and $c,d,\ldots$ for formulas
in $\Tnb(G)$.  An $\Tba G$-inequality is an inequality of the form $a
\isleq b$, with $a,b \in \Tba G$; and likewise for $\Tnb G$.
Intuitively, we obtain $\nax{\pGR}$ from $\nax$ by restricting
attention to $\Tba G$- and $\Tnb G$-inequalities, and adding the
(in)equalities of $R$ as additional axioms.

\begin{definition}
\label{d:ds1}
Given a presentation $\pGR$, we let $\nax\pGR$ denote the
  \emph{one-step derivation system} associated with $\pGR$.  The
  language of $\nax{\pGR}$ consists of $\Tba G$-inequalities, and
  $\Tnb G$-inequalities, and its axioms and rules are those of $\nax$,
  together with the set
\[
R^{\isleq} := \{ a \isleq b, b \isleq a \mid (a,b) \in R \}.
\]
A \emph{$\nax\pGR$-derivation} is a well-founded tree, labelled with
$\Tba G$- and $\Tnb G$-inequalities, such that
(i) the leaves of the tree are labelled with axioms of $\nax$ or with
inequalities in $R^{\isleq}$,
(ii) with each parent node we may associate a derivation rule of which
the conclusion labels the parent node itself, and the premisses label
its children.
\end{definition}

We will leave it for the reader to verify that in
$\nax{\pGR}$-derivations, a parent node is generally labelled with the
same type of inequality (i.e. $\Tba G$ versus $\Tnb G$) as its
children; the single exception is the rule ($\nb 1$) which links
$\Tba$-inequalities of the premises to an $\Tnb$-inequality in the
conclusion.  As a corollary, $\nax{\pGR}$-derivation trees can be
divided into a (possibly empty) upper $\Tba G$-part and a (possibly
empty) lower $\Tnb G$-part.

\begin{rem}
\label{r:sc1}
We can now clarify the syntactic interderivability notion of our
one-step soundness and completeness theorem.  Given a set $X$, recall
that $\funC \funaQ X$ is the canonical presentation of the Boolean
algebra $\funaQ X$, and observe that $\equiv_{\nax\funC\funaQ X}$ is
the associated relation of derivable equivalence of $\Tnb\funQ
X$-terms in the one-step derivation system $\nax\funC\funaQ X$.  It is
\emph{this} derivation system that Theorem~\ref{t:sc1}, stating that
the semantic equivalence relation is the same as the relation
$\equiv_{\nax\funC \funaQ X}$, is concerned with.
\end{rem}

\begin{rem}
\label{r:aiml}
Definition~\ref{d:ds1} corrects and clarifies the corresponding
definition in this paper's earlier incarnation, where the one-step
proof system $\nax\pGR$ was not properly specified.  In particular,
the sentence in~\cite[\akk{Definition~22}]{kupk:comp08}, `in which
\emph{only} elements of $X$ and $\mathfrak{L}(X)$ may be used' (where
$X$ denotes the set of generators) was not only rather vague, but in
fact \emph{mistaken}: it would not permit nontrivial applications of
the derivation rules ($\nb2$) and ($\nb3$), since these require the
use of more terms in $\Tba(X)$ than only the generators in $X$
themselves.
\end{rem}

\subsection{The functor $\funpM$ on presentations}

As we will see now, the notion of a one-step derivation system induces a functor
on the category of presentations.

\begin{definition}
\label{d:funpM}
Given a presentation $\pGR$, we let $\funpM\pGR$ denote the presentation given
as
\[
\funpM\pGR := \prs{\Tomnb\Tba(G)}{\equiv_{\nax\pGR}}.
\]
For a presentation morphism $f: \pGR \to \pGRp$, the definition
\[
\funpM f: \nb\al \mapsto \nb (\Tom \wh{f})\al
\]
provides us with a map $\funpM f: 
\Tomnb\Tba(G) \to \Tomnb\Tba(G')$.
\end{definition}

In other words, $\funpM f$ maps generators of the presentation
$\funpM\pGR$ to
generators of the presentation $\funpM\pGRp$.
We will now show that $\funpM f$ is in fact a presentation morphism from
$\funpM\pGR$ to $\funpM\pGRp$.

\begin{rem}
To be more precise, we need to compose $\funpM f$ with the unit
$\eta_{\Tomnb\Tba(G')}$ of the monad $\Tba$, instantiated at $\Tomnb\Tba(G')$,
in order to obtain a map with the right codomain, $\Tba\Tomnb\Tba(G')$.
In the sequel we will suppress this sublety.
\end{rem}

Our key tool in the proof that $\funpM f$ is a presentation morphism, 
consists of a natural way to transform $\nax\pGR$-derivations into 
$\nax\pGRp$-proofs.

\begin{prop}
\label{p:prm}
If $f: \pGR \to \prs{G'}{R'}$ is a presentation morphism, then there is a map
$(\cdot)^{f}$ transforming $\nax\pGR$-derivations into $\nax\pGRp$-derivations
such that 
\[
\D: c \isleq d \ \Rightarrow\  \D^{f}: \wh{\funpM f} c \isleq \wh{\funpM f} d.
\]
for every $\Tnb G$-inequality $c \isleq d$.
\end{prop}

\begin{proof}
As an easy auxiliary result we need that for any two terms $a,b \in \Tba G$, 
\begin{equation}
\label{eq:x1}
a \sqleq_{\nax\pGR} b \iff a \sqleq_{R} b,
\end{equation}
where $a \sqleq_{R} b$ means that $a \equiv_{R} a \land b$.
From \eqref{eq:x1} and the fact that $f$ is a presentation morphism it is easy
to derive that
\begin{equation}
\label{eq:x2}
a \sqleq_{\nax\pGR} b  \ \mbox{ only if } \ \wh{f}a \sqleq_{\nax\pGRp} \wh{f}b.
\end{equation}

We now turn to the proof of the Proposition proper, which will be based on a 
straightforward induction on the complexity of $\D: c \isleq d$, where $c$ and
$d$ are $\Tnb$-formulas.
We make a case distinction as to the last rule applied in $\D$.

First assume that the last applied rule in $\D$ was $(\nb1)$.
That is, the formulas $c$ and $d$ in $\D: c\isleq d$ are of the form $c = 
\nb\al$ and $d = \nb\be$, for some $\al$ and $\be$ in $\Tom\Tba G$,
respectively, and
we may assume that $\D$ is of the following form:
\[
\D: \AXC{$\{ \D_{ab}: a \isleq b \mid (a,b) \in Z \}$}
\UIC{$\nb\alpha \isleq \nb\beta$}
\DisplayProof 
\]
Here $Z \sse \Base(\al) \times \Base(\be)$ is some set with $(\al,\be) \in
\Tb Z$, and such that for every pair $(a,b) \in Z$, there is a depth zero
derivation $\D_{ab}: a \isleq b$.

Define $Z' := \{ (\wh{f}a,\wh{f}b) \mid (a,b) \in Z \}$, or, equivalently, 
$Z' := \converse{(\wh{f\,})};Z;\wh{f}$.
Then it follows from \eqref{eq:x2} that for each $(a',b') \in Z'$, there is a
derivation $\D^{f}_{a'b'}: a' \isleq b'$.
Using the properties of relation lifting we find that $\Tb Z' = 
\converse{(\T \wh{f}\,)}; \Tb Z; \T \wh{f}$, and from this it is immediate that
$(\T \wh{f} \al,\T \wh{f} \be) \in \Tb Z'$.
Combining these observations, we may transform the derivation $\D$ into
\[
\D^{f}: \AXC{$\{ \D_{a'b'}: a' \isleq b' \mid (a',b') \in Z' \}$}
\UIC{$\nb\T \wh{f}\al \isleq \nb\T \wh{f}\be$}
\DisplayProof 
\]
But then we are done, since $\funpM f (\nb\al) = \nb\T \wh{f}\al$, and 
likewise for $\be$.
\medskip

Second, suppose that the last applied rule in $\D$ was $(\nb2)$.
That is, $\D$ ends with 
\[
\D: \AXC{$\{ \D_{\Phi}: \nb (\T\bw)\Phi \isleq d \mid \Phi \in \SRD(A) \}$}
\UIC{$\bw \{ \nb\al \mid \al \in A \} \isleq d$}
\DisplayProof 
\]
We are to transform $\D$ into a derivation $\D^{f}$ of the inequality
$\bw \{ \nb\al' \mid \al' \in A'\} \isleq \wh{\funpM f} d$, where $A' :=
\{ \T\wh{f} \al \mid \al \in A \}$.
Working towards an application of ($\nb2$), we claim that 
\begin{equation}
\label{eq:trsrd}
\SRD(A') \subseteq \Big\{ \Phi' \in \Tom \left( \bigcup_{\al\in A} \Base ( \T\wh{f} \al) \right)\mid
\exists \Phi \in \SRD(A) \; \mbox{such that }  \T\funP\wh{f}(\Phi) = \Phi' \Big \}.
\end{equation}
To see why this is so, consider an arbitrary slim redistribution $\Phi'$ of $A'$.
First observe that 
\begin{equation}
\label{eq:trsrd2}
\wh{f}[\Base[A]] =
\bigcup_{\al\in A} (\funP \wh{f}) (\Base(\al)) =
\bigcup_{\al\in A} \Base((\T\wh{f})\al) =
\Base[A'],
\end{equation}
where the second identity is by the fact that $\Base: \Tom \ntrto \Pom$ is a
natural transformation (cf.~Fact~\ref{fact:basenatural}).
If we restrict $\wh{f}$ to the set $\Base[A]$, by \eqref{eq:trsrd2} we obtain a
\emph{surjective} map
\[
g: \Base[A] \to \Base[A'].
\]
From the surjectiveness of $g$ it follows that $(\funP g) \circ (\funQ g) =
\id_{\funP\Base[A']}$, and so we also find that $(\T\funP g) \circ (\T\funQ g) = 
\id_{\Tom\funP\Base[A']}$.
Hence if we define 
\[
\Phi := (\T\funQ g)\Phi',
\]
we see that $\Phi' = \T\funP g(\Phi) = \T\funP\wh{f}(\Phi)$.
Therefore, using $\in;\funP f \subseteq
f;\in$, it is easy to see that $\al   (\Tb \in) \Phi$ implies 
$\T f \al (\Tb \in) \Phi'$ for all $\alpha \in \Tom\Tba G$. 
Thus, in order to prove \eqref{eq:trsrd} it suffices to prove that $\Phi$ is a 
slim redistribution of $A$.
To see why this is the case, first observe that by definition of $g$ we have
that $\T\funQ g: \Tom\funP\Base[A'] \to \Tom\funP\Base[A]$, and so we find that 
$\Phi \in \Tom\funP\Base[A]$.
It is left to prove that every element of $A$ is a lifted member of $\Phi$.

Take an arbitrary element $\al\in A$, then $\T g (\al) \in A'$ by definition of
$A'$ and $g$, and so $\T g (\al)$ is a lifted member of $\Phi'$ by the assumption
that $\Phi' \in \SRD(A')$.
This means that $(\al,\Phi) \in (\T g); (\rl{T}{\in});(\T\funQ g)$.
The key observation now is that $(\T g); (\rl{T}{\in});(\T\funQ g) \,\sse\, 
\Tb {\in}$, which is immediate from $g;{\in};(\funQ g) \sse {\in}$ by the 
properties of relation lifting.
Applying this key observation we find that $(\al,\Phi) \in \Tb {\in}$ as
required.
This finishes the proof of \eqref{eq:trsrd}.

Returning to the construction of our derivation $\D^{f}$, consider an arbitrary 
slim redistribution $\Phi'$ of $A'$, which by \eqref{eq:trsrd} we may assume to
be of the form $(\T\funP\wh{f})\Phi$ with $\Phi \in \SRD(A)$.
Applying the inductive hypothesis to the derivation $\D_{\Phi}$ we obtain a 
proof $\D_{\Phi}^{f}: \wh{\funpM f}\nb(\T\bw)(\Phi) \isleq \wh{\funpM f}d$.
However, from $\wh{f}\circ\bw = \bw\circ(\funP \wh{f})$ we obtain that 
\[
\wh{\funpM f}\nb(\T\bwsmall)(\Phi) =
\nb(\T \wh{f})(\T\bwsmall)(\Phi) =
\nb(\T\bwsmall)(\T\funP \wh{f})(\Phi) =
\nb(\T\bwsmall)\Phi'.
\]
In other words, for any $\Phi' \in \SRD(A')$ there is a derivation of the
inequality $\nb(\T\bw)\Phi' \isleq \wh{\funpM f}q$.
Putting all these derivations together, one application of ($\nb 2$) gives the
desired derivation 
\[
\D^{f}: (\funpM \wh{f}) \Big( \bw \{ \nb \al \mid \al \in A \} \Big)
\isleq \wh{\funpM f}d.
\]
\medskip

Now suppose that the last applied rule in $\D$ was $(\nb3)$.
In this case $\D$ has the following shape:
\[
\D: \AXC{$\{ \D_{\al}: \nb\al \isleq q \mid \al \Tb (\in) \Phi \}$}
\UIC{$\nb(\T\bv)\Phi \isleq d$}
\DisplayProof 
\]

In order to see which inequality we need to derive, we first compute 
\[
\wh{\funpM f}(\nb(\T\bvsmall)\Phi) = 
\nb (\T\wh{f})(\T\bvsmall)\Phi =
\nb (\T\bvsmall) (\T\funP\wh{f})\Phi,
\]
where the latter identity follows from the fact that $\wh{f}\circ\bv =
\bv\circ\funP\wh{f}$.
We are looking for a derivation of the inequality $\nb
(\T\bvsmall)(\T\funP\wh{f})\Phi\isleq \wh{\funpM f}d$.
Since we want to apply the rule ($\nb3$), we first compute the set of
lifted members of $(\T\funP\wh{f})\Phi$.
But since ${\in};\converse{(\funP \wh{f\,})} = \converse{\wh{f\,}};{\in}$,
applying relation lifting we obtain ${\Tb \in};\converse{(\T\funP
\wh{f}\,)} = \converse{(\T \wh{f}\,)};
{\Tb {\in}}$.
This immediately shows that 
\[
(\al',(\T\funP\wh{f})\Phi) \in \Tb {\in}
\mbox{ iff } 
\al' = \T\wh{f} \al \mbox{ for some } \al \Tb {\in} \Phi.
\]
By the induction hypothesis, for each $\al \Tb {\in}\Phi$ we have a
derivation $\D_{\al}^{f}: \nb\T\wh{f}\al \isleq \wh{\funpM f}d$.  In
other words, for every lifted member $\al'$ of $(\T\funP\wh{f})\Phi$,
there is a derivation $\D_{\al'}: \nb\al' \isleq \wh{\funpM f}d$.  But
then by one application of ($\nb3$) we are done:
\[
\D^{f}: \AXC{$\{ \D_{\al'}: \nb\al' \isleq q \mid \al \Tb (\in) (\T\funP\wh{f})\Phi \}$}
\UIC{$\wh{\funpM f}(\nb(\T\bvsmall)\Phi) \isleq \wh{\funpM f} d$}
\DisplayProof 
\]

Finally, the cases where the last applied rule in $\D$ was a propositional one, 
are left as exercises to the reader.
\end{proof}

Given Proposition~\ref{p:prm} it is not difficult to prove that $\funpM$ is 
a functor.

\begin{thm}
\label{t:Mfun}
$\funpM: \Prs \to \Prs$ is a functor.
In addition, $\funpM$ maps pre-isomorphisms to pre-isomorphisms.
\end{thm}

\begin{proof}
Since it is not difficult to verify that $\funpM$ preserves identity arrows
and distributes over composition, we confine our attention to the proof that 
$\funpM$ maps presentation morphisms to presentation morphisms.

Let $f: \pGR \to \pGRp$ be a presentation morphism, and let $c,d \in 
\Tba(\Tomnb\Tba G) = \Tnb G$ be such that $c \equiv_{\nax\pGR} d$,
that is, there are $\nax\pGR$-derivations $\D_{1}: c \isleq d$ and $\D_{2}: d 
\isleq c$.
Proposition~\ref{p:prm} provides us with $\nax\pGRp$-derivations $\D_{1}^{f}: 
\wh{\funpM f} c \isleq \wh{\funpM f}d$ and $\D_{2}^{f}: \wh{\funpM f}d
\isleq \wh{\funpM f}c$.
This means that we have $\wh{\funpM f}c \equiv_{\nax\pGRp} \wh{\funpM
f}d$, as is required for $\funpM f$ to be a presentation morphism.

In order to prove that $\funpM$ maps pre-isomorphisms to pre-isomorphisms, a
routine proof will show that $\funpM$ preserves pre-inverses.
\end{proof}

\subsection{The functor $\funaM$ and its algebras}
\label{ss:functorM}

Given the intimate relation between Boolean algebras and their 
presentations, it should come as no suprise that the presentation functor
$\funpM$ naturally induces a functor on the category of Boolean algebras.

\begin{definition}
\label{d:funaM}
The functor $\funaM: \BA \to \BA$ is defined as $\funaM := 
\funB\circ\funpM\circ\funC$.
\end{definition}

To explain this functor in words, first consider the objects.
Given a Boolean algebra $\bbA$ with carrier $A := \funU\bbA$, the elements
of $\funaM\bbA$ are the equivalence classes of the form 
$[a]_{\nax\funC\bbA}$, where $a\in\Tnb A$ is a depth-one term over the carrier
of $\bbA$, and the equivalence relation $\equiv_{\nax\funC\bbA}$ is the
interderivability relation in the one-step derivation system $\nax\funC\bbA$
which takes, as its additional axioms, the \emph{diagram} $\Delta_{\bbA}$ of
$\bbA$ (listing the `true facts' about $\bbA$).
Summarizing, we find that 
\[
\funU\funaM\bbA = \Tnb A /{\equiv_{\nax\funC\bbA}}.
\]

In order to explain the action of $\funaM$ on a homomorphism $f: \bbA \to
\bbA'$, the upshot of Theorem~\ref{t:Mfun} is that the map
\begin{equation}
\label{eq:Mfun}
\funaM f: [a]_{\nax\funC\bbA} \mapsto [\Tnb \funU f (a)]_{\nax\funC\bbA'},
\end{equation}
correctly defines a homomorphism $\funaM f: \funaM\bbA \to \funaM\bbA'$.
Here $\Tnb$ is given in Definition~\ref{d:Tnb}, and the observation
\eqref{eq:Mfun} is a direct consequence of the definitions and of the 
following proposition.

\begin{prop}
\label{p:funpM-Tnb}
Let $f: \pGR \to \pGRp$ be a presentation morphism. 
If $f$ maps generators to generators (in the sense that $f[G] \sse G'$),
then 
\[
\wh{\funpM f} = \Tnb f.
\]
\end{prop}

\begin{proof}
Suppose that $f: \pGR \to \pGRp$ maps generators to generators, then it
is immediate that $\wh{f} = \Tba f$.
From this it follows that $\funpM f = \Tomnb \wh{f} = \Tomnb \Tba f$,
and since $\funpM f$ also maps generators to generators, we find that 
$\wh{\funpM f} = \Tba\funpM f = \Tba\Tomnb \Tba f = \Tnb f$.
\end{proof}

For future reference we mention the following.

\begin{definition}
\label{d:quot}
Given algebra $\bbB$, we shall denote with $\quot{\bbB}: \Tnb(\funU\bbB) \to 
\funaM\bbB$ the map
\[
\quot{\bbB}: b \mapsto [b]_{\nax\funC\bbB},
\]
that is, $\quot{\bbB} = \ti{\unitBC}_{\funpM\funC\bbB}$
is the quotient map sending a formula $b$ to its equivalence class under
$\equiv_{\nax\funC\bbB}$.
\end{definition}

\begin{prop}
\label{p:q}
The family of homomorphisms $\quot{\bbB}$, with $\bbB$ ranging over the class 
of Boolean algebras, provides a natural transformation $\quot{}: \Tnb\funU
\ntrto
\funaM$.
\end{prop}

\begin{proof}
Let $f: \bbB \to \bbB'$ be some Boolean homomorphism.
In order to prove that $\quot{}$ is a natural transformation, we need to show
that the diagram below commutes:
\begin{equation}
\label{dg:qntr}
\xymatrix{
\bbB \ar[d]_{f}
& \Tnb\funU\bbB  \ar[d]_{\Tnb\funU f}   \ar[r]^{\quot{\bbB}} 
& \funaM\bbB     \ar[d]_{\funaM f} 
\\ \bbB'&
  \Tnb\funU\bbB' \ar[r]^{\quot{\bbB'}} 
& \funaM\bbB'  
}
\end{equation}
This follows from a straightforward unfolding of the definitions: 
For any $b \in \Tnb\funU\bbB$ we have
\[
(\funaM f \circ \quot{\bbB})(b)
= \funaM f ([b]_{\nax\funC\bbB})
= [\Tnb\funU f(b)]_{\nax\funC\bbB'}
= \quot{\bbB'}(\Tnb\funU f(b))
= (\quot{\bbB'}\circ\Tnb\funU f)(b).
\]
Here the second step is by \eqref{eq:Mfun} above.
\end{proof}

It turns out that $\funaM$ has some nice properties that will be of use
later on.
In particular, we may show that $\funaM$ is \emph{finitary} and 
\emph{preserves embeddings}.
Intuitively, being finitary means proof-theoretically, that for any Boolean
algebra $\bbA$, a derivation of $\vdash_{\funaM(\bbA)}a_{1}\isleq a_{2}$ can
be carried out in a \emph{finite} subalgebra of $\bbA$. 
(Note that this is not obvious since we may be dealing with an infinitary proof
system.)
Formally, we need the following definition, referring to~\cite{adam94:loca}
for more details.
Recall that a partial order is \emph{directed} if any finite set of elements
has an upper bound.

\begin{definition}
  Given a category $\class{C}$, a \emph{directed diagram} over
  $\class{C}$ is a diagram which is indexed by a directed partial
  order.  An endofunctor on $\class{C}$ is \emph{finitary} if it
  preserves colimits of directed diagrams.
\end{definition}

\noindent In case of an endofunctor on $\Set$ this definition is
equivalent to the one of Section~\ref{s:preliminaries}.

\begin{exa}
\label{ex:subalg-colimit}
Given a Boolean algebra $\bbB$, let $\struc{\mathit{Sub}(\bbB),\sse}$ be the 
set of finite subalgebras of $\bbB$, ordered by inclusion. 
We can turn this poset into a diagram $S_{\bbB}$ by supplying, for each pair
of finite subalgebras $\bbB'$ and $\bbB''$ such that $\bbB' \sse \bbB''$,
the (unique) inclusion $\iota_{\bbB'\bbB''}$.
Since the variety $\BA$ is \emph{locally finite}, which means that every
finitely generated Boolean algebra is finite, one may easily see that every
Boolean algebra $\bbB$ is the directed colimit of its associated diagram
$S_{\bbB}$.
\end{exa}

In fact, it is a routine exercise to verify that for an endofunctor on the
category on Boolean algebras to be finitary, it suffices to preserve the
directed colimits of the subalgebra diagrams described in
Example~\ref{ex:subalg-colimit}.

\begin{prop}
\label{p:Mfinemb}
$\funaM$ is a finitary functor that preserves embeddings.
\end{prop}

\begin{proof}
Fix a Boolean algebra $\bbA$ with carrier set $A := U\bbA$.
Given two elements $a_{1},a_{2} \in \Tnb A$, consider the collection of elements
of $A$ that occur as \emph{subformulas} of $a_{1}$ and $a_{2}$.
It follows from our earlier remarks on subformulas that this is a \emph{finite}
set, which then generates a finite subalgebra $\bbA'$ of $\bbA$.
By definition we have $a_{1},a_{2} \in \Tnb A'$, where we define $A'
:= \funU\bbA'$.

We claim that 
\begin{equation}
\label{eq:2-4}
\vdash_{\nax\funC\bbA} a_{1} \isleq a_{2} \ \mbox{ iff } \ 
\vdash_{\nax\funC\bbA'} a_{1} \isleq a_{2}.
\end{equation}
The interesting direction of (\ref{eq:2-4}) is from left to right.
The key observation here is that from the fact that $\bbA'$ is a finite
subalgebra of $\bbA$, we may infer the existence of a \emph{surjective}
homomorphism $f: \bbA \to \bbA'$ such that $f(a') = a'$ for all $a'\in A'$.
(In other words, $\bbA'$ is a \emph{retract} of $\bbA$.)
There are various ways to prove this statement; here we refer to Sikorski's
theorem that complete Boolean algebras are injective~\cite{siko:theo48}.
But if $f$ is a homomorphism, by Proposition~\ref{p:prm} it follows
from $\vdash_{\nax\funC\bbA} a_{1} \isleq a_{2}$ that $\vdash_{\nax\funC\bbA'}
\wh{\funpM f}(a_{1}) \isleq \wh{\funpM f}(a_{2})$.
Since $f$ restricts to the identity on $A'$, so does $\wh{\funpM f} = \Tnb f$
on $\Tnb A'$.
As a direct consequence we find that $\wh{\funpM f}(a_{i}) = a_{i}$, for both 
$i = 1,2$.
Thus, indeed, $\vdash_{\nax\funC\bbA'} a_{1} \isleq a_{2}$, which proves 
\eqref{eq:2-4}.

It is now easy to see that $\funaM$ is a finitary functor. 
As mentioned above, it suffices to show that $\funaM\bbA$ is a directed
colimit of the image $\funaM S_{\bbA}$ under $\funaM$ of the subalgebra
diagram $S_{\bbA}$ of $\bbA$ (see Example~\ref{ex:subalg-colimit}).
Given a finite subalgebra $\bbB$ of $\bbA$, let $e_{\bbB}$ denote the 
inclusion homomorphism, $e_{\bbB}: \bbB \hookrightarrow \bbA$.
We claim that 
\begin{equation}
\label{eq:Mfin}
\struc{\funaM\bbA,\funaM e_{\bbB}}_{\bbB\in S_{\bbA}} \text{ is a colimit
of } \funaM S_{\bbA}.
\end{equation}
Since for every pair $\bbB,\bbB'$ such that $\bbB \hookrightarrow \bbB'
\hookrightarrow \bbA$, we have $e_{\bbB} = e_{\bbB'} \cof \iota_{\bbB\bbB'}$,
it is obvious from the functoriality of $\funaM$ that 
$\struc{\funaM\bbA,\funaM e_{\bbB}}_{\bbB\in S_{\bbA}}$ is a cocone over
$\funaM S_{\bbA}$.
To see why it is in fact a colimit, suppose that 
$\struc{\bbD,d_{\bbB}}_{\bbB\in S_{\bbA}}$ is another cocone over $\funaM
S_{\bbA}$, and take an arbitrary element of $\funaM\bbA$.
By definition, this element is of the form $[a]_{\nax\funC\bbA}$ for some 
formula $a \in \Tnb A$.
Let, as above, $\bbA'$ be a finite subalgebra of $\bbA$ such that $a \in
\Tnb A'$, then it follows from \eqref{eq:2-4} that the following provides a 
well-defined homomorphism $d: \funaM\bbA \to \bbD$:
\[
d([a]_{\nax\funC\bbA}) \isdef d_{\bbA'}([a]_{\nax\funC\bbA'}).
\]
We leave it as an exercise for the reader to verify that $d$ is the unique
homomorphism $d: \funaM\bbA \to \bbD$ such that for all $\bbB \hookrightarrow
\bbA$, the following diagram commutes:
\[
\xymatrix{ 
    \funaM\bbB \ar[r]^{\funaM e_{\bbB'}} \ar[rd]_{d_{\bbB'}}
  & \funaM\bbA \ar[d]^{d}
\\& \bbD
}\]
This proves \eqref{eq:Mfin}, and as mentioned this suffices to establish
that $\funaM$ is finitary.

For the second part of the Proposition, let $e: \bbA \to \bbB$ be an
embedding.
Without loss of generality we will assume that $e$ is actually an inclusion
(that is, $\bbA$ is a subalgebra of $\bbB$).
In order to prove that $\funaM e: \funaM\bbA \to \funaM\bbB$ is also injective, it
suffices to prove the following, for all $a_{1},a_{2} \in A$:
\begin{equation}
\label{eq:2-5}
\vdash_{\nax\funC\bbB} a_{1} \isleq a_{2} \ \mbox{ implies } \  
\vdash_{\nax\funC\bbA} a_{1} \isleq a_{2}.
\end{equation}
But the proof of (\ref{eq:2-5}) simply follows from two applications
of 
(\ref{eq:2-4}).
\end{proof}

In the sequel we will be interested in algebras for the functor $\funaM$.
Recall that these are pairs of the form $\struc{\bbA,f}$, where $\bbA$ is
some Boolean algebra, and $f$ is a homomorphism from $\funaM\bbA$ to
$\bbA$.
First of all, we will see that such $\funaM$-algebras are Moss algebras in
disguise.

\begin{definition}
\label{d:funV}
Given an $\funaM$-algebra $\struc{\bbA,f}$, we let
$\funV\struc{\bbA,f}$ denote the Moss algebra
\[
\funV\struc{\bbA,f} \isdef
\struc{\funU\bbA,\neg^{\bbA},\bvsmall^{\bbA},\
\bwsmall^{\bbA},\nb^{\funV\struc{\bbA,f}}}.
\]
Here we define $\nb^{\funV\struc{\bbA,f}}: \Tom\funU\bbA \to \funU\bbA$ by
recalling that $\Tom\funU\bbA$ is a subset of $\Tnb\funU\bbA$, and putting
\[
\nb^{\funV\struc{\bbA,f}}\al \isdef f(\quot{\bbA}(\nb\al)),
\]
where $\quot{\bbA}$ is as in Definition~\ref{d:quot}.
In addition, given an $\funaM$-morphism $g: \struc{\bbA,f} \to 
\struc{\bbA',f}$, we define $\funV g$ to be the morphism $\funV g: \funV\bbA 
\to \funV\bbA'$ given by
\[
\funV g \isdef \funU g.
\]
That is, as a map, $\funV g$ is simply the same as $g$.
\end{definition}

We leave it for the reader to verify that with this definition, $\funV$
actually defines a functor transforming $\funaM$-algebras into Moss algebras.

\begin{prop}
The operation $\funV$ defines a functor
\[
\funV: \Alg_{\BA}(\funaM) \to \Alg_{\Set}(\Mossfun).
\]
\end{prop}

Because $\funaM$ is a finitary functor we can define the initial 
$\funaM$-algebra to be the colimit of the first $\omega$ steps of the initial
sequence of $\funaM$.

\begin{definition}
\label{d:initial_sequence}
The \emph{initial sequence}
\begin{equation}
\label{dg:initseq}
\xymatrix{
  \bbtwo             \ar@{->}[r]^{j_{0}} 
& \funaM\bbtwo       \ar@{->}[r]^{j_{1}} 
& \funaM^{2}\bbtwo   \ar@{->}[r]^{j_{2}}
& \ldots 
& \funaM^{k}\bbtwo   \ar@{->}[r]^{j_{k}}
& \funaM^{k+1}\bbtwo \ar@{->}[r]^{j_{k+1}}
& \ldots
}
\end{equation}
results from starting with $j_0$ as the 
unique homomorphism from $\bbtwo$ to $\funaM \bbtwo$, and defining $j_{k+1}
\isdef \funaM j_k$, for all $k \in \bbN$.
We let $\funaM^{\om}\bbtwo$, with the collection of maps $(i_{k}: 
\funaM^{k}\bbtwo \to \funaM^{\om}\bbtwo)_{k\in\om}$, denote the colimit of
this sequence.
\end{definition}

In the following Proposition we gather some facts about these structures.

\begin{prop}
\label{p:minit}\hfill
\begin{enumerate}[\em(1)]
\item 
For each $k \in \om$, the map $j_{k}:\funaM^{k}\bbtwo \to \funaM^{k+1}\bbtwo$
is an embedding, and so is the map $i_{k}: \funaM^{k}\bbtwo \to 
\funaM^{\om}\bbtwo$.
\item
There is a map $j_{\om}: \funaM^{\om}\bbtwo \to \funaM^{\om+1}\bbtwo$ such
that the following diagram commutes, for every $k\in\om$:
\[ \xymatrix{
\funaM^{k}\bbtwo  \ar[d]_{i_{k}} \ar[r]^{j_{k}} 
   & \funaM^{k+1}\bbtwo \ar[d]^{\funaM i_{k}} 
\\
\funaM^{\om}\bbtwo \ar[r]_{j_{\om}} 
   & \funaM^{\om+1}\bbtwo
} \]
\item
The map $j_{\om}$ has an inverse $\hs^{\Minit}: \funaM^{\om+1}\bbtwo \to 
\funaM^{\om} \bbtwo$.
\item
The structure $\struc{\funaM^{\om} \bbtwo, \hs^{\Minit}}$ is an initial 
$\funaM$-algebra.
\item
For all $k\in\om$ we have that $i_{k+1} = \hs^{\Minit}\cof\funaM i_{k}$.
\end{enumerate}
\end{prop}

\begin{proof}
Part~1 is immediate by Proposition~\ref{p:Mfinemb} and basic category 
theory.
Part~2 follows from $\funaM^{\om} \bbtwo$ being a colimit of the initial 
sequence~\eqref{dg:initseq}.
The inverse of $j_{\om}$, mentioned in part~3, exists by the facts that the 
initial sequence is a chain, and hence directed, and that $\T$ preserves
directed colimits.

For part~4, consider an arbitrary $\funaM$-algebra $\xymatrix{\bbA & 
\funaM\bbA \ar[l]_{\al}}$, and define the co-cone
$\struc{\bbA,\al_{k}: \bbM^{k}\bbtwo \to \bbA}$ as follows: $\al_{0}: 
\bbtwo \to \bbA$ is given by initiality, and for $k\in\om$ we put
$\al_{k+1} \isdef \la\cof\al_{k}$.
Then by $\funaM^{\om}\bbtwo$ being the colimit of the initial sequence, 
there is a unique map $\al_{\om}:  \bbM^{\om}\bbtwo \to \bbA$ such that 
$\al_{k} = \al_{\om}\cof i_{k}$, for all $k\in\om$.
Now consider the following diagram:
\begin{equation}
\label{dg:7-1}
\xymatrix{
\funaM^{\om}\bbtwo  \ar[d]_{\al_{\om}} \ar[r]^{j_{\om}} 
   & \funaM^{\om+1}\bbtwo \ar[d]^{\funaM\al_{\om}} 
\\
\bbA & 
\funaM\bbA \ar[l]_{\al}
} \end{equation}
This diagram commutes by $\funaM^{\om}\bbtwo$ being the colimit of the
initial sequence.
Finally, consider the map $\hs^{\Minit}$ of part~3.
Then 
\begin{align*}
\al^{\om} \cof \hs^{\Minit} 
   &= (\al\cof\funaM\al_{\om}\cof j_{\om}) \cof\hs^{\Minit}
   & \text{(diagram \eqref{dg:7-1} commutes)}
\\ &= \al\cof \al_{\om}
   & \text{($j_{\om}$ and $\hs^{\Minit}$ are converses)}
\end{align*}
and from this part~4 is immediate.

Finally, for part~5, fix $k \in \om$.
By definition, $\struc{\Minit,i_{n}}_{n\in\om}$ is a co-cone of the initial
sequence, and so we have $i_{k} = j_{k} \cof i_{k+1}$.
From this it follows by (the diagram of) part~2 of this Proposition that 
$j_{\om}\cof i_{k+1} = \funaM i_{k}$, and from this we easily derive by 
part~3 that $i_{k+1} = j_{\om}^{-1}\cof \funaM i_{k} =
\hs^{\Minit}\cof \funaM i_{k}$.
\end{proof}

The above Proposition justifies the following Definition.

\begin{definition}
We let $\Minit$ denote the $\funaM$-algebra $\struc{\funaM^{\om} \bbtwo,
\hs^{\Minit}}$, and we will refer to this structure as the \emph{initial
$\bbM$-algebra}.
\end{definition}

\begin{rem}
\label{r:VMinit}
In the sequel, we will be interested in the Moss algebra $\funV\Minit$.
Observe that the nabla operation $\nb^{\funV\Minit}$ of this structure is
defined as $\nb^{\funV\Minit}(\al) =
\hs^{\Minit}(\quot{\funaM^{\om}\bbtwo}(\nb\al))$, and so by definition of
$\hs^{\Minit}$ we find that
\[
\nb^{\funV\Minit}(\al) = j_{\om}^{-1}(\quot{\funaM^{\om}\bbtwo}(\nb\al)).
\]
\end{rem}

\subsection{Proof of One-Step Soundness}

In this subsection we will establish one-step soundness of the one-step 
derivation system; that is, we prove the direction from right to left of
Theorem~\ref{t:sc1}.

\begin{prop}
\label{p:1sts}
For any set $X$, and for any pair of formulas $c,d \in \Tnb\funP X$ we have
\begin{equation}
\label{eq:1sts}
c \eqsem d \ \mbox{ if }  \ c   \equiv_{\nax\funC\funaP X}  d.
\end{equation}
\end{prop}

\begin{proof}
We argue by induction on derivations, so that clearly it suffices to show
that each of the rules $(\nabla 1) -  (\nabla 3)$ is sound.
Fix a set $X$.

\medskip\noindent \emph{Case $(\nabla 1)$}.
Let ${\sqsubseteq_{0}} \sse \Tba\funQ X \times \Tba\funQ X$ be
the relation of `provable inequality' $a \sqsubseteq_{0} b$ if the inequality 
$a \isleq b$ is derivable.
It is straightforward to see that for all $a,b \in \Tba\funQ X$, it follows
from $a \sqsubseteq b$ that $\semzero{a} \sse \semzero{b}$.
(This boils down to showing that our Boolean axioms of Table~\ref{tb:clax} 
are sound.)
Hence it remains to show that for all $\al,\be \in \Tnb\funQ X$, we have
\begin{equation}
\label{eq:s-nb1}
\text{ if } \al \rel{\Tb Z} \be \text{ for some } Z \sse  {\sqsubseteq_{0}},
\text{ then } \semone{\nb\al} \sse \semone{\nb\be}.
\end{equation}
For this purpose, assume that $\al \rel{\Tb Z} \be$ for some $Z \sse  
{\sqsubseteq_{0}}$, and take an arbitrary element $\xi \in \T X$ such that
$\T X,\xi \forces_{1} \nb\al$.
Then by definition of $\forces_{1}$, we have $\xi \rel{\Tb{\forces_{0}}}\al$,
so that by the properties of relation lifting we obtain that $\xi 
\rel{\Tb({\forces_{0}}\corel Z)} \be$.
However, it is straightforward to verify that ${\forces_{0}}\corel Z 
\sse {\forces_{0}}\corel {\sqsubseteq_{0}} \sse {\forces_{0}}$, and so we 
obtain that $\xi \rel{\Tb\forces_{0}} \beta$.
From this it is immediate that $\T X, \xi \forces_{1} \nb\be$.

\medskip\noindent 
\emph{Case $(\nabla 2)$. } 
Given a set $A\subseteq\Tom\Tba(\funP X)$ and an element $\xi \in \T X$,
assume that $\T X, \xi \forces_{1} \nb\al$ for each $\al \in A$.
We need to prove that $\T X, \xi \forces_{1} \nb (\T\bw)\Phi$ for some
$\Phi \in \SRD(A)$.
To come up with a suitable $\Phi$, let $B \isdef \bigcup \Base[A]$ and
consider the map $\phi: X \to \Pom B$ given by 
\[
\phi: x \mapsto \{ b \in B \mid X,x \forces_{0} b \}.
\]
We claim that the set 
\[
\Phi \isdef (\T\phi)(\xi)
\]
fulfills our requirements.

First of all, in order to prove that $\T X,\xi \forces \nb(\T\bw)(\Phi)$, 
observe that by definition of $\phi$, we have $\phi\corel\bw \sse {\forces_{0}}$.
Hence by the properties of relation lifting, it follows that
${\T\phi}\corel {\T{\bw}} \sse \Tb{\forces_{0}}$.
In particular, we find that $(\xi,(\T{\bw})(\Phi)) \in \Tb{\forces_{0}}$;
but then it is immediate from the definitions that  $\T X,\xi \forces
\nb(\T\bw)(\Phi)$.

Second, by definition we have $\Phi \in \Tom \Pom B$ and so, in order to show
that $\Phi\in \SRD(A)$, it suffices to prove that $\al\in \nbsem_{\Pow X}(\Phi)$
for all $\alpha \in A$.
For this purpose, observe that $\phi\corel{\cv{{\in}}} =
{\forces_{0}}\rst{X\times B}$.
Then by the properties of relation lifting we obtain 
$\T\phi\corel{\cv{(\Tb{\in}})} = \Tb{\forces_{0}}\rst{\T X\times \T B}$.
In particular, since $\xi \rel{\Tb{\forces_{0}}\rst{\T X\times \T B}} \al$
by assumption, it follows that $\al \Tin \T\phi(\xi) = \Phi$, as required.
\medskip

\noindent 
\emph{Case $(\nabla 3)$. } 
We could prove the soundness of $(\nb3)$ analogously to our proof of
Proposition~\ref{p:nb3s}, but we prefer to give a different proof here, 
stressing the role of the distributive of $\nbsem$ over the power set 
\emph{monad}, cf.~Fact~\ref{fact:distriblaw}.

Fix an element $\Phi\in\Tom\Pom\Tba\funP X$.
Given Remark~\ref{r:onestep2}, it suffices to show that 
\begin{equation}\label{equ:almostnabla3}
\semone{\nabla(T\bvsmall)(\Phi)} =
\bigcup\{\nbsem_X (\T \semzero{\cdot}(\alpha)) \mid \alpha \Tin \Phi \}.
\end{equation}
The point is now that \eqref{equ:almostnabla3} can be read off the following
diagram, where we tacitly use the fact that $\nbsem$ restricts to a 
natural transformation $\nbsem_{X}: \Tom\Pom \to \funP\Tom$ (see
Proposition~\ref{p:nbsem}).

\begin{equation}
\label{dg:nb3}
\xymatrix{ 
    \Tom\Pom\Tba\funP X  \ar[rr]^{\nbsem_{\Tba\funP X}}  \ar[d]_{\Tom\bv}
                         \ar[dr]^{\Tom\Pom(\semzero{\cdot})}
  & & \funP\Tom\Tba\funP X \ar[d]^{\funP\T\semzero{\cdot}}
\\  \Tom\Tba\funP X      \ar[d]_{\Tom\semzero{\cdot}}
  & \Tom\Pom\funP X      \ar[dl]^{\Tom\bigcup}    \ar[r]^{\nbsem_{\funP X}}
  & \funP\Tom\funP X     \ar[dd]^{\funP\nbsem_X}
\\  \Tom\funP X          \ar[d]_{\nbsem_X} 
  &&
\\   \funP\Tom X
  && \funP\funP \Tom X   \ar[ll]_{\bigcup_{TX}} }
\end{equation}
To see this, first observe that the left hand side of \eqref{equ:almostnabla3}
corresponds to the left edge of the diagram, where an arbitrary element $\Phi
\in \Tom\Pom\Tba\funP X$ is mapped to 
$$
\nbsem_X\left( \Tom \semzero{\cdot} (\Tom \bvsmall (\Phi))\right) =
\semone{\nabla(T\bvsmall)(\Phi)}.
$$ 
Similarly, the right hand side of \eqref{equ:almostnabla3} corresponds to
clockwise following $\Phi \in \Tom\Pom\Tba\funP X$ along the outer edges of
the diagram, from the upper left to the lower left corner, arriving at the
object $\bigcup\{\nbsem_X (\T \semzero{\cdot} (\alpha)) \mid \al \Tin \Phi \}$.

Therefore in order to show (\ref{equ:almostnabla3}) it suffices to show that 
the diagram commutes.
But this is fairly straightforward.
First observe that 
\begin{equation}
\label{equ:d}
\semzero{\cdot} \cof \bv = \bigcup \circ \Pom \semzero{\cdot}, 
\end{equation}
as a straightforward verification will reveal.
After applying the functor $\Tom$ to \eqref{equ:d}, we immediately obtain that
the left quadrangle of \eqref{dg:nb3} commutes.
The right-hand quadrangle commutes since $\nbsem$ is natural.
And finally, the pentagon commutes since $\nbsem$ is a distributive law over 
the power set monad, see Fact~\ref{fact:distriblaw}.
As a consequence, the diagram \eqref{dg:nb3} itself commutes.
\end{proof}

\subsection{Proof of One-Step Completeness}

We now turn to the one-step completeness of our derivation system.
Our proof is based on properties of algebras of the form $\funaM\bbB$,
with $\bbB$ an arbitrary finite Boolean algebra.  With $\At\bbB$
denoting the set of atoms of $\bbB$, we can formulate our key insight
by stating that the Boolean algebra $\funaM\bbB$ is join-generated by
its `lifted atoms', that is, its elements of the form $[\nb\al]$ with
$\al \in \T (\At\bbB)$.  That is to say, we can prove that every
element $x$ of $\funaM\bbB$ is the join of the elements in $\T
(\At\bbB)$ below it:
\begin{equation*}
x = \bv \Big\{ [\nb\al] \mid \al\in\Tom(\At\bbB), [\nb\al] \leq x \Big\}.
\end{equation*}

Here, as elsewhere in this subsection, the join is taken in the
algebra $\funaM\bbB$, and may be happen to be taken over an infinite
set; in that case, the statement should be read as saying that `the
join on the righthandside exists, and it is equal to the
lefthandside'.  As we will see, in the case that the functor does not
preserve finite sets, this is a convenient way of treating infinitary
rules as identities.

Arriving at the proof details, in order to establish the one-step completeness
of $\nax$, we need to prove the direction from left to right of \eqref{eq:1stsc}.
We will reason by contraposition, showing that for arbitrary $a',b' \in 
\Tnb\funQ X$:
\[
a' \not\equiv_{\funpM\funC\funaP X} b' \mbox{ implies }
\semone{a'} \neq \semone{b'}.
\]
Given the fact that our logic extends classical propositional logic, we may 
confine ourselves to the case where $b'= \bot$.

Fix an element $a' \in \Tnb\funQ X$, and assume that $a'$ is one-step
consistent: $a' \not\eqax \bot$, or, equivalently, $[a'] > \bot^{\funaM\bbB}$.
We will prove that $a'$ is one-step satisfiable: $\semone{a'} \neq \nada$.
Let $\{ \al_{1},\ldots, \al_{n} \}$ be the (finite!) set of elements  
$\al \in \Tom\funQ X$ such that $\nb\al$ occurs in $a'$, and 
define 
\[
\Base(a') := \bigcup_{1 \leq i \leq n} \Base(\al_{i}).
\]
This is a finite subset of $\Tba\funQ X$, that is, a finite set of Boolean
formulas in which the subsets of $X$ are the formal generators.
Let $D \sse_{\om} \funP X$ be the collection of those subsets of $X$ that 
actually occur (as a formal object) in one of the formulas in $\Base(a')$, and
let $\bbB$ be the subalgebra of $\funaQ X$ that is generated by $D$.
Then both $D$ and $\bbB$ are finite (whereas their elements may themselves be
infinite subsets of $X$).
The point is that $\bbB$ is a finite subalgebra of $\funaQ X$ such that $a' 
\in \Tnb(\funU\bbB)$.

It follows by the key lemma in the one-step completeness proof,
Theorem~\ref{l:1st2} below, that 
\begin{equation}
\label{eq:1stc1}
[a'] = \bv{}^{\funaM\bbB} \{ [\nb\al] \mid 
   \al \in \Tom (\At\bbB), \nb\al \sqleq a' \}.
\end{equation}

But since $a'$ is consistent, we have that $[a'] > \bot$, and so we may conclude 
that there actually exists an $\al \in \Tom \At\bbB$ such that 
$\nb\al \sqleq a'$ --- if there were no such $\al$, then the righthandside of
(\ref{eq:1stc1}) would evaluate to $\bot$.
By Proposition~\ref{p:1st1} we obtain for this $\al$ that $\semone{\nb\al} \neq 
\nada$, and so by soundness we may conclude that $\semone{a'} \supseteq 
\semone{\nb\al} \neq \nada$.
In other words, we find that $\semone{a'}$ is one-step satisfiable, as
required.

\begin{prop}
\label{p:1st1}
Fix a set $X$ and let $\al\in  \Tom(\At\bbB)$ for some finite subalgebra
$\bbB$ of $\funaP X$.
Then $\semone{\nb\al} \neq \nada$.
\end{prop}

\begin{proof}
Clearly the set $\At\bbB \sse \funP X$ forms a partition of $X$.
Let $h: \At\bbB \to X$ be a choice function, that is, $h(a)\in a$ for each
$a \in \At\bbB$.
Using the properties of relation lifting, it is not hard to derive from this
that $(Th)(\al) \Tb (\in_{X}) \al$ for each lifted atom $\al$.
It follows immediately that $(Th)(\al) \in \semone{\nb\al}$.
\end{proof}

The following is the key lemma in the one-step completeness proof.

\begin{thm}
\label{l:1st2}
Let $\bbB$ be a finite Boolean algebra.
\begin{enumerate}[\em(1)]
\item
For any two elements $\al,\be \in \Tom(\At\bbB)$, we have
\begin{equation}
\label{eq:atoms1}
[\nb\al] \land [\nb\be] > \bot \mbox{ iff } \al = \be.
\end{equation}
\item
The top element of $\funaM\bbB$ satisfies
\begin{equation}
\label{eq:atoms2}
\top^{\funaM\bbB} = \bv \{ [\nb\al] \mid \al\in\Tom(\At\bbB) \}.
\end{equation}
\item
The set $\{ [\nb\al] \mid \al \in \Tom(\At\bbB) \}$ join-generates $\funaM\bbB$;
that is, for all $a' \in \Tnb\funU\bbB$:
\begin{equation}
\label{eq:5:4:0}
[a'] = \bv \{ [\nb\al] \mid \al\in\Tom(\At\bbB), [\nb\al] \leq [a'] \}.
\end{equation}
\end{enumerate}
Summarizing, the algebra $\funaM\bbB$ is atomic, with $\At(\funaM\bbB) =
\{ [\nb\al] \mid \al \in \Tom(\At\bbB) \}$.
\end{thm}

\begin{proof}
Throughout the proof we will abbreviate $A := \At\bbB$ and $B := \funU\bbB$.

The proof of first two statements is immediate by Proposition~\ref{p:negder}
(take for $\phi$ the set $A$).
Concerning the third statement of the Theorem, observe that the inequality
`$\geq$' of (\ref{eq:5:4:0}) always holds, so it will be the opposite inequality
that we need to establish.
Our proof will be by induction on the complexity of $a'$ (as a boolean formula
over the set $\Tomnb\Tba B$).
\smallskip

In the base case of the induction, $a'$ is of the form $\nb\be$, with $\be\in 
\Tom\Tba B$.
Our first claim is that without loss of generality, we may assume that 
$\nb\be$ actually belongs to $\Tom B$.
The justification for this claim is that for any $b \in \Tba B$ there is a
$b_{0}\in B$ such that the equation $b_{0} \is b$ is derivable in the proof
system $\funC \bbB$ associated with the canonical presentation of $\bbB$: 
simply let $b_{0} := \ti{\id}_{B}(b)$ be the element of $B$ to which the term
$b$ evaluates.
(For the definition of $\ti{\id}_{B}$ we refer to \ref{d:ind_extension}.)
Thus an application of ($\nb1$) shows that for any $\be \in \Tom\Tba B$ there
is a $\beta_{0} \in \Tom B$ such that $\vdash_{\nax\funC\bbB} \nb\be \is 
\nb\be_{0}$: simply take $\beta_{0} := \T\ti{\id}_{B}(\beta)$.

Hence, assume that indeed, $\be \in \Tom B$.  Think of the finitary
join as a map $\bv: \Pom A \to B$.  As such it is a bijection, and
this property is inherited by the map $\T\bv: \Tom\Pom A \to \Tom B$.
Furthermore, it is easy to verify that for any $\phi \in \Pom A$ and
any $a \in A$, we have that
\begin{equation}
\label{eq:5:4:1}
a \in_{X} \phi  \ \yiff \ a \leq \bv\phi,
\end{equation}
which can be succinctly formulated as $\ni_{X} = \bv;{\geq}$ (where $\bv$ now
denotes the graph of the disjunction function).
By the properties of relation lifting, this implies $\Tb (\ni_{X}) =
\T\bv;\Tb {\geq}$, which can again be reformulated as stating that for
any $\Phi \in \Tom\funP A$ and any $\al \in \Tom A$ it holds that
\begin{equation}
\label{eq:5:4:2}
\al \Tb (\in_{X}) \Phi \yiff \al \Tb {\leq} (\T\bvsmall)\Phi.
\end{equation}

Now consider an arbitrary element $\be\in \Tom B$, and let $\Phi$ be the
(unique) element of $\Tom\Pom A$ such that $\be = (\T\bv)(\Phi)$.
Then (\ref{eq:5:4:2}) reads that
$\al \Tb (\in_{X}) \Phi \yiff \al \rl{T}(\leq) \be$,
for all $\al\in \Tom A$, and so axiom ($\nb3$) instantiates to
\begin{equation}
\label{eq:5:4:3}
[\nb\be] = \bv \{ [\nb\al] \mid \al \in \Tom A \mbox{ and } \al \Tb (\leq) \be \}.
\end{equation}
But since by the nature of the one-step derivation system we have ${\leq} =
{\sqleq}$ on elements of $\funP X$, we also have $\Tb (\leq) = \Tb 
(\sqleq)$.
So if $\al \Tb (\leq) \be$ then one application of ($\nb1$) gives that 
$\nb\al \sqleq \nb\be$, which implies that $[\nb\al] \leq [\nb\be]$.
From this and (\ref{eq:5:4:3}) is immediate that 
\[
[\nb\be] \leq \bv \{ [\nb\al] \mid \al \in \Tom(\At\bbB),
[\nb \al] \leq [\nb \be] \}.
\]
This finishes the base case of the inductive proof of (\ref{eq:5:4:0}).
\smallskip

For the inductive step of the proof there are three cases to consider.
First, assume that $a'$ is of the form $\bv_{i\in I} a_{i}'$ for some finite
index set $I$.
Then we may compute
\begin{align*}
[a'] &= \bv \{ [a_{i}'] \mid i\in I \}
  \tag*{(assumption)}
\\ &= \bv \Big\{ \bvsmall \{ [\nb\al] \mid \al \in \Tom A, [\nb\al] \leq [a_{i}'] \}
    \mid i \in I \Big\}
  \tag*{(induction hypothesis)}
\\ &= \bv \Big\{ [\nb\al] \mid \al \in \Tom A, [\nb\al] \leq [a_{i}']
    \mbox{ for some } i\in I \Big\}
  \tag*{(associativity of $\bv$)}
\\ &\leq \bv \Big\{ [\nb\al] \mid \al \in \Tom A, [\nb\al] \leq
    \mbox{$\bv_{i\in I}$} [a_{i}'] = a' \Big\}
  \tag*{(properties of $\bv$)}
\end{align*}

Second, consider the case that $a'$ is a conjunction $\bw_{i\in I} a_{i}'$
for some finite $I$.
Now we have
\begin{align*}
[a'] &= \bw \{ [a_{i}'] \mid i\in I \}
  \tag*{(assumption)}
\\ &= \bw \Big\{ \bvsmall \{ [\nb\al] \mid \al \in \Tom A, [\nb\al] \leq [a_{i}'] \}
    \mid i \in I \Big\}
  \tag*{(induction hypothesis)}
\\ &= \bv \Big\{ \bwsmall_{i\in I} [\nb\ga(i)]
    \mid \ga: I \to \Tom A \mbox{ such that } [\nb\ga(i)] \leq
    [a_{i}'] \mbox{ for all } i \Big\}
  \tag*{(distributivity)}
\\ &= \bv \Big\{ [\nb\ga]
    \mid \ga \in \Tom A \mbox{ such that } [\nb\ga] \leq
  \tag*{(part 1)}
    [a_{i}'] \mbox{ for all } i \Big\}
\\ &= \bv \Big\{ [\nb\ga] \mid \ga \in \Tom A, [\nb\ga] \leq
    \mbox{$\bw_{i\in I}$} [a_{i}'] = a' \Big\}
  \tag*{(properties of $\bv$)}
\end{align*}
Here `distributivity' refers to the fact that in any Boolean algebra, finite 
meets distribute over arbitrary joins, and `part~1' refers to the first 
statement of this Theorem.
The point here is that we only need to consider those meets 
$\bw_{i\in I} [\nb\ga(i)]$ for which $\ga(i) = \ga(j)$ for all $i,j \in I$,
since the other meets will reduce to $\bot$.

Finally, suppose that $a'$ is a negation, say $a' = \neg b'$.
We first claim that 
\begin{equation}
\label{eq:neg1}
\mbox{for all } \al \in \Tom A \mbox{ either } \nb\al \sqleq b' \mbox{ or } 
\nb\al \sqleq \neg b'.
\end{equation}
To see this, assume that $\nb\al \not\sqleq \neg b'$; then by propositional
logic,
\[
[\nb\al] \land [b'] > \bot.
\]
By the inductive hypothesis, we have
$[b'] = \bv \{ [\nb\be] \mid \be\in\Tom A, [\nb\be] \leq [b']\}$,
and so by distributivity we obtain
\[
\bv \{ [\nb\al] \land [\nb\be] \mid \be\in\Tom A, [\nb\be] \leq [b']\}
\;>\; \bot.
\]
But then there must be at least one $\be\in\Tom A$ with $[\nb\al] \land [\nb\be]
> \bot$ and $[\nb\be] \leq [b']$.
By the first statement of this Theorem, we can only have $[\nb\al] \land [\nb\be]
> \bot$ if $\al$ is \emph{identical} to $\be$, and so indeed we find that
$[\nb\al] \leq [b']$. 
This proves (\ref{eq:neg1}).

Because of this we can rewrite $[\neg b']$ as follows:
\begin{align*}
[\neg b'] &= [\neg b'] \land \bv \{ [\nb\al] \mid \al \in \Tom A \}
  \tag*{(part 2)}
\\ &=  \bv \{ [\neg b'] \land [\nb\al] \mid \al \in \Tom A \}
  \tag*{(distributivity)}
\\ &=  \bv \Big(
   \{ [\neg b' \land \nb\al] \mid b' \sqgeq \nb\al, \al \in \Tom A \}
   \cup
   \{ [\neg b' \land \nb\al] \mid \neg b' \sqgeq \nb\al,  \al \in \Tom A \}
  \Big)
  \tag*{(\ref{eq:neg1})}
\\ &=  \bv \Big(
   \{ [\bot] \mid b' \sqgeq \nb\al, \al \in \Tom A \}
   \cup
   \{ [\nb\al] \mid \neg b' \sqgeq \nb\al,  \al \in \Tom A \}
  \Big)
  \tag*{(immediate)}
\\ &=  \bv \{ [\nb\al] \mid  [\neg b'] \geq [\nb\al], \al \in \Tom A \}
  \tag*{(immediate)}
\end{align*}
This settles the remaining inductive case, and thus finishes the proof of the 
third part of the Theorem.
\end{proof}

\subsection{Connecting algebra and coalgebra}

Now that we have proved the one-step soundness and completeness of our logic, 
we will show how to connect the algebraic functor $\funaM$ to the coalgebraic
functor $\T$ by defining a natural transformation 
\[
\delta: \funaM \funaQ \ntrto \funaQ \T
\]
which in fact provides an embedding $\de_{X}$ for each set $X$.

For the definition of $\de$, note that given a set $X$, it follows from 
one-step soundness that $\semone{a} = \semone{b}$ for all $a,b \in 
\Tnb\funQ X$ such that $[a]_{\nax\funC\funaQ X} = [b]_{\nax\funC\funaQ X}$.
This ensures that the following is well-defined.

\begin{definition}
\label{d:ntrde}
Given a set $X$, let
\[
\de_{X}([a]_{\nax\funC\funaQ X}) := \semone{a}
\]
define a map $\de_{X}: \funaM\funaQ X \to \funaQ\T X$.
\end{definition}

\begin{prop}\label{p:3}
The family of maps $\de_{X}$, with $X$ ranging over the category $\Set$,
provides a natural transformation $\de: \funaM\funaQ \ntrto \funaQ\T$.
Furthermore, each $\de_{X}: \funaM\funaQ X \to \funaQ\T X$ is an embedding.
\end{prop}

\begin{proof}      
In order to demonstrate that $\delta$ is a natural transformation, we have
to prove that for any function $f: X \to Y$ the following diagram commutes:
\[
\xymatrix{ 
  \funaM\funaQ X \ar[r]^{\de_X} 
  & \funaQ\T X 
\\\funaM\funaQ Y \ar[u]^{\funaM\funaQ f} \ar[r]_{\de_Y} 
  & \funaQ\T Y \ar[u]_{\funaQ \T f} 
}
\]
In order to see that the above diagram commutes it suffices to show that it
commutes on the generators of $\funaM \funaQ Y$.
Consider such a generator $\nb \alpha \in\Tomnb\Tba \funQ Y$.
Then
\[\begin{array}{lclll}
\delta_X(\funaM \funaQ(f) (\nb \alpha))  
  &=&  \delta_X ([\Tomnb\Tba \funQ (f)(\nb \alpha)]) 
  &=& \semone{\Tomnb\Tba \funQ (f)(\nb \alpha)}
\\& \stackrel{\mbox{\tiny Remark~\ref{r:onestep2}}}{=} 
  & \nbsem_X( \T \semzero{\cdot} ( \T\Tba \funQ (f) (\alpha)) ) 
  &=&  \nbsem_X(\T (\semzero{\cdot} \circ \Tba  \funQ(f))(\alpha)) 
\\& \stackrel{\mbox{\tiny $\semzero{\cdot}$ natural, Lem.~\ref{p:semzeronat}}}{=} &\nbsem_X(\T(\funQ f \circ \semzero{\cdot})(\alpha)) 
  &=&  \nbsem_X (\T\funQ f \circ \T\semzero{\cdot} (\alpha)) 
\\& \stackrel{\mbox{\tiny $\lambda$ natural}}{=} 
  & \funQ \T f (\nbsem_Y (\T \semzero{\cdot} (\alpha))) 
  &=&   \funQ \T f (\semone{\nb \alpha})
\\ &=& \funQ \T f (\delta_Y([\nb \alpha])) 
\end{array}\]
Let us finally show that $\delta_X$ is injective for an arbitrary set $X$.
Suppose that $\de_X([a]) = \de_X([b])$ for some $a,b \in \Tba\Tomnb\Tba \funQ
X$.
By definition of $\de_X$ that means that $\semone{a} = \semone{b}$ which by 
one-step completeness of the logic entails that $[a] = [a']$ in $\funaM \funaQ 
X$.
\end{proof}

On the basis of this natural transformation we can define a second notion of 
complex algebra of a coalgebra, next to the Moss complex algebra of
Definition~\ref{d:cplxalg1}.

\begin{definition}
\label{d:cplxalg2}
Let $\T:\Set \to \Set$ be a standard, weak pullback preserving functor,
and let $\bbX = \struc{X,\xi}$ be a $\T$-coalgebra.
We define the \emph{complex $\funaM$-algebra} of $\bbX$ as the pair
$\bbX^{*} \isdef \struc{\funaQ X, \de_{X}\cof\funaQ\xi}$.
\end{definition}

The link between the two kinds of complex algebras is given by the
functor $\funV$ from Definition~\ref{d:funV} which allows us to see $\funaM$-algebras as Moss
algebras.

\begin{prop}
\label{p:plusstar}
Let $\T:\Set \to \Set$ be a standard, weak pullback preserving functor.
Then 
\[
\bbX^{+} = \funV\bbX^{*}.
\]
for any $\T$-coalgebra $\bbX$. Therefore, for any $\T$-coalgebra $\bbX$ 
and any formula $a \in \Lmoss$ we have 
	\[ \mng_{\funV\bbX^*}(a) = \mng_{\bbX^+} (a) =  \{ x \in X \mid x \Vdash a \} .\]
\end{prop}


%% file: sec-completeness.tex
\section{Soundness and completeness}
\label{s:completeness}

\newcommand{\sprs}[1]{\prs{G_{#1}}{\equiv_{#1}}}
\newcommand{\Lstr}[1]{\mathbb{L}_{#1}}
\newcommand{\eqn}[1]{[#1]_{n}}

In this section we will apply Pattinson's \emph{stratification
method}~\cite{patt:coal03}
in order to prove the soundness and completeness of our axiom system $\nax$ 
with respect to the coalgebraic semantics.
This stratification method consists in showing that not only the 
\emph{language} of our system, but also its \emph{semantics} and our
\emph{logic} can be stratified in $\om$ many layers.
As we will see further on, the results in the previous section will then 
serve to glue these layers nicely together.

In order to understand the idea of the proof, first assume that a final
$\T$-coalgebra $\bbZ = \struc{Z, \zeta:Z\to \T Z}$ exists.
Then we could prove that the unique Moss morphism $\mng_{\bbZ}$ from the 
initial Moss algebra $\Lmoss$ to the algebra $\bbZ^{+}$ actually factors as
$\mng_{\bbZ} = \funV \mng^*_\bbZ \cof q$, where $q: \Lmoss \to \Minit$ is the quotient
map modulo derivability (in the sense that $\ker(q)$ is the relation 
$\equiv_{\nax}$ of interderivability in $\nax$), and $\mng^*_\bbZ$ is an 
\emph{injective} $\funaM$-algebra morphism from $\Minit$ to $\bbZ^{*}$:
\begin{equation*}
\xymatrix{
& \Lmoss       \ar[dr]_{\mng_{\bbZ}}  \ar[r]^{q}
& \funV\Minit  \ar@{>->}[d]^{\funV \mng^*_\bbZ}
\\
& & \funV\bbZ^{*} = \bbZ^{+}
}
\end{equation*}
On the basis of this we would prove that $a \not\sqleq_{\nax}b$ implies
that $q(a) \not\leq_{\Minit} q(b)$, and so by injectivity of $m$ we would
conclude that $\mng_{\bbZ}(a)\not\sse\mng_{\bbZ}(b)$, providing a state 
$z\in Z$ such that $z\forces_{\bbZ} a$ and $z\not\forces_{\bbZ} b$.

Since our set functor $\T$ generally does not admit a final coalgebra, we
replace the final coalgebra with the \emph{final sequence}.

\begin{definition}
\label{d:finseq}
The \emph{final $\T$-sequence} is defined as follows.
\begin{equation}
\label{dg:finseq}
\xymatrix{
  1             
& \T 1       \ar@{->}[l]_{h_{0}} 
& \T ^{2}1   \ar@{->}[l]_{h_{1}} 
& \ldots 
& \T ^{n}1   \ar@{->}[l]_{h_{n}} 
& \T ^{n+1}1 \ar@{->}[l]_{h_{n+1}} 
& \ldots
}
\end{equation}
We denote by $1=\T^01$ the final object in $\Set$. 
The map $h_0:\T{1}\to{1}$ is given by finality and inductively, $h_{n+1}:
\T(\T^n{1})\to \T^n{1}$ is defined to be the map $\T^{n}h_{0} = \T h_n$.  
\end{definition}

The reader may think of the $\T^n{1}$ as approximating the final coalgebra.
Indeed, if we let the final sequence run through all ordinals, we obtain
the final coalgebra as a limit if it exists~\cite{adam:grea95}.  
Intuitively, where the states of the final coalgebra provide all possible
$\T$-behaviors, the elements of $\T^{n}1$ represent all `$n$-step
behaviors'.
Given a $\T$-coalgebra $\bbX = \struc{X,\xi}$, for each $n\in\om$ we may 
canonically define a map $\xi_{n}: X\to \T^{n}1$ providing the $n$-step
behavior of the states of $\bbX$.

\begin{definition}
Given a $\T$-coalgebra $\bbX = \struc{X,\xi}$, we define the arrows $\xi_{n}:
X\to \T^{n}1$, for $n \in \omega$, to the approximants of the final coalgebra
by the following induction: $\xi_0:X\to 1$ is given by finality of $1$ in
$\Set$, and $\xi_{n+1} \isdef \T\xi_{n}\cof\xi$ .
\end{definition}

Interestingly, every object $\T^{n}1$ in the final sequence can be equipped
with coalgebra structure.

\begin{definition}
\label{d:ncoalg}
Let, for each $n\in\om$, $\bbZ_{n}$ be the coalgebra
\[
\bbZ_{n} \isdef (\T^{n}1,\T^{n} g),
\]
where $g$ is an arbitrary but fixed map $g: 1 \to \T 1$.
\end{definition}

As we will see in a moment, these `$n$-final coalgebras' display all 
possible $n$-step behaviours, and thus act as a canonical witness for all
non-provable inequalities between formulas of depth $n$.

\subsection{A stratification of the semantics}
\label{ss:strat1}

We first show how to slice the semantics of nabla formulas into layers.
For that purpose we define the $n$-step meaning of depth-$n$ modal formulas 
as a subset of the set $\T^{n}1$.

\begin{definition}
\label{d:nstepsem}
By induction on $n$ we define maps $\mng_{n}: \Lmoss_{n} \to \funP\T^{n}1$.
For $n=0$, we define $\mng_{0}$ by initiality of $\Lmoss_{0}$, or equivalently:
\[
\mng_{0}(a) \isdef \left\{\begin{array}{ll}
  1     & \mbox{if $a$ is a tautology,}
\\\nada & \mbox{otherwise.}
\end{array}\right.\]
Inductively, assuming that $\mng_{n}: \Lmoss_{n} \to \funP\T^{n}1$ has been 
defined, we may compose $\T\mng_{n}: \T\Lmoss_{n} \to \T\funP\T^{n}1$
with $\nbsem_{\funP\T^{n}1}: \T\funP\T^{n}1 \to \funP\T^{n+1}1$ to obtain
\begin{equation*}
\nbsem_{\funP\T^{n}1}\cof\T\mng_{n}: \Tom\Lmoss_{n} \to \funP\T^{n+1}1.
\end{equation*}
Then we let $\mng_{n+1}: \Lmoss_{n+1} \to \funP\T^{n+1}1$ be the unique
$\Boole$-homomorphism from $\funaF(\Tomnb\Lmoss_{n})$ to $\funaP\T^{n+1}1$
that extends the mapping given by 
\[ \nb \alpha \mapsto \left( \nbsem_{\funP\T^{n}1}\cof\T\mng_{n}(\alpha) \right) 
\hspace{1cm} \mbox{for
$\nb \alpha \in \Tomnb \Lmoss_n$.} \vspace{-18 pt}\]
\end{definition}\medskip

\noindent The following proposition provides a clear link between the $n$-step meaning
of formulas and the $n$-step behaviour map of a coalgebra.

\begin{prop}
\label{p:nsem}
Let $\bbX$ be a coalgebra, and $a \in \Lmoss_{n}$ a formula of rank $n$.
Then
\[
\mng_{\bbX}(a) = (\funQ\xi_{n})(\mng_{n}(a)).
\]
\end{prop}
\begin{proof}
The proof of the proposition is by induction on the modal depth and on the
structure of the formula $a$. 
We only provide the induction case for $a = \nb \alpha \in \Lmoss_{n+1}$ for
some $n \in \omega$.
In this case we have
\begin{align*}
\mng_{\bbX} ( \nb \alpha)  
  & =  \funQ \xi (\lambda_X (\T \mng_{\bbX} (\alpha))) 
  & \mbox{(definition of $\mng_{\bbX}$)} 
\\&= \funQ \xi (\lambda_X (\T \funQ \xi_n (\T \mng_n (\alpha)))) 
  & \mbox{(induction hypothesis)} 
\\&= \funQ \xi \left( \funQ \T \xi_n 
      (\lambda_{\T^n 1}(\T \mng_n (\alpha)))\right) 
  & \mbox{(naturality of $\lambda$)} 
\\&= \funQ \xi_{n+1}(\mng_{n+1}(\nb \alpha)) 
  & \mbox{(definition of $\mng_{n+1}$ and $\xi_{n+1}$)} 
\end{align*}
\end{proof}
The $n$-final coalgebra of Definition~\ref{d:ncoalg} has the interesting 
property that its $n$-step behaviour map is the \emph{identity} map on
$\T^{n}1$. 
As a corollary, the $n$-step meaning of any depth-$n$ formula $a$ coincides 
with its meaning in the $n$-step coalgebra.

\begin{prop}
\label{p:ncoalg}
Let $a$ be a formula of depth $n$.
Then 
\[
\mng_{\bbZ_{n}}(a) = \mng_{n}(a).
\]
\end{prop}

\begin{proof}
It is not difficult to see that for the coalgebra $\bbZ_{n}$ (and for this 
$n$), we have
\begin{equation}
\label{eq:gh}
(\T^{n}g)_{n} \isdef \id_{\T^{n}1}.
\end{equation}
We confine ourselves to a proof sketch.
The basic idea of the proof is to prove inductively that $(\T^{n}g)_{k} =
h_{nk}$ for all $k\leq n$, where $h_{nk}: \T^{n}1 \to \T^{k}1$ is the map 
$h_{nk}\isdef h_{k} \cof h_{k+1} \cof \cdots \cof h_{n}$.
Further details can be found in~\cite[Section 4]{patt:coal03}.

The Proposition itself is immediate by Proposition~\ref{p:nsem} and
\eqref{eq:gh}.
\end{proof}

As a fairly direct corollary to the previous two propositions we can formulate
our semantic stratification theorem.
Basically it states that the meaning of depth-$n$ formulas is determined at
level $n$ of the final sequence, and in the $n$-step final coalgebra
$\bbZ_{n}$.

\begin{thm}[Semantic Stratification Theorem]
\label{t:strs}
Let $a,b \in \Lmoss_{n}$ be formulas. 
Then the following are equivalent:
\begin{enumerate}[\em(1)]
\item $a \models_{\T} b$;
\item $\mng_{n}(a) \sse \mng_{n}(b)$;
\item $\mng_{\bbZ_{n}}(a) \sse \mng_{\bbZ_{n}}(b)$.
\end{enumerate}
\end{thm}

\begin{proof}
The implication $1 \Rightarrow 3$ is immediate by the definitions, while the 
implication $2 \Rightarrow 1$ follows by Proposition~\ref{p:nsem}: given a
coalgebra $\bbX = \struc{X,\xi}$, we conclude from $\mng_{n}(a) \sse
\mng_{n}(b)$ that $\mng_{\bbX}(a) = (\funaQ\xi_{n})(\mng_{n}(a)) \sse
(\funaQ\xi_{n})(\mng_{n}(b)) = \mng_{\bbX}(b)$.
The remaining implication $3 \Rightarrow 2$ follows directly by
Proposition~\ref{p:ncoalg}.
\end{proof}

\subsection{A stratification of the logic}
\label{ss:stratification}

To see in detail how our \emph{logic} can be stratified, let us first
introduce some terminology concerning the stratification of the
language.

\begin{definition}
Let $G_{0} := \nada$, and define inductively $G_{n+1} := \Tomnb\Tba G_{n}
= \{ \nb\al \mid \al\in\Tom\Tba(G_{n}) \}$.
In addition, let $e_{0}: G_{0} \to \Tba G_{1}$ be the empty map, and define $e_{n+1}:
G_{n+1} \to \Tba G_{n+2}$ by putting $e_{n+1} := \funpM e_{n}$.
Finally, we let $d_{n}$ denote the inclusion $d_{n}: \Lmoss_{n} 
\hookrightarrow\Lmoss$.
\end{definition}

Recall that $\Lmoss_{n}$ denotes the set of formulas of rank $n$ (see
Definition~\ref{d:syntax}), and observe that $\Lmoss_{n} = \Tba
G_{n}$, for all $n$, and that each $\Lmoss_{n}$ is also the carrier of
an algebra in $\Boole$; this algebra will also be denoted as
$\Lmoss_{n}$. \akk{Consequently, $\Lmoss_{n+1}=\Lmoss_1(G_n)$, which
  is different from $\Lmoss_1(\Lmoss_n)=\Lmoss_1(\Lmoss_0(G_n)))$
  since in $\Boole$ we do not identify terms which are equivalent in
  the theory of Boolean algebras.}  Also observe that the map
$\wh{e}_{n}: \Tba G_{n} \to \Tba G_{n+1}$ is in fact the embedding of
$\Lmoss_{n}$ into $\Lmoss_{n+1}$:
\[
\wh{e}_{n}: \Lmoss_{n} \hookrightarrow\Lmoss_{n+1},
\]
and that the embedding $d_{n}: \Lmoss_{n} \hookrightarrow\Lmoss$ commutes with
the one-step embeddings, in the sense that $d_{n} = d_{n+1} \cof \wh{e}_{n}$.

We can now formulate our stratification theorem as follows.
Recall that $\Lmoss$ is the initial algebra in the category $\Boolenb$.

\begin{thm}[Axiomatic Stratification Theorem]
\label{t:strat}
Let $m \isdef \mng_{\funV\Minit}$ be the unique homomorphism $m: \Lmoss \to
\funV\Minit$ in the category of Moss algebras.
\begin{enumerate}[\em(1)]
\item
There are maps $q_{n}: \Lmoss_{n} \to \funaM^{n}\bbtwo$, with each $q_{n}$
a $\Boole$-homomorphism, such that the following diagram (in the category
$\Boole$) commutes:
\begin{equation}
\label{dg:strat1}
\xymatrix{
  \Lmoss_{0}   \ar[d]_{q_{0}}  \ar@{^{(}->}[r]^{\wh{e}_{0}} 
               \ar@/^{12mm}/[rrrrrrr]|{d_{0}}
& \Lmoss_{1}   \ar[d]_{q_{1}}  \ar@{^{(}->}[r]^{\wh{e}_{1}} 
               \ar@/^{10mm}/[rrrrrr]|{d_{1}}
& \Lmoss_{2}   \ar[d]_{q_{2}}
               \ar@/^{8mm}/[rrrrr]|{d_{2}}
& \ldots 
& \Lmoss_{n}    \ar[d]_{q_{n}} \ar@{^{(}->}[r]^{\wh{e}_{n}}
               \ar@/^{5mm}/[rrr]|{d_{n}}
& \Lmoss_{n+1} \ar[d]_{q_{n+1}}
               \ar@/^{3mm}/[rr]|{d_{n+1}}
& \ldots
& \Lmoss       \ar[d]_{m}
\\
  \bbtwo           \ar@{>->}[r]^{j_0} \ar@/_{12mm}/[rrrrrrr]|{i_{0}}
& \funaM\bbtwo     \ar@{>->}[r]^{j_1} \ar@/_{10mm}/[rrrrrr]|{i_{1}}
& \funaM^{2}\bbtwo                    \ar@/_{8mm}/[rrrrr]|{i_{2}}
& \ldots
& \funaM^{n}\bbtwo \ar@{>->}[r]^{j_n} \ar@/_{5mm}/[rrr]|{i_{n}}
& \funaM^{n+1}\bbtwo                  \ar@/_{3mm}/[rr]|{i_{n+1}}
& \ldots
& \funaM^{\om}\bbtwo
}
\end{equation}
\item
In addition, $\ker(m) = {\equiv_{\nax}}$; that is, $m(a) = m(b)$ iff $a$ and 
$b$ are provably equivalent in $\nax$.
\end{enumerate}
\end{thm}

Before turning to the proof of this result, let us briefly summarize its
meaning.
Most importantly, Theorem~\ref{t:strat} states that for each $n<\om$, the
Boolean algebra $\funaM^{n}\bbtwo$ coincides with the quotient of the
$\Boole$-algebra $\Lmoss^{n}$ under the relation $\equiv_{\nax}$ of 
provable equivalence in our derivation system $\nax$.
In addition, the quotient maps $q_{n}$ commute with the inclusions $\wh{e}_{n}$ 
of $\Lmoss_{n}$ into $\Lmoss_{n+1}$, and $j_{n}$ from $\funaM^{n}\bbtwo$ into
$\funaM^{n+1}\bbtwo$.

In order to prove Theorem~\ref{t:strat}, we will inductively define a relation
$\equiv_{n}$ of ``$n$-inter\-derivability'' between $\Lmoss_{n}$-formulas.
We will see that for every $n$, the Boolean algebra $\Lstr{n} = 
\Lmoss_{n}/_{\equiv_{n}}$ is isomorphic to $\funaM^{n}\bbtwo$, but also, that
for formulas $a,b \in \Lmoss_{n}$, we have $a \equiv_{n} b$ iff $a \equiv_{\nax}
b$.
The definition of $\equiv_{n}$ will be such that
\[
\sprs{n+1} = \funpM\sprs{n}.
\]

\begin{definition}
Let ${\equiv_{0}} \sse \Lmoss_{0} \times \Lmoss_{0}$ be the relation of
provable equivalence between closed Boolean terms.
Inductively, define the relation ${\equiv_{n+1}} \sse \Lmoss_{n+1} \times
\Lmoss_{n+1}$ as the congruence relation of the presentation $\funpM \sprs{n}$,
and let $\Lstr{n}$ denote the Boolean algebra $\funB \sprs{n}$, or equivalently, 
$\Lstr{n} = \Lmoss_{n}/_{\equiv_{n}}$.
Given a formula $a \in \Lmoss_{n}$, we let $\eqn{a}$ denote the equivalence
class of $a$ under the relation $\equiv_{n}$.
\end{definition}

\noindent
As we will see, the algebras $\Lstr{n}$ form an intermediate row in the
stratification diagram (\ref{dg:strat1}) (in the category $\Boole$):
\begin{equation}
\label{dg:strat2}
\xymatrix{
  \Lmoss_{0}   \ar@{>>}[d]_{\ti{\eta}_{0}}  \ar@{^{(}->}[r]^{\wh{e}_{0}} 
& \Lmoss_{1}   \ar@{>>}[d]_{\ti{\eta}_{1}}  \ar@{^{(}->}[r]^{\wh{e}_{1}} 
& \Lmoss_{2}   \ar@{>>}[d]_{\ti{\eta}_{2}}
& \ldots 
& \Lmoss_{n}   \ar@{>>}[d]_{\ti{\eta}_{n}}  \ar@{^{(}->}[r]^{\wh{e}_{n}}
& \Lmoss_{n+1} \ar@{>>}[d]_{\ti{\eta}_{n+1}}
& \ldots
\\
  \Lstr{0}   \ar@{>->>}[d]_{f_{0}}  \ar@{>->}[r]^{\funB e_{0}} 
& \Lstr{1}   \ar@{>->>}[d]_{f_{1}}  \ar@{>->}[r]^{\funB e_{1}} 
& \Lstr{2}   \ar@{>->>}[d]_{f_{2}}
& \ldots 
& \Lstr{n}   \ar@{>->>}[d]_{f_{n}}  \ar@{>->}[r]^{\funB e_{n}}
& \Lstr{n+1} \ar@{>->>}[d]_{f_{n+1}}
& \ldots
\\
  \bbtwo           \ar@{>->}[r]^{j_0} 
& \funaM\bbtwo     \ar@{>->}[r]^{j_1}
& \funaM^{2}\bbtwo 
& \ldots
& \funaM^{n}\bbtwo \ar@{>->}[r]^{j_n}
& \funaM^{n+1}\bbtwo & \ldots
}
\end{equation}

\noindent
We now turn to the details of the proof of Theorem~\ref{t:strat}, step by step
filling in diagram~(\ref{dg:strat2}).
Since we already discussed the embeddings $\wh{e}_{n}$, $n\in\om$, we start
with the map $\ti{\eta}_{n}$, which will denote the quotient map associated
with the congruence $\equiv_{n}$.

\begin{definition}
\label{d:nu}
Let $\eta_{n}: G_{n} \to \Lmoss_{n}/_{\equiv_{n}}$ be the map given by
$\eta_{n}: g \mapsto \eqn{g}$.
\end{definition}

We may see the map $\eta_{n}$ as a presentation morphism from $\sprs{n}$ to
$\funC(\Lstr{n})$ --- as such it is the unit $\unitBC_{\sprs{n}}$ of the
adjunction $\funB \dashv \funC$, and hence, a pre-isomorphism 
(cf.~Theorem~\ref{t:BCadj}).
This function extends to a homomorphism in $\Boole$:
\[
\ti{\eta}_{n}: \Lmoss_{n} \to \Lstr{n}
\]
which maps a formula $a \in \Lmoss_{n}$ to its $n$-equivalence class:
\[
\ti{\eta}_{n}: a \mapsto \eqn{a}.
\]

Concerning the maps $\funB e_{n}: \Lstr{n} \to \Lstr{n+1}$, it is easy to see 
that they are indeed well-typed, but in order to prove that each $\funB e_{n}$
is an embedding, some work will be needed.
The embeddings $j_{n}: \funaM^{n}\bbtwo \to \funaM^{n+1}\bbtwo$ have been
defined in Definition~\ref{d:initial_sequence}.

Finally, the isomorphisms $f_{n}$ of diagram~(\ref{dg:strat2}) will be defined
inductively.

\begin{definition}
By induction on $n$ we define Boolean homomorphisms $f_{n}: \Lstr{n} \to
\funaM^{n}\bbtwo$.
For $n=0$, we let $f_{0}$ be the (unique) isomorphism from $\Lstr{0}$ to 
$\bbtwo$.
For $n=k+1$, we first define $p_{n+1}: \Lstr{n+1} \to \funaM\Lstr{n}$ by putting
$p_{n+1} := \funB\funpM\eta_{n}$.
Then we compose the maps
\[
\Lstr{n+1} \stackrel{p_{n+1}}{\longrightarrow} \funaM\Lstr{n}
\stackrel{\funaM f_{n}}{\longrightarrow} \funaM^{n+1}\bbtwo,
\]
and define $f_{n+1} := (\funaM f_{n})\cof p_{n+1}$.
\end{definition}

The following proposition gathers all the facts about the maps defined until now
that are needed to prove that diagram~(\ref{dg:strat2}) commutes:

\begin{prop}
\label{l:str1}\hfill
\begin{enumerate}[\em(1)]
\item
In the category $\Prs$ of presentation each map $e_{n}$ is a morphism $e_{n}:
\sprs{n}$  $\to \sprs{n+1}$, each map $\eta_{n}: \sprs{n+1} \to \funC\Lstr{n}$ is
a pre-isomorphism, and each of the following diagrams commutes:
\begin{equation}
\label{dg:tec0}
\xymatrix{
  \sprs{n}   \ar[d]_{\eta_{n}}   \ar[r]^{e_{n}} 
& \sprs{n+1} \ar[d]_{\eta_{n+1}} 
\\
  \funC(\Lstr{n})        \ar[r]^{\funC\funB e_{n}} 
& \funC(\Lstr{n+1})  
}
\end{equation}
\item
In the category $\Boole$, each of the following diagrams commutes:
\begin{equation}
\label{dg:tec1a}
\xymatrix{
  \Lmoss_{n}   \ar[d]_{\ti{\eta}_{n}}   \ar[r]^{\wh{e}_{n}} 
& \Lmoss_{n+1} \ar[d]_{\ti{\eta}_{n+1}} 
\\
  \Lstr{n}        \ar[r]^{\funB e_{n}} 
& \Lstr{n+1} 
}
\end{equation}
\item
In the category $\BA$ of Boolean algebras, each map $p_{n+1}$ is an isomorphism,
and each of the following diagrams commutes:
\begin{equation}
\label{dg:tec2}
\xymatrix{
  \Lstr{n+1}  \ar@{>->>}[d]_{p_{n+1}}   \ar[r]^{\funB e_{n+1}} 
& \Lstr{n+2}  \ar@{>->>}[d]_{p_{n+2}} 
\\
  \funaM(\Lstr{n})        \ar[r]^{\funaM\funB e_{n}} 
& \funaM(\Lstr{n+1})  
}
\end{equation}
\item
In the category $\Boole$, each of the following diagrams commutes:
\begin{equation}
\label{dg:tec3}
\xymatrix{
  \Lmoss_{n+1}  \ar[d]_{\Tnb\eta_{n}}   \ar[r]^{\ti{\eta}_{n+1}} 
& \Lstr{n+1}    \ar[d]_{p_{n+1}} 
\\
  \Tnb\funU(\Lstr{n})        \ar[r]^{\quot{\Lstr{n}}} 
& \funaM(\Lstr{n})  
}
\end{equation}
with $\quot{\Lstr{n}}$ as in Definition~\ref{d:quot}.

\item
In the category $\BA$ of Boolean algebras, each map $f_{n}$ is an isomorphism;
each map $\funB e_{n}: \Lstr{n} \to \Lstr{n+1}$ is an embedding; and each of the
following diagrams commutes:
\begin{equation}
\label{dg:strat-f}
\xymatrix{
  \Lstr{n}   \ar@{>->>}[d]_{f_{n}}   \ar@{>->}[r]^{Be_{n}} 
& \Lstr{n+1} \ar@{>->>}[d]_{f_{n+1}} 
\\
  \funaM^{n}\bbtwo       \ar@{>->}[r]^{j_{n}} 
& \funaM^{n+1}\bbtwo  
}
\end{equation}
\end{enumerate}
\end{prop}

\begin{proof}\hfill
\begin{enumerate}[(1)]
\item
It follows by a straightforward induction that every $e_{n}$ is a presentation
morphism.
The other statements of this item
follow from the fact that $\eta_{n} = \unitBC_{\prs{G_{n}}{\equiv_{n}}}$,
together with our earlier observation (cf.~Theorem~\ref{t:BCadj}) that 
$\unitBC: \Id_{\Prs} \to \funC\funB$ is a natural transformation of which 
each $\unitBC_{\pGR}$ is a pre-isomorphism.

\item
We claim that if $f: \pGR \to \prs{G'}{R'}$ is the presentation morphism 
represented by one of the four arrows of the diagram (\ref{dg:tec0}), then the
corresponding arrow $\hat{f}$ in (\ref{dg:tec1a}) is the \emph{unique}
$\Boole$-morphism extending $f$ (seen as a map between sets).
For instance, if $f$ is the presentation morphism $\eta_{n}: \sprs{n} \to
\funC\Lstr{n}$, then using the fact that $\Lmoss_{n} = \Tba G_{n}$ is the free
$\Boole$-algebra over $G_{n}$, it follows that $\hat{f} =\ti{\eta}_{n}$ is the
unique homomorphism in $\Boole$ from $\Lmoss_{n}$ to $\Lstr{n}$.
Or, to give a second example, $\funB e_{n}$ is clearly the only homomorphism
from $\Lstr{n}$ to $\Lstr{n+1}$ which ``extends'' $\funC\funB e_{n}:
\funC\Lstr{n} \to \funC\Lstr{n+1}$.

From this it follows that both $\ti{\eta}_{n+1}\cof\wh{e}_{n}$ and 
$\funB e_{n}\cof\ti{\eta}_{n}$ are morphisms in $\Boole$ that extend the map
$\eta_{n+1}\cof e_{n} = \funC\funB e_{n} \cof \eta_{n}$ (with the identity
holding because diagram~(\ref{dg:tec0}) commutes).
But then, again by the freeness of $\Lmoss_{n}$ over $G_{n}$ in $\Boole$,
these two extensions must be equal, which is the same as to say that 
(\ref{dg:tec1a}) commutes.

\item
It is easy to see that our definition of the map $p_{n+1}$ indeed provides an
isomorphism, because
\[
\funpM\eta_{n}: \sprs{n+1} = \funpM\sprs{n} \to \funpM\funC\Lstr{n},
\]
is a pre-isomorphism in $\Prs$, by Theorem~\ref{t:Mfun} inheriting this
property from $\eta_{n}: \sprs{n} \to \funC\Lstr{n}$, 
and $\funB$ maps pre-isomorphisms to isomorphisms, see Proposition~\ref{p:piBi}.

To prove that diagram (\ref{dg:tec2}) commutes it suffices to see that we may
obtain it from diagram (\ref{dg:tec0}) by applying the functor $\funB\funpM$.

\item
Recall that the family of presentation morphisms $\unitBC_{\pGR}: \pGR \to 
\funC\funB\pGR$, defined by (\ref{eq:defunit}), constitutes a natural 
transformation $\unitBC: \Id_{\Prs} \ntrto \funC\funB$.
Instantiating the diagram which expresses this fact for the arrow 
$\funpM\eta_{n}: \funpM\sprs{n} \to \funpM\funC\Lstr{n}$, we obtain the 
following commuting diagram:
\begin{equation}
\label{dg:tec4}
\xymatrix{
  \funpM\sprs{n}  \ar[d]_{\funpM\eta_{n}}   \ar[rr]^{\unitBC_{\funpM\sprs{n}}} 
  && \funC\funB\funpM\sprs{n} = \funC\Lstr{n+1}  \ar[d]^{\funC\funB\funpM\eta_{n}} 
\\
  \funpM\funC\Lstr{n}        \ar[rr]^{\unitBC_{\funpM\funC\Lstr{n}}} 
&& \funC\funB\funpM\funC\Lstr{n} = \funC\funaM\Lstr{n} 
}
\end{equation}
Now we can, similarly as in the proof of item~2, show that each of the arrows
in (\ref{dg:tec3}) is the unique morphism in $\Boole$ that extends the 
corresponding map in (\ref{dg:tec4}).
For example, consider the map $\Tnb\eta_{n}: \Lmoss_{n+1} \to
\Tnb\funU \Lstr{n}$.
It follows from a straightforward unravelling of the definitions that 
$\Tnb\eta_{n}$ extends $\funpM\eta_{n}$ (see Proposition~\ref{p:funpM-Tnb}).
The latter, as a function between sets, is just a map from $\Tomnb\Tba G_{n} 
= G_{n+1}$ to the set of generators of the presentation $\funpM\funC\Lstr{n}$, 
which is nothing but the set $\Tomnb\Tba\funU \Lstr{n}$. 

But then, again similar to the proof of item~2, we can prove that the maps
$p_{n+1}\cof \ti{\eta}_{n+1}$ and $\quot{\Lstr{n}}\cof\Tnb\eta_{n}$ are
identical, by noting that both are morphisms in $\Boole$ that extend the
presentation morphism $\funC\funB\funpM\eta_{n} \cof \unitBC_{\funpM\sprs{n}}
= \unitBC_{\funpM\funC\Lstr{n}} \cof \funpM\eta_{n}$ of
diagram~\eqref{dg:tec4}.

\item
This part of the Proposition is proved by induction on $n$.
For $n=0$, the map $f_{0}$ is an isomorphism by definition, and the map 
$\funB e_{0}$ is an embedding by initiality of $\bbtwo$ in $\BA$.
Finally, the following diagram commutes simply by the initiality of the algebra
$\Lstr{0}$ in the category $\BA$:
\begin{equation}
\xymatrix{
  \Lstr{0}  \ar@{>->>}[d]_{f_{0}}   \ar@{>->}[r]^{Be_{0}} 
& \Lstr{1}  \ar[d]_{f_{1}} 
\\
  \bbtwo    \ar@{>->}[r]^{j_{0}} 
& \funaM\bbtwo  
}
\end{equation}
In the inductive case for $n+1$, by hypothesis the map $f_{n}$ is an
isomorphism, and the map $\funB e_{n}$ an embedding.  From this it is
immediate that $\funaM f_{n}$ is an isomorphism as well, and since
$p_{n+1}$ is an isomorphism by Proposition~\ref{l:str1}(2), it follows
that the map $f_{n+1}$, being the composition of two isomorphisms, is
an isomorphism as well.

Now consider the following diagram:
\begin{equation}
\xymatrix{
  \Lstr{n+1}   \ar@/_{20mm}/[dd]_{f_{n+1}} \ar@{>->>}[d]_{p_{n+1}}  \ar[r]^{Be_{n+1}} 
& \Lstr{n+2}   \ar@/^{20mm}/[dd]^{f_{n+2}} \ar@{>->>}[d]_{p_{n+2}} 
\\
  \funaM(\Lstr{n})   \ar@{>->>}[d]_{\funaM f_{n}}  \ar@{>->}[r]^{\funaM\funB e_{n}} 
& \funaM(\Lstr{n+1}) \ar@{>->>}[d]_{\funaM f_{n+1}}
\\
  \funaM^{n+1}\bbtwo       \ar@{>->}[r]^{j_{n+1}} 
& \funaM^{n+2}\bbtwo  
}
\end{equation}
The upper rectangle of this diagram commutes by Proposition~\ref{l:str1}(2), and
the lower rectangle, by applying the functor $\funaM$ to the diagram 
(\ref{dg:strat-f}) which commutes by the inductive hypothesis.
As a consequence, the outer rectangle, which exactly corresponds to the diagram
(\ref{dg:strat-f}) for the case $n+1$, commutes as well.
Finally, then, the injectivity of $\funB e_{n+1}$ is immediate by that of
$j_{n+1}$, which was established in Lemma~\ref{p:minit}(1).

\end{enumerate}
\end{proof}

By Proposition~\ref{l:str1} it follows that the diagram~(\ref{dg:strat2})
commutes.
\medskip

For future reference we state the following technical fact, which links the 
quotient maps $q_{n}$ and $q_{n+1}$ to the natural transformation
$\quot{}$ of Definition~\ref{d:quot}, instantiated at the Boolean algebra
$\funaM^{n}\bbtwo$.

\begin{prop}
\label{p:qqq}
For any element $\al \in \Tom\Lmoss_{n}$, we have 
\begin{equation}
\label{eq:qqq}
q_{n+1}(\nb\al) = \quot{\funaM^{n}\bbtwo}\nb(\T q_{n}(\al)).
\end{equation}
\end{prop}

\begin{proof}
To see why this proposition holds, recall that $q_{k} = f_{k}\cof
\ti{\eta}_{k}$ for each $k\in\om$, and consider the diagram below
\begin{equation}
\label{dg:nbhom3}
\xymatrix{
   \Tom\Lmoss_{n}           \ar[d]_{\T\ti{\eta}_{n}} \ar[r]^{\nb_{G_{n}}}
&  \Tnb(G_{n})=\Lmoss_{n+1} \ar[d]_{\Tnb\eta_{n}}    \ar[r]^{\ti{\eta}_{n+1}} 
& \Lstr{n+1}                \ar[d]_{p_{n+1}} 
       \ar@/^{20mm}/[dd]^{f_{n+1}}
\\ \Tom\funU(\Lstr{n})      \ar[d]_{\T f_{n}}       \ar[r]^{\nb_{\funU(\Lstr{n})}}
&  \Tnb\funU(\Lstr{n})      \ar[d]_{\Tnb f_{n}}     \ar[r]^{\quot{\Lstr{n}}} 
& \funaM(\Lstr{n})          \ar[d]_{\funaM f_{n}}
\\ \Tom\funU(\funaM^{n}\bbtwo) \ar[r]^{\nb_{\funU\funaM^{n}\bbtwo}}
& \Tnb\funU(\funaM^{n}\bbtwo)        \ar[r]^{\quot{\funaM^{n}\bbtwo}} 
& \funaM^{n+1}\bbtwo  
}
\end{equation}
\akk{where, in order to simplify the diagram, we omit the forgetful
  functors to $\Set$ on the right-hand side of the diagram and exploit
  our ambiguous notation allowing $\Lmoss_1$ to be considered as
  $\Set$-valued or $\Boole$-valued.}

Here an arrow labelled $\nb_{G}$ represents the function mapping an object
$\al \in \Tom \Tba G$ to the corresponding formula $\nb\al \in \Tnb(G)$.
Note that in the case that $G = \funU(\Lstr{n})$ and $G =\funU\funaM^{n}\bbtwo$
we use the fact that $\Tom G \sse \Tom\Tba G$.

We claim that all squares of (\ref{dg:nbhom3}) commute.
To check this for the left squares this is simply a matter of unravelling the
definitions, and the upper right square has been shown to commute in
Proposition~\ref{l:str1}(4).
Finally, that the lower right square commutes is a consequence of the fact 
that $\quot{}$ is a natural transformation $\quot{}: \Tnb\funU \ntrto \funaM$,
cf.~Proposition~\ref{p:q}.

But if indeed all squares of (\ref{dg:nbhom3}) commute, then the identity 
(\ref{eq:qqq}) can simply be read off from the outer sides of the diagram.
\end{proof}

Continuing the proof of the Stratification Theorem, what is left to do is
link the algebras $\Lmoss$ and $\Minit$ to diagram~(\ref{dg:strat2}).
We first need a proof-theoretical result stating that on formulas in
$\Lmoss_{n}$, the notions of $n$-derivability and derivability coincide.

\begin{prop}
\label{p:str3}
Let $a$ and $b$ be two formula in $\Lmoss_{n}$.
\begin{enumerate}[\em(1)]
\item
$a \equiv_{n} b$ iff $a \equiv_{m} b$ for some $m\in\om$;
\item    
$a \equiv_{n} b$ iff $a {\equiv_{\nax}} b$.
\end{enumerate}
\end{prop}

\begin{proof}
Part~1 of the proposition is a direct consequence of diagram~(\ref{dg:strat2})
commuting.
Concerning the second part, the left-to-right direction can be proved by a 
straightforward induction on $n$.
For the opposite direction `$\Leftarrow$', it suffices to establish that for 
two formulas $a,b \in \Lmoss_{n}$ we have
\begin{equation}
\label{eq:str}
\D: \;\; \vdnax a \isleq b \mbox{ implies } a \sqleq_{n} b,
\end{equation}
where we use $a \sqleq_{n} b$ to denote that $a \equiv_{n} a\land b$.
The proof of \eqref{eq:str} is by induction on the complexity of the derivation 
$\D$.

We confine ourselves to the most difficult case of the inductive step, namely
where the last applied rule in $\D$ is the cut rule; that is, we assume $\D$ 
to be of the form
\[
\D:\hspace{10mm}
\AXC{$\D_{1}$}
\UIC{$a \isleq c$}
\AXC{$\D_{2}$}
\UIC{$c \isleq b$}
\LL{cut}
\BIC{$a \isleq b$}
\DisplayProof
\]
(This case is the most difficult one since here we may not assume $c$ to be
in $\Lmoss_{n}$.)
Let $m$ be such that $c \in \Lmoss_{m}$, and put $k := \max(m,n)$.
Then inductively, we have $a \sqleq_{k} c$ and $c \sqleq_{k} b$, from which we
easily obtain that $a \sqleq_{k} b$.
But then by the first part of the Proposition, we see that $a \sqleq_{n} b$,
as required.
\end{proof}

\begin{prop}
\label{p:str4}
The relation ${\equiv_{\nax}} \sse \Lmoss \times \Lmoss$ is the kernel of the 
unique $\Boolenb$-quotient map from $\Lmoss$ to $\funV\Minit$.
\end{prop}

\begin{proof}
Define the map $q: \Lmoss \to \funaM^{\om}\bbtwo$ as follows.
Given a formula $a \in \Lmoss$, there is some $n\in\om$ such that $a \in
\Lmoss_{n}$.
Now define
\[
q(a) := i_{n}q_{n}(a)
\]
This is well-defined by the fact that diagram~(\ref{dg:strat2}) commutes and we have
$\ker(q) = {\equiv_{\nax}}$ by Proposition~\ref{p:str3}.

Then by initiality of $\Lmoss$ in $\Boolenb$ it suffices to prove that $q$ is
an algebraic homomorphism.
For the Boolean connectives/operators this is straightforward, and so we leave
this as an exercise for the reader.
For the $\nb$ modality we need to prove that the following diagram commutes:
\begin{equation}
\label{dg:nbhom}
\xymatrix{
  \Tom\Lmoss   \ar[d]_{\T q}   \ar[r]^{\nb^{\Lmoss}} 
& \Lmoss       \ar[d]_{q} 
\\
  \Tom\funU \funaM^{\om}\bbtwo  \ar[r]^{\nb^{\funV\Minit}}
& \funU\funaM^{\om}\bbtwo
}
\end{equation}
In order to prove this, take an arbitrary element $\al\in\Tom(\Lmoss)$.
Without loss of generality, assume that $\al\in\Tom(\Lmoss_{n})$, so that 
$\nb\al \in \Lmoss_{n+1}$.
Then by definition of $q$, we have 
\begin{equation}
\label{eq:nbhom1}
(q \cof \nb^{\Lmoss})(\al) = q(\nb\al) = i_{n+1}q_{n+1}(\nb\al).
\end{equation}
Computing $(\nb^{\funV\Minit}\cof \T q)(\al)$, we first calculate
\begin{eqnarray*}
(\T q)(\al) 
&=& \T i_{n} \big( (\T q_{n})(\al) \big),
\end{eqnarray*}
where $(\T q_{n})(\al)$ belongs to the set $\Tom\funU\funaM^{n}\bbtwo$.
Now we claim that 
for all $\be \in \Tom\funU\funaM^{n}\bbtwo$:
\begin{equation}
\label{eq:yz1}
\nb^{\funV\Minit} (\T i_{n}) \be = i_{n+1} \quot{\funaM^{n}\bbtwo} (\nb\be),
\end{equation}
with $\quot{\funaM^{n}\bbtwo}$ as in Definition~\ref{d:quot}.
To see this, consider the following calculation:

\begin{align*}
\nb^{\funV\Minit} (\T i_{n}) \be 
  &=j_{\om}^{-1}\left(\quot{\funaM^{\om}\bbtwo}\left(
  \nb\left(\T i_{n}\right) (\beta) \right)
  \right)
  &\text{(Remark~\ref{r:VMinit})}
\\&=j_{\om}^{-1}\left(\quot{\funaM^{\om}\bbtwo}\left(
  \left(\Tomnb i_{n}\right) (\nb\beta) \right)
  \right)
  &\text{(definition of $\Tomnb$)}
\\&=j_{\om}^{-1}\left(\quot{\funaM^{\om}\bbtwo}\left(
  \left(\Tnb \funU i_{n}\right) (\nb\beta) \right)
  \right)
  &\text{($\Tnb \funU i_{n} \rst{\Tomnb\funU\funaM^{n}\bbtwo} = \Tomnb i_{n}$)}
\\&=j_{\om}^{-1}  \left(\left(\funaM i_{n} \circ \quot{\funaM^{n}\bbtwo}\right) (\nb\beta)
  \right)
  &\text{(naturality of $\quot{}$)}
\\&= i_{n+1}\quot{\funaM^{n}\bbtwo} (\nb\be)
  &\text{(\dag)}
\end{align*}
where the last equality (\dag) follows by
Proposition~\ref{p:minit}(5).

And so we obtain that 
\begin{equation}
\label{eq:nbhom2}
(\nb^{\funV\Minit}\cof\T q)(\al) = 
i_{n+1}\quot{\funaM^{n}\bbtwo} (\nb(\T q_{n})(\al))
\end{equation}

Thus in order to prove the commutativity of (\ref{dg:nbhom}), by
(\ref{eq:nbhom1}) and (\ref{eq:nbhom2}) it suffices to prove that 
\begin{equation}
\label{eq:nbhom3}
q_{n+1}(\nb\al) = 
\quot{\funaM^{n}\bbtwo} (\nb(\T q_{n})(\al)).
\end{equation}
But this is precisely the content of Proposition~\ref{p:qqq}.
\end{proof}

We can now prove the Stratification Theorem.

\begin{proofof}{Theorem~\ref{t:strat}}
Given the Propositions~\ref{l:str1}, \ref{p:str3} and~\ref{p:str4},
all that is left to do is prove that the following diagram commutes for
each $n \in \om$:
\begin{equation}
\label{dg:stratfin}
\xymatrix{
& \Lmoss_{n}   \ar@{>>}[d]_{q_{n}}  \ar@{^{(}->}[r]^{d_{n}}
& \Lmoss       \ar@{>>}[d]^{m}
\\
& \funaM^{n}\bbtwo \ar@{>->}[r]_{i_n}
& \funV\Minit 
}
\end{equation}
We already saw in the proof of Proposition~\ref{p:str4} that the map 
$q: \Lmoss \to \funaM^{\om}\bbtwo$, defined by putting, for $a \in 
\Lmoss_{n}$,
\[
q(a) \isdef i_{n}(q_{n}(a)),
\]
is the unique Moss homomorphism from $\Lmoss$ to $\funV\Minit$; in other
words, this map $q$ coincides with $m$.
Reformulating this in terms that explicitize the role of the inclusion map 
$d_{n}: \Lmoss_{n} \hookrightarrow \Lmoss$, we obtain that $m(d_{n}(a)) =
q(d_{n}(a)) = i_{n}(q_{n}(a))$.
In other words, the diagram \eqref{dg:stratfin} commutes indeed.
\end{proofof}    
 
As a corollary we obtain that the algebra $\funV\Minit$ is the initial 
algebra in the class of Moss algebras that satisfy the nabla-equations.
This means that we may see $\Minit$ as the \emph{Lindenbaum-Tarski} algebra
of our logic.

\begin{cor}
\label{c:Minit}
Let $\bbB = \struc{B,\neg^{\bbB},\bw^{\bbB},\bv^{\bbB},\nb^{\bbB}}$ be a 
Moss algebra such that $\bbB$ validates every instance of the axioms $(\nb1)$
-- $(\nb3)$.
Then there is a unique morphism $\mng^{*}_{\bbB}: \funV\Minit \to \bbB$
through which the meaning function $\mng_{\bbB}$ factors:
\begin{equation*}
\xymatrix{
& \Lmoss       \ar[dr]_{\mng_{\bbB}}  \ar[r]^{m}
& \funV\Minit  \ar[d]^{\mng^{*}_{\bbB}}
\\
& & \bbB
}
\end{equation*}
\end{cor}

\begin{proof}
An arbitrary element of (the carrier of) $\funV\Minit$ is of the form
$m(a)$ for some formula $a \in \Lmoss$.
We leave it as an exercise for the reader to verify that the following map
\[
\mng^{*}_{\bbB}(m(a)) \isdef \mng_{\bbB}(a)
\]
is well-defined and has the right properties.
\end{proof}

\begin{rem}
In fact, we can show that the functor $\funV$ constitutes an 
\emph{isomorphism} between the category $\Coalg_{\BA}(\funaM)$ and the 
variety of Moss algebras validating the nabla axioms.
We omit the details of this proof.
\end{rem}

\subsection{Proof of soundness and completeness}

We are almost ready to prove our main result.
What is left to do is link the final $\T$-sequence to the initial
$\funaM$-sequence.
Recall that the elements of $\T^{n}1$ intuitively correspond to the
$n$-behaviors associated with $\T$, and that $\Minit$, the initial 
$\funaM$-algebra, is the colimit of the initial sequence 
$\struc{\funaM^n\mathbbm{2},j_{n}}_{n<\om}$, where elements of 
$\funaM^n\mathbbm{2}$ correspond to (equivalence classes of)
formulas of depth $n$.

\begin{definition}
We define the sequence of maps $\finsem_n:\funaM^n\mathbbm{2}\to \funaQ\T^n 1$
as follows.
The map $\finsem_0: \bbtwo \to \funaQ 1$ is given by initiality (and is actually
the identity).
For the definition of $\finsem_{n+1}$, recall from Defintion~\ref{d:ntrde} that $\delta_{\T^{n}1}:
\funaM\funaQ \T^{n}1 \to \funaQ \T^{n+1} 1$, and assume inductively that
$\finsem_{n}: \funaM^{n}\bbtwo \to \funaQ \T^{n}1$ has been defined, so that
$\funaM \finsem_{n} : \funaM^{n+1}\bbtwo \to \funaM\funaQ \T^{n}1$.  Composing
these two maps, we obtain $\finsem_{n+1} := \delta_{\T^n 1}
\cof\funaM(\finsem_n)$. \end{definition}

Intuitively, the reader may think of the map $s_{n}$ as providing semantics 
of elements of $\funaM^{n}\bbtwo$.
This can be made more precise by proving that the following diagram 
commutes:
\begin{equation*}
\xymatrix{
  \Lmoss_{n}   \ar[dr]_{\mng_{n}} \ar[r]^{q_{n}} 
& \funaM^{n}\bbtwo \ar[d]^{s_{n}}
\\
& \funaQ\T^{n}1
}
\end{equation*}
Here $q_{n}$ is the quotient map under $n$-step derivability of
Theorem~\ref{t:strat} and $\mng_{n}$ is the $n$-step meaning function of 
Definition~\ref{d:nstepsem}.

From this perspective, the following proposition states that the semantics 
of a formula with respect to the final sequence is independent of the
particular approximant we choose. 

\begin{prop}
\label{p:fin1}
The following diagram commutes:
\begin{equation}\label{finalsequence}
  \xymatrix{
    \funaQ 1\ar[r]^{\funaQ h_0} & \ldots
    & \funaQ\T^n 1\ar[r]^{\funaQ h_n}
    & \funaQ\T^{n+1} 1 & \ldots\\
    \mathbbm{2}\ar[u]_{\finsem_0}\ar[r]_{j_0} & \ldots
    & \funaM^n\mathbbm{2}\ar[u]_{\finsem_n}\ar[r]_{j_n}
    & \funaM^{n+1}\mathbbm{2}\ar[u]_{\finsem_{n+1}} & \ldots
  }
\end{equation}
In addition, each map $s_{n}$ is injective.
\end{prop}

\begin{proof}
In order to show that diagram (\ref{finalsequence}) commutes, we will prove
that
\[
\finsem_{n+1} \cof j_n = \funaQ h_n \cof \finsem_n
\]
for all $n \in \omega$ .
The proof is by induction on $n$.
The base case 
$\finsem_1 \cof j_0 = \funaQ h_0 \cof \finsem_0$ is a consequence of the fact
that $\mathbbm{2}$ is the initial object in $\BA$. 
For the inductive case, where $n=k+1$ for some $k \in \omega$, we reason as
follows:
\begin{align*}
  \finsem_{k+2} \cof j_{k+1} 
  &= \delta_{\T^{k+1} 1} \cof \funaM (\finsem_{k+1}) \cof \funaM(j_k)
  &  \text{(unfolding definitions)}
\\&= \delta_{\T^{k+1} 1} \cof \funaM (\finsem_{k+1} \cof j_k) 
  &  \text{(functoriality of $\funaM$)}
\\&= \delta_{\T^{k+1}1} \cof \funaM( \funaQ h_k \cof \finsem_k)
  &  \text{(inductive hypothesis)}
\\&= \funaQ \T h_k \cof \delta_{\T^k 1} \cof \funaM(\finsem_k) 
  &  \text{(naturality of $\de$)}
\\&= \funaQ h_{k+1} \cof \finsem_{k+1} 
  &  \text{(definition $\finsem_{k+1}$)}
\end{align*}
Since $\delta$ is injective (Proposition~\ref{p:3}) and $\funaM$ preserves
embeddings (Proposition~\ref{p:Mfinemb}), a straightforward inductive proof
shows that all $\finsem_n$, $n\in\omega$, are injective.
\end{proof}

We are now going to demonstrate that the coalgebraic semantics and the 
semantics via the final sequence coincide.

\begin{prop}
\label{p:fin2}
For a given coalgebra $\bbX = \struc{X,\xi}$ and any formula $a \in \Lmoss_{n}$,
the following holds:
\begin{equation}\label{equ:semantics_stratify}
\mng_{\bbX} (a)=\xi_n^{-1}(\finsem_n(q_n(a))), \qquad \mbox{ for all } 
a \in \Lmoss_n \text{ and }n \in \omega.
\end{equation}
\end{prop}

\begin{proof}
First note that $\funaQ X$ together with the maps $\funaQ 
\xi_n \cof \finsem_n =\xi_n^{-1} \cof \finsem_n$ form a cocone over the initial
sequence of $\funaM$.
Therefore there is a mediating arrow 
\[
\mng^*_\bbX: \funaM^\omega \mathbbm{2} \to \funaQ X
\]
from the carrier of the initial $\funaM$-algebra $\Minit$ to $\funaQ X$ with
the property that $\mng^*_\bbX \cof i_n = \xi_n^{-1} \cof \finsem_n$.

We claim that 
\begin{equation}
\label{eq:minit}
\text{the map $\mng^*_\bbX$ is an $\funaM$-algebra morphism from $\Minit$
to $\bbX^{*}$.}
\end{equation}
In order to prove \eqref{eq:minit}, observe that by Proposition~\ref{p:minit},
for all $n \in \omega$ we have
\begin{equation}\label{equ:Minitinvmap}
 j_\omega \cof i_{n+1} = \funaM (i_n),
 \end{equation}
where $j_\omega: \funaM^\omega \mathbbm{2} \to \funaM \funaM^\omega \mathbbm{2} $ 
is the inverse of the algebra structure map $\hs^\Minit$ of the initial 
$\funaM$-algebra.
In order to prove the claim it suffices 
to show that the following diagram
commutes
\[ \xymatrix{
\funaM \funaM^\omega \mathbbm{2}  \ar[rr]^{\funaM \mng^*_\bbX} 
  & & \funaM \funaQ X \ar[d]^{\delta_X}
\\& & \funaQ T X \ar[d]^{\funaQ \xi} 
\\ \funaM^\omega \mathbbm{2}  \ar[uu]^{j_\omega} \ar[rr]_{\mng^*_\bbX}  
  & & \funaQ X 
} 
\]
We prove that the diagram commutes by showing that 
$f \coloneqq \funaQ \xi \cof \delta_X \cof \funaM (\mng^*_\bbX) \cof j_\omega$ is a
mediating arrow from $(\funaM^\omega \mathbbm{2} , \{i_n\}_{n \in \omega})$
to $(\funaQ X, \{\funaQ \xi_n \cof \finsem_n \}_{n \in \omega})$. 
Therefore $f$ has to be equal to $\mng^*_\bbX$ by the universal property of the colimit
$(\funaM^\omega \mathbbm{2} , \{i_n\}_{n \in \omega})$. 
We show that $f$ has the claimed property by proving that for all $n \in \omega$
we have
\begin{equation}\label{equ:malgmorph} 
 \funaQ (\xi_{n}) \cof s_{n}   = 
f \cof i_n
\end{equation}
For $n=0$ the equation holds by initiality of $\mathbbm{2}$.
Furthermore for an arbitrary $n \geq 0$ we have
\begin{align*}
\funaQ (\xi_{n+1}) \cof s_{n+1}  
  & = \funaQ (\T \xi_n \cof \xi) \cof \delta_{\T^n 1} \cof \funaM s_n  
  & \text{(definition of $\xi_{n+1}$ and of $\finsem_{n+1}$)} 
\\& = \funaQ (\xi) \cof \funaQ(\T \xi_n) \cof \delta_{\T^n 1} \cof \funaM s_n  
  & \text{(functoriality of $\funaQ$)}
\\& =\funaQ(\xi) \cof \delta_X \cof \funaM\funaQ \xi_n \cof \funaM s_n  
  &\text{(naturality of $\delta$)}
\\& = \funaQ(\xi) \cof \delta_X \cof \funaM(\funaQ \xi_n \cof s_n)
  &\text{(functoriality of $\funaM$)}
\\& = \funaQ(\xi) \cof \delta_X \cof \funaM(\mng^*_\bbX \cof i_n) 
  &\text{($\mng^*_\bbX$ mediating arrow)} 
\\& = \funaQ (\xi) \cof \delta_X \cof \funaM \mng^*_\bbX \cof \funaM i_n 
  &\text{(functoriality of $\funaM$)}
\\& = \funaQ (\xi) \cof \delta_X \cof \funaM \mng^*_\bbX \cof j_\omega \cof i_{n+1} .
  &\text{(equation \eqref{equ:Minitinvmap})} 
\end{align*}
Therefore equation (\ref{equ:malgmorph}) holds for all $n$, which finishes
the proof of \eqref{eq:minit}.

From this it follows that $\funV\mng^{*}: \funV\Minit \to \funV\bbX^{*}$ is 
a Moss algebra homomorphism.
Recalling from Proposition~\ref{p:plusstar} that $\funV\bbX^{*} = \bbX^{+}$,
we obtain by initiality of $\Lmoss$ as a Moss algebra, that $\funV \mng^*_\bbX 
\cof m = \mng_{\bbX}$. 
Here $\mng_{\bbX}: \Lmoss \to \bbX^{+}$ is the unique Moss algebra homomorphism
that maps an element of $\Lmoss$ to its semantics in $\bbX^{+}$, and
$m \isdef \mng_{\funV\Minit}$ is the unique homomorphism $m: \Lmoss \to
\funV\Minit$ in the category of Moss algebras.
But then by the Axiomatic Stratification Theorem~\ref{t:strat},
for all $n \in \omega$ and all formulas $a \in \Lmoss_n$ we have
$\mng_{\bbX}(a) = \mng^*_\bbX(m(a)) = \mng^*_\bbX(i_n(q_n(a))) = 
\funaQ \xi_n \cof \finsem_n(q_n(a))$, where the last identity holds by
the definition of $\mng^{*}_\bbX$ as a mediating arrow.
This shows that
(\ref{equ:semantics_stratify}) holds, and finishes the proof of the claim.
\end{proof}

On the basis of the results obtained so far, the proof of our soundness and
completeness results is now more or less immediate.

\begin{proofof}{Theorem~\ref{t:main}}
Let $a$ and $b$ be two formulas in $\Lmoss$.
Fix a natural number $n$ such that $a,b \in \Lmoss_{n}$.
Recall that $\bbF_{n} = \struc{\T^{n}1,\T^{n}g}$ denotes the `$n$-step 
coalgebra' defined in Definition~\ref{d:ncoalg}.

Now consider the following sequence of equivalences:
\begin{align*}
a \sqleq_{\nax} b
  & \iff q_{n}(a) \sse q_{n}(b)
  & \text{(Axiomatic Stratification Theorem~\ref{t:strat})} 
\\& \iff \finsem_{n}q_{n}(a) \sse \finsem_{n}q_{n}(b)
  & \text{(injectivity of $\finsem_{n}$)} 
\\& \iff (\funaQ(\T^{n}g)_{n})(\finsem_{n}q_{n}(a)) \sse
           (\funaQ(\T^{n}g)_{n})\rlap{$(\finsem_{n}q_{n}(b))$}
  & \text{(equation~\eqref{eq:gh})} 
\\& \iff \mng_{\bbF_{n}}(a) \sse \mng_{\bbF_{n}}(b)
  & \text{(Proposition~\ref{p:fin2})} 
\\& \iff a \models_{\T} b
  & \text{(Semantic Stratification Theorem~\ref{t:strs})} 
\end{align*}
From this the Theorem is immediate.
\end{proofof}


%% file: sec-conclusion.tex
\section{Conclusions}
\label{s:conclusion}

\paragraph{Summary of results}
Obviously, as the main contributions of this paper we see the definition of 
the \emph{derivation system $\nax$} for the finitary version of Moss'
coalgebraic logic, the result stating that $\nax$ provides a \emph{sound and
complete} axiomatization for the collection of coalgebraically valid 
inequalities, and the fact that all of our definitions, results and our proofs
are completely \emph{uniform} in the coalgebraic type functor $\T$

Our proof of the soundness and completeness theorem is rather elaborate and 
technical, but we believe that the effort has been worth the while, and 
that on the way we have identified some new concepts and obtained some
auxiliary results that may be of independent interest. 
Of these we list the following:
\begin{enumerate}[(1)]
\item
a survey of the properties of the notion $\Tb$ of relation lifting, induecd
by an arbitrary but fixed set functor $\T$ (section~\ref{s:relationlifting});
\item
the introduction in Definition~\ref{d:Prs} of the category $\Prs$ of Boolean
algebra presentations, and the establishment in Theorem~\ref{t:BCadj} of an
adjunction between $\Prs$ and the category $\BA$ of Boolean algebras;
\item
the introduction in section~\ref{ss:functorM} of the functor $\funaM: \BA \to
\BA$, and the results in Proposition~\ref{p:Mfinemb} that $\funaM$ is finitary
and preserves embeddings, and in Theorem~\ref{l:1st2} that it preserves
atomicity of Boolean algebras.
\item
the stratification of our logic, both semantically (Theorem~\ref{t:strs})
and syntactically (Theorem~\ref{t:strat});
\item
the identification, in Corollary~\ref{c:Minit}, of the initial $\funaM$-algebra 
$\Minit$, through the functor $\funV$, as the Lindenbaum-Tarski algebra of our
logic.
\end{enumerate}

\paragraph{Related and ongoing work}

As mentioned in the introduction, this paper replaces, extends and partly 
corrects an earlier version~\cite{kupk:comp08}.
Since the publication of the latter paper, and the preparation of the current
manuscript there have been a number of developments in the area of Moss'
logic that we would like to mention here.
First of all, based on our one-step soundness and completeness results,
Bergfeld gave a more direct version of our completeness proof in his MSc
thesis~\cite{berg:moss09}; as a corollary he established a strong completeness
theorem for Moss' logic (modulo some restrictions on the functor $\T$).
Second, B\'{\i}lkov\'a, Palmigiano \& Venema generalized their earlier result
on the power set nabla~\cite{bilk:proo08} to the general case of a standard,
weak pullback preserving functor $\T$: in~\cite{bilk:proo10} they provide a
sound, complete, and 
cut-free proof system for (the finitary version of) Moss' coalgebraic logic.
Systematically using Stone duality, Kurz \& Leal~\cite{kurz:modaxx} make
a detailed comparison between Moss' approach towards coalgebraic logic, and 
the one based on associating standard modalities with predicate liftings; 
their main contribution is a new coalgebraic logic combining features of both
approaches. 
Venema, Vickers \& Vosmaer~\cite{vene:powe10} study a variant of the 
derivation system $\nax$ in the setting of geometric logic; their main 
contribution is to generalize Johnstone's power construction on locales, to 
a functor $V_{\T}$, parametrically defined in a set functor $\T$, on the 
category of locales.
Finally, B\'{\i}lkov\'a, Velebil \& Venema~\cite{bilk:monoxx} prove that on
the (semantically defined) Lindenbaum-Tarski algebra of our logic, the nabla
modality has the interesting order-theoretic property of being a so-called 
$\mathcal{O}$-adjoint.

\paragraph{Future research}
We finish with mentioning some directions for future research.
To start with, in this paper we have studied the nabla operator in the
setting of the diagram~\eqref{diag:duality}, which is a particular 
instantiation of the general Stone duality diagram
\begin{equation}
\label{diag:Stonegen}
\xymatrix{
\Alg \ar@(dl,ul)[]^{L} \ar@/_/[r]_{S}
  & {\mathsf{Sp}^{\mathrm{op}}} \ar@/_/[l]_{P} \ar@(dr,ur)[]_{T}
}
\end{equation}
where $\Alg$ denotes a category of algebras representing the base logic, 
$\mathsf{Sp}$ is a category of spaces representing the semantics of the 
logic, $\T$ is the coalgebra functor representing all one-step behaviours,
and $L$ represents the one-step version of the coalgebraic modal logic.
Given the flexibility of the Stone duality approach we believe it to be of
interest to consider more instances of the diagram~\eqref{diag:Stonegen}
where $L$ is some version of our nabla logic.
Of particular interest are the cases where for $\Alg$ we take the variety of
distributive lattices, because this could clarify the role of the negation
in our setting.

Second, a clear drawback of the current nabla-based approach towards 
coalgebraic logic is the restriction to functors that preserve weak
pullbacks.
It would therefore be interesting to see whether this restriction can be 
removed.
A first step in this direction has been made by Santocanale \& 
Venema~\cite{sant:unif10}, who introduce a nabla-based version of monotone
modal logic, a variant of basic modal logic that is naturally interpreted 
in coalgebras for the monotone neighborhood functor of Example~\ref{ex:1} 
--- a functor that does not preserve weak pullbacks.

Finally, in the introduction we mentioned that the work of Janin \& 
Walukiewicz~\cite{jani:auto95} on automata theory and modal fixpoint logics 
is an independent source for the introduction of the cover modality 
$\nb_{\!\funP}$ as a primitive modality.
Since $\nb_{\!\funP}$ also plays a fundamental role in Walukiewicz'
completeness result for the modal $\mu$-calculus~\cite{walu:comp00}, this
naturally raises the question whether we can extend our completeness result
to the setting with fixpoint operators.

%% file: sec-appendix.tex
\section{Appendix: overview of notation}
\label{s:appendix}

Since this paper features a multitude of categories, functors and natural
transformations, for the reader's convenience we list these in the tables
below.
\bigskip

\begin{center}

\begin{tabular}{c}
\begin{tabular}{|ll|}\hline
   \multicolumn{2}{|c|}{Categories}
\\ \hline
   $\BA$           & section~\ref{ss:basics1}
\\ $\Boole$        & Definition~\ref{d:Boole}
\\ $\Prs$          & Definition~\ref{d:Prs}
\\ $\Set,\Rel$     & section~\ref{ss:basics1}
\\ $\Boolenb$ & Definition~\ref{d:Mossfun}
\\ \hline
\end{tabular}
\\[20mm]
\begin{tabular}{|ll|}\hline
   \multicolumn{2}{|c|}{Natural Transformations}
\\ \hline
   $\Base^{\T}: \Tom \ntrto \Pom$          & Definition~\ref{d:base}
\\ $\nbsem: \T\funP \ntrto \funP\T$        & Definition~\ref{def:elementlift}
\\ $\quot{}: \Tba\funU \ntrto \funaM$      & Definition~\ref{d:quot}
\\ $\de: \funaM\funaQ \ntrto \funaQ\T$  & Definition~\ref{d:ntrde}
\\ \hline
\end{tabular}
\\[12mm]
\end{tabular}
\begin{tabular}{|ll|}\hline
   \multicolumn{2}{|c|}{Functors}
\\ \hline
   $\funB: \Prs \to \BA$       & Definition~\ref{d:funB1}, \ref{d:funB2}
\\ $\funC: \BA \to \Prs$       & Definition~\ref{d:funC}
\\ $\funaF: \Set \to \Boole$      & page~\pageref{d:funaF}  
\\ $\Id,\Bag_{\om},D_{\om}: \Set \to \Set$       & Example~\ref{ex:1}
\\ $\Mossfun: \Set \to \Set$   & Definition~\ref{d:Mossfun}
\\ $\funpM: \Prs \to \Prs$     & Definition~\ref{d:funpM}  
\\ $\funaM: \BA \to \BA$       & Definition~\ref{d:funaM}  
\\ $\Tba: \Set \to \Set$       & Definition~\ref{d:BAsyntax}  
                                   \&~\eqref{eq:Tba}
\\ $\Tnb: \Set \to \Boole/\Set$   & Definition~\ref{d:Tnb}  
\\ $\funP,\Pom: \Set \to \Set$ & section~\ref{ss:basics1}
\\ $\funQ: \Set \to \Set^{\mathit{op}} $ & section~\ref{ss:basics1}
\\ $\funaQ: \Set \to \BA^{\mathit{op}} $ & Definition~\ref{d:funaQ}
\\ $\Tom:\Set \to \Set$        & page~\pageref{page:Tom}
\\ $\Tomnb:\Set \to \Set$      & Definition~\ref{d:Tomnb}
\\ $\funU:\Boole \to \Set$     & page~\pageref{d:funU} 
\\ $\funV:\Alg_{\BA}(\funaM) \to \Alg_{\Set}(\Mossfun)$    
                               & Definition~\ref{d:funV}
\\ \hline
\end{tabular}
%
%
%
\end{center}